# Role of core-collapse supernovae in explaining Solar System abundances of $p$ nuclides

C. Travaglio[1,2]

[1]INFN - Istituto Nazionale Fisica Nucleare, Turin, Italy

[2]B2FH Association - Turin, Italy

travaglio@oato.inaf.it, claudia.travaglio@b2fh.org

T. Rauscher[3,4,5]

[3]Department of Physics, University of Basel, Switzerland

[4]Centre for Astrophysics Research, University of Hertfordshire, United Kingdom

[5]UK Network for Bridging Disciplines of Galactic Chemical Evolution (BRIDGCE), https://www.bridgce.net

A. Heger[6,7,8,9,10]

[6]Monash Centre for Astrophysics, Monash University, Melbourne, Victoria, 3800, Australia

[7]Astronomy Department, Shanghai Jiao Tong University, Shanghai 200240, China

[8]School of Physics and Astronomy, University of Minnesota, Minneapolis, Minnesota 55455, USA

[9]The NuGrid Collaboration, http://www.nugridstars.org

[10]Joint Institute for Nuclear Astrophysics—Center for the Evolution of the Elements, USA

M. Pignatari[11,9,10]

[11]E.A. Milne Centre for Astrophysics, University of Hull, HU6 7RX, United Kingdom

and

C. West[12,8,10]

[12]Center for Academic Excellence, Metropolitan State University, St. Paul, MN, 55106, USA

## ABSTRACT

The production of the heavy stable proton-rich isotopes between $^{74}$Se and $^{196}$Hg – the $p$ nuclides – is due to the contribution from different nucleosynthesis processes, activated in different types of stars. Whereas these processes have been subject to various studies, their relative contributions to Galactic Chemical Evolution (GCE) are still a matter of debate. Here we investigate for the first time the nucleosynthesis of $p$ nuclides in GCE by including metallicity and progenitor mass-dependent yields of core-collapse supernovae (ccSNe) into a chemical evolution model. We used a grid of metallicities and progenitor masses from two different sets of stellar yields and followed the contribution of ccSNe to the Galactic abundances as function of time. In combination with previous studies on $p$-nucleus production in thermonuclear supernovae (SNIa), and using the same GCE description, this allows us to compare the respective roles of SNeIa and cc-SNe in the production of $p$-nuclei in the Galaxy. The $\gamma$ process in ccSN is very efficient for a wide range of progenitor masses ($13\,\mathrm{M}_\odot - 25\,\mathrm{M}_\odot$) at solar metallicity. Since it is a secondary process with its efficiency depending on the initial abundance of heavy elements, its contribution is strongly reduced below solar metallicity. This makes it challenging to explain the inventory of the $p$ nuclides in the Solar System by the contribution from ccSNe alone. In particular, we find that ccSNe contribute less than 10 % of the solar $p$ nuclide abundances, with only a few exceptions. Due to the uncertain contribution from other nucleosynthesis sites in ccSNe, such as neutrino winds or $\alpha$-rich freeze out, we conclude that the light $p$-nuclides $^{74}$Se, $^{78}$Kr, $^{84}$Sr, and $^{92}$Mo may either still be completely or only partially produced in ccSNe. The $\gamma$-process accounts for up to twice the relative solar abundances for $^{74}$Se in one set of stellar models and $^{196}$Hg in the other set. The solar abundance of the heaviest $p$ nucleus $^{196}$Hg is reproduced within uncer-tainties in one set of our models due to photodisintegration of the Pb isotopes $^{208,207,206}$Pb. For all other $p$ nuclides, abundances as low as 2 % of the solar level were obtained.

## 1. Introduction

The pioneering works of Cameron (1957) and Burbidge et al. (1957) realized that the production of 35 stable nuclides between $^{74}$Se and $^{196}$Hg on the proton-rich side of the valley of stability, called $p$ nuclides, cannot proceed via the $s$ and $r$ neutron-capture processes required for the synthesis of the bulk of the remaining nuclides beyond Fe (for this reason they were also called *excluded isotopes* by Cameron 1957). Their astrophysical origin is still under debate. Overviews of possible production sites, observations and uncertainties, have been presented by various authors (e.g., Arnould & Goriely (2003); Rauscher et al. (2013); Pignatari et al. (2016a), and references therein).

Solar System $p$ abundances have been derived from geological and meteoritic data. Understanding the origin of the $p$ nuclides is challenging because they cannot be directly observed in stars and supernova remnants, as their contribution to elemental abundances is small and no element is dominated by a $p$ isotope. The synthesis of $p$ nuclei has to be studied in models without the possibility of direct verification, with the exception of presolar stellar dusts. Signatures of enrichments in proton-rich isotopes compared to the solar composition and/or with respect to more neutron-rich isotopes of the same element have been identified for Xe in bulk measurements from presolar nano-diamonds (Xe-HL component, e.g., Lewis et al. 1987), and more recently for Mo and Ru possibly in single SiC-X grains (e.g., Pellin et al. 2006, Pignatari et al. 2016a and references therein) and SiC AB grains (Savina et al. 2003). They all show, however, a non-solar pattern, likely carrying the signature of not well-mixed ejecta from single core-collapse supernovae (ccSNe). On the other hand, terrestrial and meteoritic $p$ abundances have to be derived from galactic chemical evolution (GCE) models, integrating the production of different sites over the history of the Galaxy. The solar composition might also not be representative of the average galactic composition as calculated in GCE models.

It is the current paradigm that the largest fraction of $p$ nuclides is synthesized by photodisintegrations of pre-existing seed nuclei and subsequent $\beta$ decays, the so-called $\gamma$ process (see, e.g., Woosley & Howard 1978; Rayet et al. 1990). This would occur during explosive O/Ne burning during ccSN explosions. Some authors (e.g., Woosley & Howard 1990; Woosley & Weaver 1995; Rayet et al. 1995; Rauscher et al. 2002) claimed that these sources could reproduce the solar abundances of the intermediate and heavy $p$ nuclides within factors of $2 - 3$. Rauscher et al. (2002) presented detailed nucleosynthesis calculations in massive stars from the onset of central H-burning through the supernova explosion for Population I stars with progenitor masses between $15 \leq M/\mathrm{M}_{\odot} \leq 25$. Serious deficiencies in the production of Mo and Ru isotopes were found in that work, consistent with earlier studies (e.g., Woosley & Howard 1978; Woosley & Howard 1990; Woosley & Weaver 1995; Rayet et al. 1995). Neutrino-induced spallation reactions or weak interactions were found to be efficient only for a few isotopes (e.g., Heger et al. 2005, Austin et al. 2014). Ritter et al. (2017) recently considered the impact of CO-shell mergers on the synthesis of $p$ nuclei, including multi-dimensional hydrodynamics simulations for the onset and evolution of the merging shells, confirming enhancements of $p$ nuclide yields up to about an order of magnitude. What still needs to be studies in detailed supernova simulation is what fraction of the $p$ process is “relocated” or “reset” in the supernova explosion – much of the $p$ process made in pre-supernova stage is located in deep layers at low radii is destroyed by the heat of supernova shock and then “re-created” further out where temperature during the shock becomes adequate to make these $p$ nuclei but not destroy then (Heger et al. 2005), only portions of the $p$ that is transported outward may survive from the pre-supernova stage. The occurrence of shell mergers and their relevance in term of GCE need to be studied in more detail, taking into account also its strong impact on intermediate-mass elements like Cl, K, and Sc. Its relevance for GCE when integrating over the entire initial mass function (hereafter IMF) is uncertain. However, it is interesting that CO-shell mergers would increase the production of both p-nuclei and intermediate mass elements like K and Sc, underproduced by previous galactic chemical evolution simulations compared to observations (Kobayashi et al. 2011).

More recently, $p$ nucleosynthesis calculations in ccSNe have been performed by Farouqi et al. (2009). These authors re-calculated $p$ nucleosynthesis in the explosion of a $15\,M_\odot$ progenitor and compared with Rauscher et al. (2002). Farouqi et al. (2009) claim that their yields of the $p$ nuclei up to the Ru region reproduced quite well the solar System composition. They further claim that in their model no initial solar, $s$, or $r$ process seed composition was introduced and that thus the obtained nucleosynthesis result is primary. Similar to the neutrino-wind model of Hoffman et al. (1996) and the more recent electron-capture SN model of Wanajo et al. (2009), the $\alpha$-component of their high-entropy wind model is a primary process. Their conclusion is that the classical light $p$ nuclei production does not require any assumptions about the initial composition of the SN progenitor star. Hayakawa et al. (2006, 2008) analyzed the Solar System distribution of the $p$ nuclides, deriving phenomenological constraints without the use of GCE simulations.

Common to all previous work on GCE is that the contribution of different stellar generations to the Solar System composition was not taken into account. It was also not pondered that the solar composition may not be representative of the average galactic composition which is calculated in chemical evolution models. It has been shown that a large variation in ejected $p$ nuclei is found even between ccSNe models from progenitors with the same initial metallicity but different initial mass (e.g., Pignatari et al. 2016a). Therefore, GCE simulations are required in order to compare ccSNe yields with the solar system distribution.

So far it has not been demonstrated that it is possible to reproduce the solar abundances of all $p$ nuclides by a single stellar process. For instance, ccSN models suffer from a strong underproduction of the most abundant $p$ nuclides, $^{92,94}$Mo and $^{96,98}$Ru. Alternative processes and sites for the production of these nuclei have been proposed by many authors, e.g., a $\nu$p-process in the deepest layers of ccSN ejecta and in neutrino driven winds of ccSN (Fröhlich et al. 2006; Pruet et al. 2006; Farouqi et al. 2009; Wanajo 2006; Roberts et al. 2010; Wanajo et al. 2011a, 2011b; Arcones & Janka 2011; Arcones & Montes 2011; Fischer et al. 2011) or rapid proton-captures in hot, proton-rich matter accreted onto the surface of a neutron star (e.g., Schatz et al. 2001). It has been known for a long time that the $\nu$ process in ccSN contributes to the abundances of $^{138}$La and $^{180m}$Ta (Woosley & Howard 1990; Arnould & Goriely 2003; Heger et al. 2005, Rauscher et al. 2013).

Production of $p$ nuclides has also been suggested to occur in the outermost layers of thermonuclear supernovae (Type Ia supernovae, hereafter SNIa): in the Chandrasekhar-mass white dwarf delayed-detonation model (Howard & Meyer 1993: Travaglio et al. 2011, hereafter TRV11), in the sub-Chandrasekhar-mass white dwarf He-detonation model (Goriely et al. 2005; Arnould & Goriely 2006), and in the carbon-deflagration model (Kusakabe et al. 2011). The assumed $s$ process seed distribution plays a fundamental role for the $p$ production in these models and different assumptions are made by the various authors.

Travaglio et al. (2015) (hereafter TRV15) explored single-degenerate SNIa in the framework of two-dimensional delayed-detonation models, with $s$ process seeds at different metallicities in the WD as well as in the accreted material on the surface of the WD. Travaglio et al. (2014) performed a similar study for radiogenic $p$ nuclides. Both investigations demonstrated that explosions of Chandrasekhar-mass single-degenerate systems can provide a considerable contribution to the Solar System composition contributing to a large amount (more than 50 %) of $p$ nuclei in our Galaxy, both in the range of light ($A \leq 120$) and heavy $p$ nuclides, at almost flat average production factors (within a factor of about three).

Exceptions are the lightest $p$ nuclides, $^{74}$Se and $^{78}$Kr, for which a low production efficiency was obtained.

A major challenge for the single-degenerate SNIa scenario comes from the difficulty to reach the Chandrasekhar mass by accretion from a typical $0.6\,M_\odot$ WD. At present there are no one-dimensional stellar models that were successfully reaching that critical mass limit, once He-fusion runaway and the consequent mass loss was considered (e.g., Cassisi et al. 1998, Denissenkov et al. 2017). At least one Chandrasekhar-mass SNIa object have been identified observationally, based on the element ratios observed in the ejecta and the required extreme conditions to produce those same ratios (Yamaguchi et al. 2015).

Coupling $p$ nucleosynthesis to chemical evolution models, several questions have to be dealt with: the occurrence frequency of the astrophysical sources, their spatial distribution and yields, dependence of the yields on metallicity, and mixing of the nucleosynthesis products with the interstellar medium. There are insufficient observables for $p$ nuclides to constrain GCE models or even determine single production sites, as the isotopic abundances of $p$ nuclides cannot be separately determined in stellar spectra. This underlines the importance of analyzing meteoritic material. Combining the isotopic information, e.g., of extinct radioactives, with GCE predictions allows to put severe constraints on the possible astrophysical sources and nucleosynthetic processes (see, e.g., Travaglio et al. 2014; Lugaro et al. 2016 and references therein). Meteoritic isotope anomalies also have been reported for the $p$ nuclides $^{184}$Os (Reisberg et al. 2009) and $^{180}$W (Schulz, Muenker,& Peters 2013 and references therein) but further work is needed before their origin can be understood.

In this work we investigate the production of $p$ nuclides in ccSNe on a grid of masses and metallicities, using two different sets of stellar data, one set of calculations obtained with a recent version of the KEPLER code (Weaver et al. 1978; Rauscher et al. 2002; Heger & Woosley 2010; West & Heger 2013) and the NuGrid data set (Pignatari et al. 2016a).

The paper is organized as follows. In Section 2, we present the set of KEPLER models and discuss the role of different masses and metallicities for the production of $p$ nuclei. A similar discussion for the NuGrid models is presented in Section 3. Section 4 introduces the GCE model and discusses the results obtained with the nucleosynthesis sets of the previous sections and their implications for the role of ccSN contributions to the Solar System $p$ abundances. Section 5 briefly considers possible uncertainties in the GCE model as well as in the nuclear input to the stellar models which could affect our conclusions. Finally, conclusions and an outlook on work in progress are presented in Section 6.

## 2. Production of $p$ nuclides in the KEPLER models

In this work we use non-rotating stellar models of West & Heger (2018, in preparation) computed with the KEPLER stellar evolution, nucleosynthesis, and SN code (Weaver, Zimmerman & Woosley 1978; Rauscher et al. 2002). The progenitor models were calculated using the physics setup, opacities, and nuclear reaction rates as described in Woosley & Heger (2007) and in West, Heger & Austin (2013). The nucleosynthetic yields for seven different initial masses, from $13\,\mathrm{M}_{\odot}$ to $30\,\mathrm{M}_{\odot}$, and 14 different metallicities, from $Z = 1.5 \times 10^{-6}$ up to $Z = 0.3$ were calculated (see also Table 1 for Model *xi45*, i.e., one of the models considered in this work). The initial composition of the models used the Galactic Chemical History model of West & Heger (2013) which is based on Lodders (2009) solar abundances. Details of KEPLER models will be published in a forthcoming paper (West & Heger 2018, in preparation), including the impact of choosing proper metallicity-dependent initial compositions. We exploded all KEPLER models using a piston model such that a final kinetic energy of the ejecta of 1.2 B ($1\ B = 10^{51}$ erg) was achieved (see, e.g., Rauscher et al. 2002; Heger & Woosley 2010), and we computed fallback based on the 1D hydrodynamics in KEPLER. This also included mixing prior to fallback using our “standard” mixing width of 0.1 helium core masses as outlined in the references before. The explosion energy and mixing used are a good match for typical supernovae such as SN 1087A (though some light curve fitting may suggest larger mixing for SN 1987A, see., e.g., Utrobin et al. 2015). In nature, we would anticipate significant variation of explosion energies for the different models (e.g., Müller et al 2016) rather than a fixed value, however, for the models in the mass range studies here, determination of reliable explosion energies from simulation or first principle for a given progenitor model is not possible at the present. Whether a model would actually explode of collapse to a black hole we then decided independently of that based on the different criteria listed below.

Different criteria to determine whether a successful explosion would actually occur have been employed, based on different criteria in the literature for the explodability given the pre-SN structure of the star at onset of core collapse. The first simple case was to assume that all stars explode (*no-cutoff* model). In this case, the entire non-fallback mass of all stars including winds contributes to the yields. Next, we explored different prescriptions for explodability based on formulae from the literature. As one choice we used the compactness parameter $\xi$ of 0.25 (model *xi25*) as suggested by O'Connor & Ott (2011) and 0.45 (Model *xi45*) as suggested by Sukhbold & Woosley (2014; we chose this model for Table 1 and for the figures in this paper), as possible cut-offs for black hole formation, as well as the prescription by Ertl et al. (2016; model *ertl*). The description by Ertl et al. gives cut-offs for black hole formation not too dissimilar to the more refined semi-analytic model by Müller et al. (2016), as shown by Sukhbold et al. (2017), hence we do not need to explore it here separately. When the criteria for black-hole formation were fulfilled, we assumed that the entire star would collapse to a black hole without providing further nucleosynthesis. We neglect the possibility that some of the outer layers of the star may still escape due to neutrino losses prior to black hole formation (e.g., Lovegrove & Woosley 2013). Only the contribution from mass loss due to winds prior to collapse would be present in this case. In the cases for which no black hole is formed, the full yields are used. See also discussions in West & Heger (2013) and in Côté et al. (2016). In Table 3 we provide the results from the chemical evolution calculations for all the mentioned models mentioned above.

A large, adaptive nuclear reaction network allowed us to follow nucleosynthesis self-consistently throughout the hydrostatic burning phases and the explosion. The network size adapted itself to its to the requirements and thus also allowed to include the weak $s$ process and the $\gamma$ process with all participating nuclides. For details of the network and the reaction rate data used, see Rauscher et al. (2002).

Table 1 (Model *xi45*) summarizes the mass fractions of $^{16}$O, $^{56}$Fe, and all the $p$ nuclides from $^{74}$Se up to $^{196}$Hg (in solar masses) for our grid of metallicities and masses (14 metallicities, 7 masses). Solar values used are from Lodders (2009) and assuming $Z_\odot = 0.015$.

For our GCE studies described in Section 4, we chose yields from the *xi45* series and therefore discuss those yields in more detail below. In Figure 1 we show the overproduction normalized to $^{16}$O for the $15\,\mathrm{M}_\odot$ *xi45* model and three different metallicities ($Z = 0.019$, 0.006, 0.0015), starting from nuclear mass-number $A \geq 70$. The $p$ nuclides are shown as filled triangles and the different isotopes of an element are connected with a solid line. In the upper panel ($Z = 0.019$) we show that the solar composition is reproduced within a factor of three for almost all the $p$ nuclides. Exceptions are the light $p$ nuclides $^{84}$Sr, $^{92,94}$Mo, and $^{96,98}$Ru. Various authors have discussed the problem of the production of Mo and Ru isotopes in the past (see, e.g., Rauscher et al. 2013 and references therein). The Mo and Ru isotopes are produced entirely in the O-shell during explosion. The distribution of $^{92}$Mo within the star is shown in Figure 2 for the $15\,\mathrm{M}_\odot$ model with $Z = 0.019$ (see also Section 4). In the models presented in this section, Mo and Ru isotopes are mostly due to photodisintegration reactions (see also Rauscher et al. 2016 for a detailed investigation of production paths) and therefore are strongly sensitive to the initial metallicity.

A very high production is shown for the heaviest $p$ nuclide $^{196}$Hg. This isotope comes from the decay of $^{206}$Pb, and the abundance of explosive $^{206}$Pb derives from the pre-explosive content of $^{207}$Pb and $^{208}$Pb, see Figures 3 and 4 for pre- and post-explosive compositions. According to these two maps, the Pb isotopes are converted to $^{196}$Hg during explosive nucleosynthesis. As pointed out by, e.g., Arnould & Goriely (2003); Dillmann et al. (2008); TRV11; Rauscher et al. (2013), if a significant fraction of the seed abundances is present in the form of Pb and Bi isotopes, they are converted to lighter nuclei through photodisintegration processes. TRV11 specified that about 60 % of $^{196}$Hg comes from $^{208}$Pb and the remaining 40 % mainly from the other Pb and Bi isotopes. The nuclide $^{208}$Pb is produced in explosive conditions at a different location inside the star than $^{196}$Hg. See also Figures 7 and 8 in Rauscher et al. (2016) for a detailed overview of the $p$ nuclide production in the mass zones of a star. Thus, the seed for the $^{196}$Hg production is the pre-explosive Pb and Bi content of the mass zones producing $^{196}$Hg.

The nucleus $^{180}$Ta and its isomer $^{180\mathrm{m}}$Ta is partly synthesized by the $\nu$ process (Woosley & Howard 1990) but also receives considerable contributions from the photodisintegration of heavier nuclei in the $\gamma$-process (Heger et al. 2005). For the latter, the final $^{180\mathrm{m}}$Ta abundance depends on the freeze-out from thermal equilibrium between excited nuclear states, as pointed out by Rauscher et al. (2002); Belic et al. (2002); Mohr et al. (2007). Above the freeze-out temperature, the ground state, isomeric state, and further excited states of $^{180}$Ta are in communication and can be converted into each other through $\gamma$ transitions. This was not explicitly followed in the current calculations and thus the $^{180}$Ta abundances shown in the figures and tables of this paper only provide an upper limit of the final $^{180\mathrm{m}}$Ta abundance. According to Rauscher et al. (2002); Mohr et al. (2007); Hayakawa et al. (2010), only about 35 % of the synthesized $^{180}$Ta survives in the form of

$^{180m}$Ta. Consequently, in all supernova results presented here, a more realistic estimate of the $^{180m}$Ta yield would require to decrease the shown $^{180}$Ta yield accordingly.

Reducing the metallicity in the models, the *p* yields are drastically reduced as soon as half of the solar metallicity is reached (Figure 1, middle panel). For this and lower metallicities almost none of the *p* nuclides is any longer within the factor of three from solar. The exception is $^{196}$Hg, as mentioned above. This will be further discussed in more detail below. With a metallicity almost a factor of 10 below solar (Figure 1, lower panel) we can see that all isotopes are at least $10^{-2}$ below the solar abundances (still with the exception of $^{196}$Hg). The relevance of these yields in a GCE study will be discussed in Section 4.

Similar to Figure 1, Figures 5 and 6 show results for the $20\,M_\odot$ and the $25\,M_\odot$ star, respectively. For the $25\,M_\odot$ progenitor one can see a slightly different behavior for the light *p* nuclides: they have a higher abundance with respect to the $15\,M_\odot$ and the $20\,M_\odot$ models at solar metallicity but still exhibit a drastic drop in the abundances from $^{84}$Sr up.

## 3. Production of *p* nuclides in the NuGrid models

The second set of *p*-nuclide yields used in this work are taken from Pignatari et al. (2016a). The one-dimensional stellar progenitors were calculated using the stellar evolution code GENEC (Eggenberger et al. 2008) for massive stars.

CCSN explosive conditions are obtained from a semi-analytic treatment of shock heating, based on the hydrodynamics simulations by Fryer et al. (2012). The shock velocity beyond fallback is initially $2 \times 10^9$ cm s$^{-1}$. The obtained average ccSN explosion energy is $4 - 5 \times 10^{51}$ ergs. The nucleosynthesis in the preSN and ccSN stages was followed consistently with the same post-processing parallel code, MPPNP (see Pignatari et al. 2016a for details). The NuGrid results were included in our GCE study, using a grid of three progenitor masses ($15\,M_\odot$, $20\,M_\odot$, and $25\,M_\odot$) and two metallicities ($Z = 0.02, 0.01$). In particular, we considered two models for the $15\,M_\odot$ progenitor (*15d*, *15r2*), one model for the $20\,M_\odot$ progenitor (*20d*), and one for the $25\,M_\odot$ progenitor (*25d*). The only difference between models *15d* and *15r2* is that the ccSN shock velocity was reduced by a factor of two in the second case, resulting in a reduction of the total explosion energy.

In Table 2 we summarize the yields (in $M_\odot$) of $^{16}$O, $^{56}$Fe, and all $p$ nuclides from $^{74}$Se up to $^{196}$Hg (the same as in Table 1 for KEPLER models) for the two metallicities and the three masses we took into account.

As in Figures 1, 5, 6, in Figures 7−9 we show the overproduction factors normalized to $^{16}$O for the $15\,M_\odot$, $20\,M_\odot$, and $25\,M_\odot$ models, respectively, for two different metallicities ($Z = 0.02, 0.001$). The $p$ nuclides are shown as filled triangles and the isotopes of the same element are connected with a solid line.

Contrary to the KEPLER models discussed in the previous section, the middle and lower panels (model series d, with fast shock velocities and metallicities of $Z = 0.02, 0.01$, respectively) of Figure 7 show for the $15\,M_\odot$ model that the lightest $p$ nuclides $^{74}$Se, $^{78}$Kr, $^{84}$Sr, and $^{92}$Mo are largely overproduced. This is due to a strong $\alpha$-process (Woosley & Hoffman 1992) in these models which appears because of $\alpha$-rich freeze-out conditions in the $15\,M_\odot$ models of Pignatari et al. (2016a) with fast shocks and slightly neutron-rich progenitor composition. This freeze-out from nuclear statistical equilibrium is a primary process and therefore the resulting yields of light $p$ nucleido not show strong sensitivity to metallicity. The impact of this component in producing radioactive species including the unstable nuclide $^{92}$Nb has been discussed by Lugaro et al. (2016). This component has also important consequences for GCE (see Section 4).

For comparison, the upper panel of Figure 7 shows the same results for model 15r2, in which more $^{56}$Ni is synthesized but no $\alpha$-rich freeze-out occurs. In this model, the $p$ abundances are products of the $\gamma$ process, which is a secondary process, i.e., depending on initial metallicity.

Figures 8 and 9 show the results for the $20\,M_\odot$ and $25\,M_\odot$ models, respectively. These more massive progenitors have larger fall-back, preventing the $\alpha$-process to contribute to the ejected light $p$ nuclei. This leads to the obvious secondary-like behavior of the $p$ abundances, also for the light species, and a pronounced variation in the amount of ejected $p$ nuclei with varying metallicity. The contribution from neutrino spallation of matter during the ccSN explosion, producing $^{138}$La and $^{180m}$Ta, among other species, is not considered in the nucleosynthesis calculations by Pignatari et al. (2016a). The high production of $^{196}$Hg seen in KEPLER models and discussed in the previous section is much lower in the NuGrid models, is maybe due to the seed composition in the progenitor.

## 4. Galactic Chemical Evolution: the role of ccSN in $p$-nucleus enrichment

Since no $p$ isotope dominates an elemental abundance, terrestrial and meteoritic $p$-abundances have to be explained purely through GCE models, integrating the production of different sites over the history of the Galaxy. The main goal of this work is to clarify the contribution of ccSN to the measured Solar System abundances of $p$ nuclides and to discuss the interplay between different stellar sources (ccSNe from this work and SNIa predictions from TRV15).

For this purpose, we implement metallicity-dependent ccSN yields of $p$ nuclei in a GCE code (Travaglio et al. 1999; Travaglio et al. 2004) for the first time. The same GCE code has been used by TRV15 to study the role of thermonuclear supernovae in $p$ nucleus production and their contribution to the Solar System abundance. The model follows the evolution of the Galaxy in three interconnected zones: halo, thick disk, and thin disk. The original set of nuclides within the GCE code was chosen to cover all the light nuclei up to the Fe group and all the heavy nuclei along the *s*-process path up to $^{209}$Bi. In TRV15 and for the present work, the nuclide set was extended to include *p* nuclides and allowed to follow their evolution over time and metallicity until solar metallicity is reached. Here, we implemented the *p* yields presented in Sections 2 and 3 and interpolated between masses and metallicities where necessary.

The GCE results for *p*-nucleus yields using the KEPLER models (*xi45, xi25, nocutoff, ertl*) as well as the NUGRID model are presented in Table 3 at the epoch of the Solar System formation and compared to the Solar composition by Lodders et al. (2009) in the first column.

In Figures 10 and 11 we show the resulting *p* nuclei production factors taken at the epoch of Solar System formation for nuclei in the nuclear mass-number range $70 \leq A \leq 210$ for the two sets of models described in Sections 2 and 3, respectively. We note that for the results presented here, we only included ccSN in the mass range $13\,\mathrm{M}_{\odot} - 30\,\mathrm{M}_{\odot}$.

Using the yields of the KEPLER models, Figure 10 shows that the Solar System composition is reproduced within a factor of about three for the light *p* nuclides $^{74}$Se and $^{84}$Sr. Due to the $\nu$ process contribution, ccSNe produce $^{138}$La and $^{180\mathrm{m}}$Ta a factor of about three below solar. As mentioned in Section 2, the abundance of $^{180\mathrm{m}}$Ta in this figure should be reduced by 1/3. A small production of $^{184}$Os is also found. The nuclide $^{196}$Hg is obtained at a level of 10 % of the solar abundance. The contribution from ccSNe to all the other *p* nuclides is very similar when using yields from KEPLER models.

Figure 11 shows GCE results obtained with the post-processed NuGrid yields presented in Section 3. The appearance of material ejected from a strong $\alpha$ process occurring in the $15\,\mathrm{M}_{\odot}$ model has important consequences for the light *p* nuclides $^{74}$Se, $^{78}$Kr and $^{84}$Sr, reproducing their solar abundances within a factor of about two. This only occurs in the 15 $M_\odot$ model, however. The only other significant contribution to the Solar System abundance (within a factor of about three) coming from these models is to $^{184}$Os.

Finally, Figures 12 and 13 compare GCE results using the *xi45* KEPLER model (shown in Figure 10) and the NUGRID model, respectively, to TRV15 results where only SNIa were considered as *p*-nuclei sources. It can be clearly seen that, starting from $^{92}$Mo, SNIa play the dominant role in explaining the Solar System abundances of *p* nuclides, provided the existence of single-degenerate thermonuclear supernovae and the validity of TRV15's hypothesis regarding the enrichment in *s*-seeds during mass accretion.

## 5. Model uncertainties

Concerning the impact of uncertainties in numerical predictions within our GCE model, we refer to the discussion from Côté et al. (2016). These authors identified the following basic parameters as main sources of uncertainties in GCE calculations: the lower and upper mass limits of the stellar IMF, the slope of the high-mass end of the stellar IMF, the slope of the delay-time distribution function of SNIa, the number of SNIa per $M_\odot$ formed, the total stellar mass formed, and the final mass of gas. They conclude that the slope of the IMF and the number of SNIa are the two main sources of uncertainty. These uncertainties are not relevant for the results presented in this work for the following reasons. First, changing the slope of the IMF only impacts progenitor masses lower than $\sim 10\,M_\odot$ which are not important for the *p* nuclei production. Second, the fact that the production of *p* nuclei in Type II supernovae is secondary is dependent the rate of Type II supernovae through galactic history.

A crucial source of uncertainty for the relevance of SNIa in the production of p-nuclei in GCE is the fraction of SNIa made via the single-degenerate scenario, compared to double-degenerate scenario, where only the first type are most likely source of $p$ nuclei. TRV15, following Li et al. (2011), found that if about $50-70\,\%$ of all SNIa are made via single degenerate scenario, they can be responsible for at least $50\,\%$ of the $p$ nuclei abundances in the Solar System. By looking at the chemical evolution of Mn with respect to Fe, Seitenzahl et al. (2013) obtained that $50\,\%$ of SNIa are from near-Chandrasekhar mass explosions, likely made via single-degenerate channel. Single degenerate channel, according Seitenzahl et al., is needed to reach the observed [Mn/Fe] in the Sun. Matteucci et al. (2009) reaches a similar conclusion reproducing [O/Fe] as a function of [Fe/H] and the metallicity distribution of G-type stars in the solar neighborhood. These authors conclude that both single-degenerate as well as double-degenerate progenitors must contribute to the Galactic population of SNe Ia. For a better understanding of the role of single degenerate scenario in chemical enrichment of the Galaxy we urgently need a detailed investigation of the dependence of the yields to metallicity (dependency on metallicity of the Mn yields invoked by Cescutti et al. (2008). A large analysis of single degenerate yields as a function of metallicity will be published in a forthcoming paper (Travaglio et al. 2018, in preparation). On the other hand, consistently with previous theoretical predictions by Woods & Gilfanov (2013), Johansson et al. (2016) found that the single-degenerate SNIa should be less than $3-6\,\%$ of the total SNIa population, based on observations on Sloan Digital Sky Survey spectra from a large sample of early-type galaxies. In conclusion, the total contribution to the SNIa population of the single-degenerate channel is still an open puzzle, with contradicting indications from observational constraints, stellar population constraints as galactic chemical evolution. In particular, by using the upper limits provided by Johansson et al. (2016), the contribution to the inventory of the $p$-process nuclei from the single-degenerate SNIa channel would be marginal, while the production of Fe-group elements like Fe and Mn should be also revised. There are no investigations in the literature concerning the contribution to $p$ nuclei from the WD-mergers SNIa channel, therefore to attach this open problem is urgently needed.

The yields of the stellar models bear uncertainties due to the details in numerical treatment and the implementations of explosion mechanism and fallback. We used yields from two different stellar evolution codes, and implementing different explosions and fallback in each code, to obtain an estimate of the differences to expect between results obtained with different codes. We also have only a limited set of models (see, e.g., Müller et al. 2016; Sukhbold et al. 2017) and do not cover the lowest mass or very high masses. The low masses may not contribute much material, though they my have their own peculiar nucleosynthesis. This higher masses are disfavored by the IMF but may contribute more intermediate-mass elements.

A further source of uncertainty are the astrophysical reaction rates used in the stellar evolution codes. The $\gamma$ process proceeds by sequences of initial ($\gamma$,n) reactions photodisintegrating the previously existing seed nuclei at temperatures between 2 GK and 4 GK, thereby producing proton-richer isotopes. Towards the proton-rich side of the nuclear chart, the ($\gamma$,n) reactions are competing with ($\gamma$,$\alpha$) reactions (above the $N = 50$ neutron shell) or ($\gamma$,p) reactions (for $N \leq 50$), deflecting the reaction path to lower charge numbers $Z$. Reactions occur at stability and a few mass number units off stability towards the proton-rich side and thus the majority of nuclei involved are unstable. Furthermore, at $\gamma$-process temperatures contributions of thermally excited states in the target nuclei are dominating the stellar reaction rate. This implies that none of the rates are constrained experimentally, not even those on stable nuclides.

Several studies have been performed in the past to explore the effect of rate variations on the production of $p$ nuclei (e.g., Arnould & Goriely 2003; Rapp et al. 2006; Rauscher 2006) and neutrino cross sections (e.g., Heger 2005). These were restricted to testing the isolated impact of particular rate variations. Very recently, Rauscher et al. (2016); Nishimura et al (2017) have quantified the *combined* uncertainties in the final yields of stable and radiogenic $p$ nuclei stemming from the uncertainties of *all* rates. This was achieved by performing a Monte Carlo approach of unprecedented scale, varying all rates in a network within their bespoke, temperature-dependent nuclear physics uncertainties. These temperature-dependent nuclear physics uncertainties were constructed from a combination of theoretical and experimental errors, individually for each rate in the network. Relevant to this work, Rauscher et al. (2016) explored these uncertainties for $15\,\mathrm{M}_\odot$ and $25\,\mathrm{M}_\odot$ KEPLER models with initial solar metallicity. Similar main uncertainties can be expected also for the relative $\gamma$ process yields at lower metallicity. Nuclear uncertainties in the $\gamma$ process at lower metallicity, however, have a smaller relevance because of the secondary-like nature of the process. Despite of larger uncertainties in predicted astrophysical reaction rates, the final uncertainties in the $p$ yields found by Rauscher et al. (2016) were lower than a factor of two, with a few exceptions: $^{113}$In, $^{115}$Sn, $^{168}$Yb, and $^{174}$Hf. The asymmetric uncertainty distributions derived favor increased yields. Even for the exceptions with particularly large uncertainties, the upper limit remains below a factor of about 3.5. On the other hand, particularly small uncertainties were found for $^{184}$Os. Also the uncertainties for $^{190}$Pt and $^{196}$Hg are below 20 % in the $15\,\mathrm{M}_\odot$ as well as the $25\,\mathrm{M}_\odot$ model. We consider the uncertainties in the $\gamma$ process yields calculated by Rauscher et al. (2016) as a realistic estimate that to first approximation can also be applied to the other set of $p$ process yields dominated by the $\gamma$ process, i.e., for the NuGrid yields of the models *15r2*, *20d* and *25d*.

Nishimura et al. (2017) applied the Monte Carlo procedure of Rauscher et al. (2016) to $p$ production in the SNIa model also used by TRV11. The final uncertainties found for the production of $p$ nuclei in this SNIa model were even lower than the ones for the ccSN cases, below $30 - 40$ %. The only exceptions were $^{162}$Er with a factor of two uncertainty and $^{180\mathrm{m}}$Ta with a factor of 1.8 as upper limit. As mentioned above, however, different production mechanisms are contributing to $^{180\mathrm{m}}$Ta, including the $\nu$ process which does not appear in SNIa.

Given the comparatively small nuclear uncertainties, we conclude that they do not affect our conclusions regarding the relevance (or irrelevance) of ccSN for the Solar System $p$-nuclide abundances, even though we did not perform a GCE calculation including the propagation of the yield uncertainties.

## 6. Discussion and Conclusions

We have presented results of detailed nucleosynthesis calculations for the production of $p$ nuclides in two sets of ccSN models (four sets of models using the KEPLER code and two sets of models using the NUGRID code), on a fine grid of masses and metallicities (seven initial masses from 13 $M_\odot$ up to 30 $M_\odot$, and 14 metallicities for the KEPLER sets; three masses at $15\,\mathrm{M}_\odot$, $20\,\mathrm{M}_\odot$, and $25\,\mathrm{M}_\odot$, and two metallicities for the NUGRID sets). They are all non-rotating, one-dimensional models. Nucleosynthesis in the KEPLER models has been followed coupled to the stellar evolution and explosion, whereas nucleosynthesis has been calculated in a post-processing approach for the NUGRID models.

Between KEPLER and NUGRID models we find few significant differences, with the exception of the considerable production of light $p$ nuclei using the $15\,\mathrm{M}_\odot$ NUGRID progenitor and the production level of $^{196}$Hg in the KEPLER models. The change in initial abundance distribution with metallicity is not the same in the two models discussed in the paper. In the KEPLER models, the abundances are redistributed according to the GCE considerations as explained in West & Heger (2013). In the Nugrid models the initial abundances are scaled to $Z$ =0.01 and $Z$ =0.02 from the solar values from Grevesse & Noels (1993), with the solar isotopic percentage for each element given by Lodders (2003). Therefore the abundances are solar abundances scaled with metallicity in the Nugrid models whereas in the KEPLER the initial abundances are not only scaled but also redistributed, leading to higher relative abundances in the seed region relevant to the production of $^{196}$Hg. This explains the difference in the resulting $^{196}$Hg. We also notice that there is a small difference in neutron capture and photodisintegration rates between Bao et al. (2000) (used in the KEPLER models) and KADONIS v0.2 (used in the Nugrid models). This leads to a slightly lower production of heavy $p$-nuclei because of lower (n,$\gamma$) and ($\gamma$,n) rates on the heavy seed nuclei. This effect was explored in Dillmann et al. (2008) where only a tiny difference in the production of 196Hg was found when comparing calculations with the two rate sets. Therefore the change in reaction rates cannot account for the large difference between the two models discussed in this work. Only the difference in the treatment of abundances at different metallicities remains as the cause for the production difference in $^{196}$Hg.

Making use of the $p$ yields from the KEPLER and NUGRID models, a detailed GCE study was performed to investigate the production of $p$ nuclei over the course of Galactic history until Solar System formation. For the first time in literature, yields from ccSN originating from progenitors of different masses and metallicities were coupled to GCE. Being able to self-consistently trace the Galactic $p$ nucleus enrichment with time also allowed to clarify the role of ccSN contributions to the abundance distribution of $p$ nuclei in the Solar System. For both sets of ccSN nucleosynthesis analyzed in this paper, we conclude that ccSN can give a significant contribution to the Solar System abundances only for few $p$ nuclides: (a) the light $p$ nuclei up to $^{92}$Mo, if the contribution can come from a primary process in the massive star (such as, e.g., an $\alpha$ process as discussed in Section 3 or a $\nu$p-process, etc.); (b) $^{138}$La and $^{180\mathrm{m}}$Ta that have a strong contribution from the $\nu$ process; (c) $^{184}$Os; (d) $^{196}$Hg, if significant abundances of Pb isotopes come from the progenitor, as in the KEPLER models. These findings do not depend on details of the GCE model used here.

Contributions from a $\nu$p-process or from neutrino driven winds in the deepest layers of ccSN (e.g., Fröhlich et al. 2006; Arcones & Janka 2011; Wanajo et al. 2011a, and references therein) were not considered. These processes would also be primary (i.e., independent to metallicity) but only light *p* nuclei can be created in such processes and the proton-rich environments required were not found in more recent multi-D hydrodynamic ccSN explosion studies.

Assuming that the Solar System *p* nucleus abundances are typical for the local Galaxy, this therefore implies a necessity to find different astrophysical site(s) for the synthesis of the majority of *p* nuclides. TRV15 showed single-degenerate SNIa to be a viable alternative site, pending a closer investigation of the assumed synthesis of *s* process seeds in the accreted layer (e.g., Battino, et al. 2018, in preparation) and confirmation of the assumed single-degenerate SNIa occurrence frequency.

We thank Tyrone Woods for discussion on Supernova Type Ia rates. C.T. thanks B2FH Association (http://www.b2fh.org) and University of Frankfurt for the support in computer power for numerical calculations. C.T. also thanks JINA-CEE for travelling support that made possible this collaboration. A.H. was supported by an ARC Future Fellowship (FT120100363). MP acknowledges support to NuGrid from NSF grants PHY 02-16783 and PHY 08-22648 (Joint Institute for Nuclear Astrophysics, JINA), NSF grant PHY-1430152 (JINA Center for the Evolution of the Elements) and EU MIRG-CT-2006-046520. MP also acknowledges support by the University of Hulls High Performance Computing Facility viper. NuGrid uses services of the Canadian Advanced Network for Astronomy Research (CANFAR) which in turn is supported by CANARIE, Compute Canada, University of Victoria, the National Research Council of Canada, and the Canadian Space Agency.

## REFERENCES

Angulo, C., et al. 1999, Nuclear Physic A, 656, 3

Arcones, A., & Janka, H.-T. 2011, A&A, 526, 160

Arcones, A., & Montes, F. 2011, ApJ, 731, 5

Arnould M. & Goriely S., 2003, Phys. Rep., 384, 1

Arnould, M., & Goriely, S. 2006, Nucl. Phys. A, 777, 157

Austin, S.M., West, C., Heger, A. 2014, Phys. Rev. Lett.112, 1101

Bao, Z.Y., Beer, H., Käppeler, F., Voss, F., Wisshak, K., & Rauscher, T. 2000, Atomic Data Nucl. Data Tables, 76, 70

Belic, D. et al. 2002, Physical Review C, 65.035801

Burbidge, E.M., Burbidge, G.R., Fowler, W.A., & Hoyle, F. 1957, Rev. Mod. Phys., 29, 547

Cameron, A.G.W. 1957, AJ, 62, 9

Cassisi, S., Iben, I.Jr., & Tornambé A. 1998, ApJ, 496, 376

Cescutti, G., Matteucci, F., Lanfranchi, G.A., & McWilliam, A. 2008, A&A, 491, 401

Côté, B., West, C., Heger, A., Ritter, C., O'Shea, B.W., Herwig, F., Travaglio, C., Bisterzo, S. 2016, MNRAS, 463, 3755

Denissenkov, P. A., Herwig, F., Battino, U., Ritter, C., Pignatari, M., Jones, S., & Paxton, B. 2017, ApJ, 834, L10

Dillmann, I., Rauscher, T., Heil, M., Käpeler, F., Rapp, W., & Thielemann, F.-K. 2008, J. Phys. G, 35, 014029

Eggenberger, P., Meynet, G., Maeder, A., Hirschi, R., Charbonnel, C., Talon, S., & Ekströ, S. 2008, Astrophysics and Space Science, 316, 43

Ertl, T., Janka, H.-Th., Woosley, S. E., Sukhbold, T., & Ugliano, M. 2016, ApJ, 818, 124

Farouqi, K., Kratz, K.-L., & Pfeiffer, B. 2009, PASA, 26, 194

Fischer, T., et al. 2011, ApJS, 194, 39

Fryer, C.L., Belczynski, K., Wiktorowicz, G., Dominik, M., Kalogera, V., Holz, & Daniel E. 2012, ApJ, 749, 91

Fröhlich, C., et al. 2006, ApJ, 637, 415

Goriely, S., García-Senz, D., Bravo, E., & José, J. 2005, A&A, 444, L1

Grevesse, N., & Noels, A. 1993, Symposium 'Origin and evolution of the elements', p. 15 - 25

Hayakawa, T., Iwamoto, M., Kajino, T., Shizuma, T., Umeda, H., & Nomoto, K. 2006, ApJ, 648, L47

Hayakawa, T., Iwamoto, M., Kajino, T., Shizuma, T., Umeda, H., & Nomoto, K. 2008, ApJ, 685, 1089

Hayakawa, T., Mohr, P., Kajino, T., Chiba, S., & Mathews, G. J. 2010, Physical Review C, vol. 82, Issue 5

Heger, A., Kolbe, E., Haxton, W.C., Langanke, K., Martínez-Pinedo, G., Woosley, S.E. 2005, Phys. Lett. B, 606, 258

Heger, A., Woosley, S.E. 2010, ApJ, 724, 341

Hoffman, R. D., Woosley, S. E., Fuller, G. M., & Meyer, B. S. 1996, ApJ, 460, 478

Howard, W.M., & Meyer, B.S. 1993, in Proceedings of the 2nd International Symposium on Nuclear Astrophysics, held at Karlsruhe, Germany. Ed. by F. Kaeppeler and K. Wisshak, (Bristol: IOP Publishing) p.575

Johansson, J., Woods, T.E., Gilfanov, M., Sarzi, M., Chen, Y.-M., & Oh, K. 2016, MNRAS, 461, 4505

Kobayashi, C., Karakas, A.I., & Umeda, H. 2011, MNRAS, 414, 3231

Kusakabe, M., Iwamoto, N., & Nomoto, K. 2005, Nucl. Phys. A, 758, 459

Kusakabe, M., Iwamoto, N., & Nomoto, K. 2011, ApJ, 726, 25

Lewis, R.S., Ming, T., Wacker, J.F., Anders, E., & Steel, E 1987, Nature, 326, 160

Li, W., et al. 2011, MNRAS, 412, 1441

Limongi, M., Straniero, O., & Chieffi, A. 2000, ApJS, 129, 625

Lodders, K., Palme, H., & Gail, H.-P. 2009, Landolt-Börnstein - Group VI Astronomy and Astrophysics Numerical Data and Functional Relationships in Science and Technology, Edited by J.E. Trümper, 4B: solar system, 4.4

Lovegrove, E., Woosley, S.E. 2013, ApJ, 769, 109

Lugaro, M., Pignatari, M., Ott, U., Zuber, K., Travaglio, C., Gyürky, G., & Fülop, Z. 2016, PNAS, 113, 907

Matteucci, F., Spitoni, E., Recchi, S., & Valiante, R. 2009, A&A, 501, 531

Mohr, P., Käppeler, F., & Gallino, R. 2007, Phys. Rev. C, 75, 012802

Müller, B., Heger, A., Liptai, D., Cameron, J.B. 2016, MNRAS, 460, 742

Nishimura, N., Rauscher, T., Hirschi, R., Cescutti, G., Murphy, A. St.J., Travaglio, C., 2017, MNRAS, submitted

Nomoto, K., Thielemann, F.-K., & Yokoi, K. 1984, ApJ, 286, 644

O'Connor, E., & Ott, C.D. 2011, ApJ, 730, 700

Paxton, B., Bildsten, L., Dotter, Aa., Herwig, F., Lesaffre, P., & Timmes, F. 2011, ApJS, 192, 3

Pellin, M. J., Savina, M. R., Calaway, W. F., Tripa, C. E., Barzyk, J. G., Davis, A. M., Gyngard, F., Amari, S., Zinner, E., Lewis, R. S., & Clayton, R. N. 2006, Lunar and Planetary Science LPI, 37, 2041

Pignatari, M., et al. 2016a, ApJS, 225, 24

Pignatari, M., Göbel, K., Reifarth, R., & Travaglio, C. 2016b, Int. J. Mod. Phys. E, 25, 1630003

Pruet, J., Hoffman, R.D., Woosley, S.E., Janka, H.-T., & Buras, R. 2006, ApJ, 644, 1028

Rapp, Rapp, W., Görres, J., Wiescher, M., Schatz, H., & Käppeler, F. 2006, ApJ, 653, 474

Rauscher, T., Heger, A., Hoffmann, R.D., & Woosley, S.E. 2002, ApJ, 576, 323

Rauscher, T. 2006, Phys. Rev. C, 73

Rauscher T., Dauphas N., Dillmann I., Fröhlich C., Fulop Zs., Gyurky Gy., 2013, Rep. Prog. Phys., 76, 066201

Rauscher T., Nishimura, N., Hirschi, R., Cescutti, G., Murphy, A. St. J., & Heger, A. 2016, MNRAS, 463, 4153

Rayet, M., Arnould, M., & Prantzos, N. 1990, A&A, 227, 271

Rayet, M., Arnould, M., Hashimoto, M., Prantzos, N., & Nomoto, K. 1995, A&A, 298, 517

Reisberg, L., Dauphas, N., Luguet, A., Pearson, D.G., Gallino, R., & Zimmermann, C. 2009, Earth and Planetary Science Letters, 277, 334

Ritter, C., Andrassy, R., Côté, B.; Herwig, F., Woodward, P. R., Pignatari, M., & Jones, S 2017, MNRAS submitted

Roberts, L.F., Woosley, S.E., & Hoffmann, R.D. 2010, ApJ, 722, 954

Savina, M. R., Davis, A. M., Tripa, C. E., Pellin, M. J., Gallino, R., Lewis, R. S., & Amari, S. 2003, Geochimica et Cosmochimica Acta Supplement, 67, 418

Schatz, H., Aprahamian, A., Barnard, V., Bildsten, L., Cumming, A., Ouellette, M., Rauscher, T., Thielemann, F.-K., & Wiescher, M. 2001, Physical Review Letters, vol. 86, Issue 16, pp. 3471-3474

Schulz, T., Muenker, C., & Peters, S.T.M. 2013, Earth and Planetary Science Letters, 362, 246

Seitenzahl, I.R., Cescutti, G., Röpke, F.K., Ruiter, A.J., & Pakmor, R. 2013, A&A, 559, L5

Sukhbold, T., & Woosley, S.E. 2014, ApJ, 783, 10

Sukhbold, T., Woosley, S.E., & Heger, A. 2017, arXiv:1710.03243

Travaglio, C., Galli, D., Gallino, R., Busso, M., Ferrini, F., & Straniero, O. 1999, ApJ, 521, 691

Travaglio, C., Gallino, R., Arnone, E., Cowan, J., Jordan, F., & Sneden, C. 2004, ApJ, 601, 864

Travaglio, C., Röpke, F.K., Gallino, R., & Hillebrandt, W. 2011, ApJ, 739, 93 (TRV11)

Travaglio, C., Gallino, R., Rauscher, T., Dauphas, N., R??pke, F., & Hillebrandt, W. 2014, ApJ, 795, 141

Travaglio, C., Gallino, R., Rauscher, T., Röpke, F.K., & Hillebrandt, W. 2015, ApJ, 799, 54 (TRV15)

Utrobin V. P., Wongwathanarat A., Janka H.-T., Müller E. 2015, A&A, 581, 40

Wanajo, S. 2006, ApJ, 647, 1323

Wanajo, S., Nomoto, K., Janka, H.-T., Kitaura, F. S., & Müller, B. 2009, ApJ, 695, 208

Wanajo, S., Janka, H.-T., & Müller, B. 2011a, ApJ, 726, L15

Wanajo, S., Janka, H.-T., & Kubono, S. 2011b, ApJ, 729, 46

Weaver, T. A., Zimmerman, G. B., & Woosley, S. E. 1978, ApJ, 225, 1021

West, C., Heger, A., & Austin, S.M. 2013, ApJ, 769, 2

West, C., Heger, A. 2013, ApJ, 774, 75

Woods, T.E., & Gilfanov, M. 2013, MNRAS, 432, 1640

Woosley, S. E., & Hoffman, R. D., 1992, ApJ, 395, 202

Woosley, S. E., & Howard, W. M. 1978, ApJS, 36, 285

Woosley, S. E., & Howard, W. M. 1990, ApJ, 354, L21

Woosley, S. E., & Weaver, T. A. 1995, ApJS, 101, 181

Woosley, S.E., & Heger, A. 2007, Physics Reports, 442, 269

Yamaguchi, H. et al. 2015, ApJ, 801, L31

---

Table 1. KEPLER xi45 ccSN model: ejected mass

| $Z$ | 2.425e-02 | 1.930e-02 | 1.530e-02 | 9.655e-03 | 6.092e-03 | 3.844e-03 | 2.425e-03 | 1.530e-03 | 4.839e-04 | 1.530e-04 | 4.839e-05 | 1.530e-05 | 1.530e-06 |
|---|---|---|---|---|---|---|---|---|---|---|---|---|---|
| **13** $M_{\odot}$ | | | | | | | | | | | | | |
| $^{16}$O | 6.08037e+00 | 6.35707e+00 | 8.20406e+00 | 8.30079e+00 | 9.26554e+00 | 9.71062e+00 | 1.06918e+01 | 1.10877e+01 | 1.11725e+01 | 1.07598e+01 | 1.09339e+01 | 1.07299e+01 | 9.99283e+00 |
| $^{56}$Fe | 2.98414e+00 | 2.71080e+00 | 5.36772e+00 | 6.35314e+00 | 6.73086e+00 | 7.38476e+00 | 4.46390e+00 | 3.79939e+00 | 3.62655e+00 | 3.79444e+00 | 3.81015e+00 | 3.73790e+00 | 4.27102e+00 |
| $^{74}$Se | 3.42237e-06 | 3.06218e-06 | 2.64840e-06 | 1.35830e-06 | 1.25927e-06 | 9.02767e-07 | 4.41461e-07 | 2.25410e-07 | 5.99328e-08 | 3.11291e-08 | 2.32187e-08 | 2.15791e-08 | 1.99634e-08 |
| $^{78}$Kr | 7.14244e-07 | 5.82897e-07 | 4.55772e-07 | 2.55535e-07 | 1.65108e-07 | 1.02033e-07 | 6.12784e-08 | 3.60656e-08 | 1.12226e-08 | 4.31160e-09 | 2.13535e-09 | 1.44612e-09 | 1.14551e-09 |
| $^{84}$Sr | 5.15743e-07 | 3.99276e-07 | 3.28973e-07 | 1.84947e-07 | 1.21424e-07 | 8.13639e-08 | 4.81828e-08 | 2.78227e-08 | 9.09888e-09 | 3.10376e-09 | 1.15163e-09 | 6.22047e-10 | 3.43517e-10 |
| $^{92}$Mo | 1.44853e-06 | 1.16981e-06 | 9.35743e-07 | 5.95657e-07 | 3.82251e-07 | 2.44913e-07 | 1.58420e-07 | 1.01985e-07 | 3.42867e-08 | 1.22203e-08 | 5.08863e-09 | 2.33602e-09 | 1.21784e-09 |
| $^{94}$Mo | 9.85087e-07 | 7.86624e-07 | 6.28021e-07 | 3.96880e-07 | 2.51720e-07 | 1.60586e-07 | 1.03225e-07 | 6.67499e-08 | 2.22576e-08 | 7.66869e-09 | 2.67158e-09 | 9.21539e-10 | 1.27015e-10 |
| $^{96}$Ru | 4.22456e-07 | 3.43455e-07 | 2.73270e-07 | 1.73771e-07 | 1.12364e-07 | 7.22734e-08 | 4.65724e-08 | 2.96757e-08 | 9.99501e-09 | 3.65190e-09 | 1.67465e-09 | 8.65289e-10 | 5.69841e-10 |
| $^{98}$Ru | 1.61649e-07 | 1.28883e-07 | 1.04691e-07 | 6.66337e-08 | 4.26709e-08 | 2.75767e-08 | 1.75303e-08 | 1.12144e-08 | 3.79827e-09 | 1.41023e-09 | 6.51350e-10 | 3.34632e-10 | 2.19228e-10 |
| $^{102}$Pd | 1.08039e-07 | 8.43923e-08 | 6.94928e-08 | 4.07907e-08 | 2.61816e-08 | 1.72781e-08 | 1.00026e-08 | 5.86985e-09 | 2.05754e-09 | 1.08432e-09 | 9.67342e-10 | 7.49308e-10 | 7.11671e-10 |
| $^{106}$Cd | 1.76464e-07 | 1.34280e-07 | 1.15123e-07 | 6.51702e-08 | 4.26894e-08 | 2.86634e-08 | 1.55872e-08 | 9.04738e-09 | 3.35114e-09 | 1.87626e-09 | 1.68931e-09 | 1.39969e-09 | 1.25328e-09 |
| $^{108}$Cd | 1.40988e-07 | 1.05897e-07 | 8.87340e-08 | 5.13444e-08 | 3.09847e-08 | 1.94486e-08 | 1.12850e-08 | 7.02858e-09 | 2.22975e-09 | 8.77876e-10 | 4.58844e-10 | 3.25924e-10 | 2.44144e-10 |
| $^{113}$In | 5.63487e-08 | 4.21961e-08 | 3.40207e-08 | 1.86418e-08 | 1.09987e-08 | 6.37271e-09 | 3.49716e-09 | 2.06554e-09 | 6.09095e-10 | 2.16437e-10 | 1.12816e-10 | 9.10390e-11 | 6.80127e-11 |
| $^{112}$Sn | 3.66695e-07 | 2.75566e-07 | 2.47845e-07 | 1.33670e-07 | 8.94763e-08 | 5.40201e-08 | 2.85396e-08 | 1.78312e-08 | 6.67390e-09 | 3.08625e-09 | 2.07767e-09 | 1.88510e-09 | 1.42829e-09 |
| $^{114}$Sn | 2.48052e-07 | 1.79053e-07 | 1.58227e-07 | 9.09762e-08 | 5.42691e-08 | 3.33339e-08 | 1.82903e-08 | 1.19709e-08 | 4.02422e-09 | 1.43157e-09 | 6.54663e-10 | 4.95053e-10 | 3.50602e-10 |
| $^{115}$Sn | 8.50277e-08 | 6.35910e-08 | 4.82863e-08 | 2.70209e-08 | 1.51651e-08 | 8.59580e-09 | 4.85126e-09 | 2.79641e-09 | 7.13630e-10 | 1.92854e-10 | 5.69458e-11 | 2.28705e-11 | 1.00600e-11 |
| $^{120}$Te | 5.75882e-08 | 4.11268e-08 | 3.65677e-08 | 2.14512e-08 | 1.33496e-08 | 8.53189e-09 | 4.73132e-09 | 3.17358e-09 | 1.01808e-09 | 3.94226e-10 | 1.85261e-10 | 1.19262e-10 | 7.08896e-11 |
| $^{124}$Xe | 1.41616e-07 | 9.50474e-08 | 9.34778e-08 | 5.23708e-08 | 3.06614e-08 | 1.89930e-08 | 1.04159e-08 | 6.87245e-09 | 2.53493e-09 | 1.11932e-09 | 5.67764e-10 | 4.40861e-10 | 3.12938e-10 |
| $^{126}$Xe | 1.10892e-07 | 8.18001e-08 | 7.49165e-08 | 4.30741e-08 | 2.89972e-08 | 1.84698e-08 | 1.03673e-08 | 6.91470e-09 | 2.36593e-09 | 1.02106e-09 | 5.52956e-10 | 4.02277e-10 | 2.61405e-10 |
| $^{130}$Ba | 1.89259e-07 | 1.22158e-07 | 1.12778e-07 | 7.11949e-08 | 4.33344e-08 | 2.98374e-08 | 1.59734e-08 | 1.11508e-08 | 2.96014e-09 | 7.58986e-10 | 2.85393e-10 | 1.99985e-10 | 1.39512e-10 |
| $^{132}$Ba | 1.07797e-07 | 9.47031e-08 | 8.05110e-08 | 4.88722e-08 | 3.78890e-08 | 2.56938e-08 | 1.40731e-08 | 9.21915e-09 | 2.26476e-09 | 6.65911e-10 | 2.87227e-10 | 2.04631e-10 | 1.40468e-10 |
| $^{138}$La | 1.32614e-08 | 1.06777e-08 | 9.95245e-09 | 6.51506e-09 | 4.39368e-09 | 2.75994e-09 | 1.90468e-09 | 1.11398e-09 | 2.32161e-10 | 3.74478e-11 | 6.74772e-12 | 1.65583e-12 | 1.40925e-13 |
| $^{136}$Ce | 3.89419e-08 | 2.63369e-08 | 2.19689e-08 | 1.26555e-08 | 7.17707e-09 | 4.23822e-09 | 2.31646e-09 | 1.40677e-09 | 3.82605e-10 | 1.69747e-10 | 1.00379e-10 | 8.52756e-11 | 6.25757e-11 |
| $^{138}$Ce | 4.29050e-08 | 3.37314e-08 | 2.73037e-08 | 1.58194e-08 | 9.77824e-09 | 5.71868e-09 | 3.18107e-09 | 1.82373e-09 | 4.00891e-10 | 1.24979e-10 | 5.38724e-11 | 3.50265e-11 | 2.25276e-11 |
| $^{144}$Sm | 1.82104e-07 | 1.31327e-07 | 1.21143e-07 | 6.97760e-08 | 4.34973e-08 | 2.92640e-08 | 1.66982e-08 | 1.29911e-08 | 7.63841e-09 | 5.56840e-09 | 3.77564e-09 | 3.56419e-09 | 3.03884e-09 |
| $^{152}$Gd | 1.74564e-08 | 1.31076e-08 | 9.93236e-09 | 5.46579e-09 | 3.03534e-09 | 1.69232e-09 | 9.49204e-10 | 5.18334e-10 | 9.36386e-11 | 1.54055e-11 | 3.67111e-12 | 1.30978e-12 | 2.48048e-13 |
| $^{156}$Dy | 3.75766e-09 | 2.63708e-09 | 2.18340e-09 | 1.22435e-09 | 7.06726e-10 | 4.25023e-10 | 2.41591e-10 | 1.52185e-10 | 5.45399e-11 | 2.56472e-11 | 1.32308e-11 | 6.70864e-12 | 2.39913e-12 |
| $^{158}$Dy | 6.65812e-09 | 4.84148e-09 | 4.05775e-09 | 2.25434e-09 | 1.29958e-09 | 7.54811e-10 | 4.36496e-10 | 2.84496e-10 | 9.60665e-11 | 4.45878e-11 | 2.55093e-11 | 1.36392e-11 | 5.97514e-12 |
| $^{162}$Er | 8.73516e-09 | 6.65188e-09 | 5.72946e-09 | 3.39775e-09 | 2.27977e-09 | 1.51140e-09 | 9.03374e-10 | 6.64488e-10 | 3.08832e-10 | 1.94571e-10 | 1.23401e-10 | 9.59747e-11 | 7.42156e-11 |
| $^{164}$Er | 6.76465e-08 | 4.97299e-08 | 3.65612e-08 | 1.93896e-08 | 1.06854e-08 | 6.11357e-09 | 3.36580e-09 | 2.08255e-09 | 7.07921e-10 | 3.43623e-10 | 1.82866e-10 | 1.24782e-10 | 8.76254e-11 |
| $^{168}$Yb | 9.37747e-09 | 6.81209e-09 | 6.39738e-09 | 4.05573e-09 | 2.40367e-09 | 1.50659e-09 | 9.41061e-10 | 6.72724e-10 | 3.40016e-10 | 2.20062e-10 | 1.06965e-10 | 6.61653e-11 | 4.26977e-11 |
| $^{174}$Hf | 9.66964e-09 | 8.67544e-09 | 8.44166e-09 | 7.02535e-09 | 3.47358e-09 | 1.86461e-09 | 1.35804e-09 | 1.05068e-09 | 6.94910e-10 | 3.21951e-10 | 1.19662e-10 | 6.74327e-11 | 4.06784e-11 |
| $^{180}$Ta | 2.82265e-10 | 2.28490e-10 | 2.16785e-10 | 1.48305e-10 | 8.72849e-11 | 5.54805e-11 | 3.62073e-11 | 2.79665e-11 | 2.16373e-11 | 8.30962e-12 | 2.65395e-12 | 6.51177e-13 | 6.46555e-14 |
| $^{180}$W | 8.55567e-09 | 8.29016e-09 | 8.15766e-09 | 8.21236e-09 | 3.97921e-09 | 2.54190e-09 | 1.96471e-09 | 1.73113e-09 | 1.09665e-09 | 3.60856e-10 | 1.45988e-10 | 9.49486e-11 | 6.12113e-11 |

Table 1—Continued

| $Z$ | 2.425e-02 | 1.930e-02 | 1.530e-02 | 9.655e-03 | 6.092e-03 | 3.844e-03 | 2.425e-03 | 1.530e-03 | 4.839e-04 | 1.530e-04 | 4.839e-05 | 1.530e-05 | 1.530e-06 |
|---|---|---|---|---|---|---|---|---|---|---|---|---|---|
| $^{184}$Os | 6.78966e-09 | 6.14436e-09 | 6.31042e-09 | 4.98058e-09 | 2.55455e-09 | 1.65012e-09 | 1.09252e-09 | 8.96607e-10 | 3.96266e-10 | 7.53169e-11 | 2.73539e-11 | 1.49816e-11 | 4.82631e-12 |
| $^{190}$Pt | 4.34172e-09 | 3.84691e-09 | 3.27645e-09 | 2.34713e-09 | 1.36615e-09 | 9.21537e-10 | 5.58787e-10 | 4.09676e-10 | 1.38459e-10 | 3.54472e-11 | 2.69295e-11 | 1.57482e-11 | 2.84759e-12 |
| $^{196}$Hg | 4.85644e-07 | 3.82757e-07 | 4.16815e-07 | 3.05545e-07 | 2.64996e-07 | 2.38996e-07 | 1.74134e-07 | 1.64938e-07 | 1.40631e-07 | 1.38342e-07 | 1.25542e-07 | 1.25854e-07 | 1.09300e-07 |
| **15** $M_\odot$ | | | | | | | | | | | | | |
| $^{16}$O | 1.15751e+01 | 1.24597e+01 | 1.28733e+01 | 1.32564e+01 | 1.59543e+01 | 1.69097e+01 | 2.03436e+01 | 1.90898e+01 | 1.75980e+01 | 1.88401e+01 | 1.85671e+01 | 1.71341e+01 | 1.61794e+01 |
| $^{56}$Fe | 6.29704e+00 | 5.81038e+00 | 5.31856e+00 | 7.49352e+00 | 7.43371e+00 | 6.15745e+00 | 7.25111e+00 | 6.84046e+00 | 8.76815e+00 | 8.31504e+00 | 9.90240e+00 | 7.83629e+00 | 5.41446e+00 |
| $^{74}$Se | 7.20844e-06 | 5.86128e-06 | 4.37027e-06 | 3.76162e-06 | 1.51355e-06 | 6.77746e-07 | 7.12400e-07 | 3.32068e-07 | 7.42872e-08 | 5.23870e-08 | 5.37770e-08 | 2.20415e-08 | 1.83437e-08 |
| $^{78}$Kr | 1.12781e-06 | 8.42294e-07 | 5.95857e-07 | 3.60095e-07 | 1.69327e-07 | 1.40914e-07 | 7.61908e-08 | 3.97087e-08 | 1.27599e-08 | 5.41237e-09 | 3.24562e-09 | 1.43594e-09 | 1.02012e-09 |
| $^{84}$Sr | 6.20061e-07 | 4.79622e-07 | 3.88484e-07 | 2.42879e-07 | 2.62048e-07 | 3.79320e-07 | 5.52765e-08 | 4.60805e-08 | 1.05935e-08 | 4.79256e-09 | 1.74401e-09 | 8.41264e-10 | 3.83461e-10 |
| $^{92}$Mo | 1.65670e-06 | 1.32469e-06 | 1.05580e-06 | 6.87159e-07 | 4.14345e-07 | 2.65209e-07 | 1.82968e-07 | 1.10492e-07 | 4.03179e-08 | 1.29003e-08 | 5.80178e-09 | 2.98284e-09 | 1.57687e-09 |
| $^{94}$Mo | 1.10766e-06 | 8.82411e-07 | 7.09162e-07 | 4.53124e-07 | 2.89449e-07 | 1.83613e-07 | 1.18875e-07 | 7.61683e-08 | 2.54214e-08 | 8.83942e-09 | 3.11935e-09 | 1.08426e-09 | 1.52942e-10 |
| $^{96}$Ru | 4.87931e-07 | 3.89869e-07 | 3.10949e-07 | 2.02625e-07 | 1.17531e-07 | 7.49708e-08 | 5.37194e-08 | 3.11324e-08 | 1.19820e-08 | 3.59679e-09 | 1.80666e-09 | 1.04440e-09 | 6.67654e-10 |
| $^{98}$Ru | 1.90605e-07 | 1.51323e-07 | 1.20956e-07 | 7.95069e-08 | 4.57743e-08 | 2.75250e-08 | 2.10409e-08 | 1.16775e-08 | 4.57594e-09 | 1.37146e-09 | 6.99652e-10 | 4.08522e-10 | 2.60345e-10 |
| $^{102}$Pd | 1.42943e-07 | 1.12057e-07 | 8.63085e-08 | 5.38696e-08 | 1.73198e-08 | 9.00252e-09 | 1.43774e-08 | 3.30665e-09 | 3.41126e-09 | 3.77572e-10 | 9.23969e-10 | 9.49499e-10 | 8.68236e-10 |
| $^{106}$Cd | 2.32801e-07 | 1.83132e-07 | 1.40145e-07 | 8.65413e-08 | 2.36211e-08 | 1.22615e-08 | 2.24762e-08 | 4.01550e-09 | 5.52919e-09 | 5.45124e-10 | 1.63813e-09 | 1.75734e-09 | 1.58378e-09 |
| $^{108}$Cd | 1.75320e-07 | 1.35514e-07 | 1.07467e-07 | 6.52507e-08 | 3.24563e-08 | 1.62209e-08 | 1.40509e-08 | 6.05037e-09 | 2.78509e-09 | 7.14447e-10 | 5.42350e-10 | 4.42827e-10 | 3.30464e-10 |
| $^{113}$In | 7.01303e-08 | 5.19357e-08 | 3.81855e-08 | 2.23924e-08 | 1.08804e-08 | 5.74690e-09 | 4.80871e-09 | 1.86866e-09 | 8.73343e-10 | 1.93413e-10 | 1.56846e-10 | 1.52219e-10 | 1.08565e-10 |
| $^{112}$Sn | 5.30415e-07 | 3.90629e-07 | 2.82694e-07 | 1.79910e-07 | 5.60911e-08 | 2.45961e-08 | 4.51224e-08 | 8.54937e-09 | 1.02186e-08 | 1.79072e-09 | 2.43496e-09 | 2.67896e-09 | 2.01159e-09 |
| $^{114}$Sn | 3.22065e-07 | 2.41752e-07 | 1.85110e-07 | 1.18039e-07 | 5.75223e-08 | 2.03218e-08 | 2.60808e-08 | 8.89308e-09 | 5.57636e-09 | 1.28265e-09 | 8.64498e-10 | 7.80893e-10 | 4.84050e-10 |
| $^{115}$Sn | 9.70318e-08 | 7.24341e-08 | 5.44152e-08 | 3.09243e-08 | 1.72946e-08 | 9.45534e-09 | 5.74554e-09 | 3.12507e-09 | 8.67804e-10 | 2.44239e-10 | 8.18132e-11 | 3.84528e-11 | 1.58546e-11 |
| $^{120}$Te | 7.16112e-08 | 5.46763e-08 | 4.35520e-08 | 2.85797e-08 | 1.51072e-08 | 5.54961e-09 | 7.05097e-09 | 2.75968e-09 | 1.47620e-09 | 3.63484e-10 | 2.31105e-10 | 1.86310e-10 | 9.59240e-11 |
| $^{124}$Xe | 1.96533e-07 | 1.46615e-07 | 1.09624e-07 | 7.07941e-08 | 3.47800e-08 | 7.72867e-09 | 1.64879e-08 | 4.62859e-09 | 3.39954e-09 | 8.84955e-10 | 7.53646e-10 | 7.13085e-10 | 4.29406e-10 |
| $^{126}$Xe | 1.48540e-07 | 1.16666e-07 | 9.20727e-08 | 6.24500e-08 | 3.64471e-08 | 1.25933e-08 | 1.55031e-08 | 5.78917e-09 | 3.19076e-09 | 9.31547e-10 | 6.73408e-10 | 5.83697e-10 | 3.42894e-10 |
| $^{130}$Ba | 2.37594e-07 | 1.81872e-07 | 1.58091e-07 | 1.04585e-07 | 7.22880e-08 | 2.12568e-08 | 2.79511e-08 | 1.23428e-08 | 4.91639e-09 | 9.55714e-10 | 4.11424e-10 | 3.46787e-10 | 2.12371e-10 |
| $^{132}$Ba | 1.78768e-07 | 1.50747e-07 | 1.07277e-07 | 8.37061e-08 | 5.72620e-08 | 2.72158e-08 | 2.22708e-08 | 1.08058e-08 | 4.37871e-09 | 8.81079e-10 | 3.76917e-10 | 3.23648e-10 | 2.05046e-10 |
| $^{138}$La | 2.45862e-08 | 2.10219e-08 | 1.70170e-08 | 1.06836e-08 | 7.92588e-09 | 5.03150e-09 | 3.40994e-09 | 1.82544e-09 | 3.88010e-10 | 5.92507e-11 | 9.50039e-12 | 2.29780e-12 | 2.09136e-13 |
| $^{136}$Ce | 4.89772e-08 | 3.68211e-08 | 2.91204e-08 | 1.66194e-08 | 1.02554e-08 | 3.98999e-09 | 3.98080e-09 | 1.63527e-09 | 5.51839e-10 | 1.40836e-10 | 1.24738e-10 | 1.37991e-10 | 9.49492e-11 |
| $^{138}$Ce | 6.09477e-08 | 4.97494e-08 | 3.49025e-08 | 2.18834e-08 | 1.37312e-08 | 6.99404e-09 | 5.29704e-09 | 2.35364e-09 | 6.30463e-10 | 1.32469e-10 | 7.00457e-11 | 6.35375e-11 | 3.89081e-11 |
| $^{144}$Sm | 2.49363e-07 | 1.90166e-07 | 1.41204e-07 | 9.41567e-08 | 4.49864e-08 | 1.54216e-08 | 2.62112e-08 | 7.70666e-09 | 1.17512e-08 | 5.12938e-09 | 5.69078e-09 | 6.74433e-09 | 4.58795e-09 |
| $^{152}$Gd | 1.94954e-08 | 1.45842e-08 | 1.10115e-08 | 6.16665e-09 | 3.04219e-09 | 1.59472e-09 | 8.14891e-10 | 4.35376e-10 | 9.24068e-11 | 2.03741e-11 | 5.33194e-12 | 2.11393e-12 | 3.81376e-13 |
| $^{156}$Dy | 3.93910e-09 | 3.19753e-09 | 2.40952e-09 | 1.55548e-09 | 8.75265e-10 | 5.12650e-10 | 3.47453e-10 | 1.75858e-10 | 7.05709e-11 | 2.57646e-11 | 1.47949e-11 | 1.15753e-11 | 3.60410e-12 |
| $^{158}$Dy | 7.28954e-09 | 5.70876e-09 | 4.39591e-09 | 2.73397e-09 | 1.52334e-09 | 8.59752e-10 | 6.06058e-10 | 2.91358e-10 | 1.41924e-10 | 4.33600e-11 | 2.75182e-11 | 2.09625e-11 | 8.32091e-12 |
| $^{162}$Er | 1.09488e-08 | 9.12608e-09 | 6.84228e-09 | 4.64556e-09 | 2.80049e-09 | 1.45900e-09 | 1.13769e-09 | 5.49783e-10 | 4.16337e-10 | 2.02903e-10 | 1.59922e-10 | 1.55068e-10 | 1.08181e-10 |
| $^{164}$Er | 8.06111e-08 | 5.93179e-08 | 4.19630e-08 | 2.32844e-08 | 1.25406e-08 | 6.51854e-09 | 3.86245e-09 | 1.99546e-09 | 9.89848e-10 | 3.84206e-10 | 2.31054e-10 | 1.99571e-10 | 1.27977e-10 |
| $^{168}$Yb | 1.26042e-08 | 1.02123e-08 | 7.87462e-09 | 5.10201e-09 | 3.07120e-09 | 1.68757e-09 | 1.22995e-09 | 5.45724e-10 | 5.20080e-10 | 2.06005e-10 | 1.41656e-10 | 1.06131e-10 | 6.61506e-11 |

Table 1—Continued

| $Z$ | 2.425e-02 | 1.930e-02 | 1.530e-02 | 9.655e-03 | 6.092e-03 | 3.844e-03 | 2.425e-03 | 1.530e-03 | 4.839e-04 | 1.530e-04 | 4.839e-05 | 1.530e-05 | 1.530e-06 |
|---|---|---|---|---|---|---|---|---|---|---|---|---|---|
| $^{174}$Hf | 1.93692e-08 | 1.46672e-08 | 1.14670e-08 | 5.84146e-09 | 3.68374e-09 | 1.98263e-09 | 1.63248e-09 | 6.35628e-10 | 1.45962e-09 | 2.79238e-10 | 1.49928e-10 | 1.19217e-10 | 7.71436e-11 |
| $^{180}$Ta | 7.57039e-10 | 5.87470e-10 | 5.11977e-10 | 3.29046e-10 | 1.54251e-10 | 9.43903e-11 | 1.02462e-10 | 3.24751e-11 | 4.59879e-11 | 1.69260e-11 | 4.28490e-12 | 2.36887e-12 | 1.36794e-13 |
| $^{180}$W | 1.91563e-08 | 1.62134e-08 | 1.21533e-08 | 6.87473e-09 | 4.09180e-09 | 2.52597e-09 | 2.10421e-09 | 9.54314e-10 | 2.61327e-09 | 3.56186e-10 | 1.56703e-10 | 1.38969e-10 | 8.73042e-11 |
| $^{184}$Os | 1.19437e-08 | 9.48864e-09 | 8.11842e-09 | 5.41609e-09 | 2.89864e-09 | 1.67601e-09 | 1.18822e-09 | 5.43513e-10 | 1.71064e-09 | 1.14227e-10 | 3.09571e-11 | 2.40990e-11 | 8.78522e-12 |
| $^{190}$Pt | 6.51899e-09 | 4.95781e-09 | 4.01846e-09 | 3.07590e-09 | 1.45933e-09 | 8.49438e-10 | 6.85990e-10 | 2.91926e-10 | 6.42230e-10 | 3.81406e-11 | 2.17337e-11 | 2.16247e-11 | 5.61705e-12 |
| $^{196}$Hg | 7.69886e-07 | 6.94179e-07 | 5.58473e-07 | 5.06263e-07 | 3.55247e-07 | 1.64876e-07 | 2.86778e-07 | 1.43618e-07 | 2.20630e-07 | 1.52889e-07 | 1.59761e-07 | 1.93832e-07 | 1.51146e-07 |
| **17** $M_{\odot}$ | | | | | | | | | | | | | |
| $^{16}$O | 1.92752e+01 | 1.96987e+01 | 2.03317e+01 | 2.37326e+01 | 2.35823e+01 | 2.51673e+01 | 2.62200e+01 | 2.35696e+01 | 2.37114e+01 | 2.67925e+01 | 2.35392e+01 | 2.19643e+01 | 2.24669e+01 |
| $^{56}$Fe | 7.33083e+00 | 7.19398e+00 | 6.90195e+00 | 6.73991e+00 | 4.10421e+00 | 5.90557e+00 | 5.30308e+00 | 7.07942e+00 | 8.04633e+00 | 3.07959e+00 | 8.00178e+00 | 6.75242e+00 | 4.81463e+00 |
| $^{74}$Se | 6.92724e-06 | 4.29399e-06 | 2.88508e-06 | 4.81063e-06 | 3.82566e-06 | 1.52151e-06 | 3.94160e-06 | 1.01060e-06 | 1.03017e-07 | 4.46709e-08 | 2.67500e-08 | 2.14913e-08 | 1.89965e-08 |
| $^{78}$Kr | 2.54952e-06 | 1.56872e-06 | 1.02871e-06 | 3.77119e-07 | 2.91623e-07 | 2.93138e-07 | 2.83955e-07 | 1.62233e-07 | 1.94188e-08 | 7.75903e-09 | 2.78378e-09 | 2.02930e-09 | 1.08296e-09 |
| $^{84}$Sr | 8.96598e-06 | 6.04707e-06 | 3.82601e-06 | 4.26681e-07 | 1.83114e-07 | 7.56473e-07 | 4.06074e-07 | 2.62048e-07 | 3.41215e-08 | 2.13609e-08 | 4.25286e-09 | 3.74079e-09 | 1.21427e-09 |
| $^{92}$Mo | 1.75262e-06 | 1.40185e-06 | 1.11862e-06 | 7.50279e-07 | 4.90411e-07 | 2.97650e-07 | 2.57373e-07 | 1.29914e-07 | 4.28166e-08 | 1.96473e-08 | 5.15604e-09 | 9.22357e-09 | 7.43011e-09 |
| $^{94}$Mo | 1.25433e-06 | 9.93933e-07 | 7.90663e-07 | 5.13156e-07 | 3.24445e-07 | 2.06329e-07 | 1.65123e-07 | 8.42395e-08 | 2.81145e-08 | 1.10438e-08 | 3.41134e-09 | 1.31363e-09 | 8.40827e-10 |
| $^{96}$Ru | 4.96380e-07 | 3.96540e-07 | 3.16109e-07 | 2.17885e-07 | 1.43188e-07 | 8.34501e-08 | 6.71743e-08 | 3.66490e-08 | 1.18990e-08 | 5.25581e-09 | 1.38843e-09 | 2.31162e-09 | 1.28862e-09 |
| $^{98}$Ru | 1.79999e-07 | 1.43056e-07 | 1.14241e-07 | 8.71040e-08 | 5.68388e-08 | 3.11238e-08 | 4.35237e-08 | 1.38426e-08 | 4.33091e-09 | 2.78771e-09 | 5.32583e-10 | 1.20833e-09 | 1.20815e-09 |
| $^{102}$Pd | 8.60860e-08 | 6.52620e-08 | 4.95297e-08 | 5.97609e-08 | 3.99977e-08 | 1.02783e-08 | 5.63304e-08 | 6.87516e-09 | 1.38830e-09 | 4.30431e-09 | 1.49042e-10 | 4.25187e-09 | 4.02294e-09 |
| $^{106}$Cd | 1.37064e-07 | 1.02519e-07 | 7.65376e-08 | 1.06574e-07 | 6.72058e-08 | 1.38552e-08 | 6.92345e-08 | 1.01628e-08 | 1.44332e-09 | 5.03712e-09 | 1.34984e-10 | 5.78181e-09 | 3.33985e-09 |
| $^{108}$Cd | 1.65352e-07 | 1.23829e-07 | 9.38073e-08 | 8.04981e-08 | 4.41339e-08 | 1.90368e-08 | 7.91445e-08 | 8.01242e-09 | 1.63758e-09 | 4.32541e-09 | 2.38348e-10 | 1.44936e-09 | 1.66218e-09 |
| $^{113}$In | 6.32289e-08 | 4.69448e-08 | 3.51543e-08 | 2.76310e-08 | 1.47849e-08 | 6.33233e-09 | 1.57235e-08 | 2.33955e-09 | 4.92098e-10 | 9.01954e-10 | 3.97585e-11 | 4.21630e-10 | 2.42847e-10 |
| $^{112}$Sn | 2.74450e-07 | 2.05277e-07 | 1.53250e-07 | 2.38300e-07 | 1.35630e-07 | 2.79716e-08 | 2.60515e-07 | 2.00158e-08 | 2.32675e-09 | 1.33035e-08 | 2.67045e-10 | 7.23648e-09 | 4.87820e-09 |
| $^{114}$Sn | 2.05263e-07 | 1.51142e-07 | 1.13347e-07 | 1.39918e-07 | 7.68047e-08 | 2.42401e-08 | 3.14421e-07 | 1.32779e-08 | 1.66919e-09 | 1.16590e-08 | 3.61766e-10 | 1.80830e-09 | 1.94036e-09 |
| $^{115}$Sn | 1.04528e-07 | 7.81968e-08 | 5.85981e-08 | 3.50923e-08 | 1.93490e-08 | 1.06037e-08 | 1.07684e-08 | 3.49603e-09 | 8.37117e-10 | 3.74549e-10 | 6.77243e-11 | 6.71138e-11 | 3.78532e-11 |
| $^{120}$Te | 4.42702e-08 | 3.00903e-08 | 2.29322e-08 | 3.32563e-08 | 1.86413e-08 | 6.36397e-09 | 2.20768e-08 | 3.23024e-09 | 3.87289e-10 | 1.77532e-09 | 1.46707e-10 | 2.00134e-10 | 2.43037e-10 |
| $^{124}$Xe | 6.71788e-08 | 4.92063e-08 | 3.68356e-08 | 8.95798e-08 | 4.82799e-08 | 1.03408e-08 | 1.66280e-07 | 7.46943e-09 | 5.44751e-10 | 8.30321e-09 | 3.27306e-10 | 1.16997e-09 | 1.72015e-09 |
| $^{126}$Xe | 8.63200e-08 | 4.73186e-08 | 3.84599e-08 | 7.73996e-08 | 4.47944e-08 | 1.39591e-08 | 1.14387e-07 | 8.26786e-09 | 7.99831e-10 | 8.23401e-09 | 3.69129e-10 | 6.63389e-10 | 1.42276e-09 |
| $^{130}$Ba | 7.88192e-08 | 3.72535e-08 | 3.05272e-08 | 1.26607e-07 | 7.46464e-08 | 2.68738e-08 | 2.31902e-07 | 1.26327e-08 | 7.67535e-10 | 1.28041e-08 | 2.65476e-10 | 5.72738e-10 | 1.25763e-09 |
| $^{132}$Ba | 1.81625e-07 | 5.83221e-08 | 6.37122e-08 | 1.01071e-07 | 6.20227e-08 | 2.69679e-08 | 1.94342e-07 | 1.09088e-08 | 2.28125e-09 | 1.12156e-08 | 2.37077e-10 | 3.23819e-10 | 9.56991e-10 |
| $^{138}$La | 4.21108e-08 | 3.24311e-08 | 2.67141e-08 | 1.81623e-08 | 1.21118e-08 | 6.38265e-09 | 1.19841e-09 | 2.31371e-09 | 6.11897e-10 | 3.27994e-11 | 1.55447e-11 | 1.88838e-12 | 5.03318e-13 |
| $^{136}$Ce | 3.01947e-08 | 1.90624e-08 | 1.46979e-08 | 2.19171e-08 | 1.32712e-08 | 5.34894e-09 | 3.70873e-08 | 1.98839e-09 | 2.26740e-10 | 2.29361e-09 | 3.07693e-11 | 2.30252e-10 | 1.11951e-09 |
| $^{138}$Ce | 6.43363e-08 | 3.47684e-08 | 2.90944e-08 | 2.94811e-08 | 1.82266e-08 | 8.29093e-09 | 3.66655e-08 | 2.60596e-09 | 4.75583e-10 | 1.26242e-09 | 3.72746e-11 | 4.75239e-11 | 4.34639e-10 |
| $^{144}$Sm | 1.27748e-07 | 8.48700e-08 | 6.46392e-08 | 1.12745e-07 | 6.24753e-08 | 1.74050e-08 | 4.88027e-07 | 1.26002e-08 | 1.31320e-09 | 2.14791e-07 | 2.62491e-09 | 9.18634e-09 | 1.44548e-07 |
| $^{152}$Gd | 1.93400e-08 | 1.36445e-08 | 9.77190e-09 | 4.93795e-09 | 2.54364e-09 | 1.43723e-09 | 9.20127e-10 | 3.59635e-10 | 7.15055e-11 | 2.07501e-10 | 5.17706e-12 | 2.93456e-12 | 1.76660e-10 |
| $^{156}$Dy | 4.79617e-09 | 3.42935e-09 | 2.63034e-09 | 1.91760e-09 | 1.25974e-09 | 5.84210e-10 | 3.37120e-08 | 2.01387e-10 | 5.44833e-11 | 2.51425e-08 | 1.59771e-11 | 2.10601e-11 | 2.28071e-08 |
| $^{158}$Dy | 8.40067e-09 | 6.18295e-09 | 4.70867e-09 | 3.46221e-09 | 2.22668e-09 | 9.78995e-10 | 6.50132e-09 | 3.32551e-10 | 9.09585e-11 | 6.38430e-09 | 2.61010e-11 | 1.41107e-11 | 8.19305e-09 |

Table 1—Continued

| $Z$ | 2.425e-02 | 1.930e-02 | 1.530e-02 | 9.655e-03 | 6.092e-03 | 3.844e-03 | 2.425e-03 | 1.530e-03 | 4.839e-04 | 1.530e-04 | 4.839e-05 | 1.530e-05 | 1.530e-06 |
|---|---|---|---|---|---|---|---|---|---|---|---|---|---|
| $^{162}$Er | 1.28722e-08 | 7.75931e-09 | 6.40366e-09 | 5.37406e-09 | 3.11401e-09 | 1.36729e-09 | 3.31253e-08 | 6.37572e-10 | 1.93496e-10 | 3.34352e-08 | 1.47418e-10 | 1.02585e-10 | 2.46144e-08 |
| $^{164}$Er | 8.97111e-08 | 5.88349e-08 | 4.41499e-08 | 2.57966e-08 | 1.35750e-08 | 6.65317e-09 | 6.47813e-09 | 2.16912e-09 | 6.98323e-10 | 4.76827e-09 | 2.53971e-10 | 1.35020e-10 | 4.22849e-09 |
| $^{168}$Yb | 1.71831e-08 | 1.15937e-08 | 9.70677e-09 | 6.19452e-09 | 3.55117e-09 | 1.55651e-09 | 1.41516e-08 | 6.28678e-10 | 2.57840e-10 | 2.00374e-08 | 1.42869e-10 | 6.71547e-11 | 1.36738e-08 |
| $^{174}$Hf | 2.20311e-08 | 1.56406e-08 | 1.24874e-08 | 8.49533e-09 | 4.97237e-09 | 1.77971e-09 | 3.99060e-09 | 6.40236e-10 | 3.34103e-10 | 6.75728e-09 | 1.41375e-10 | 3.77573e-11 | 5.35501e-09 |
| $^{180}$Ta | 1.36217e-09 | 9.42184e-10 | 7.43228e-10 | 4.19977e-10 | 3.75115e-10 | 1.30479e-10 | 1.75499e-11 | 3.51158e-11 | 1.74393e-11 | 1.41068e-11 | 7.21638e-12 | 4.32778e-13 | 1.12445e-11 |
| $^{180}$W | 2.23446e-08 | 1.57623e-08 | 1.19163e-08 | 8.23907e-09 | 5.31203e-09 | 1.82766e-09 | 1.43828e-09 | 8.99114e-10 | 6.27149e-10 | 2.39566e-09 | 1.84226e-10 | 7.24569e-11 | 2.47393e-09 |
| $^{184}$Os | 1.41395e-08 | 1.07494e-08 | 8.81491e-09 | 5.67072e-09 | 3.02073e-09 | 1.22682e-09 | 5.57807e-09 | 4.82898e-10 | 4.74947e-10 | 1.20368e-08 | 4.13266e-11 | 1.92286e-11 | 1.04957e-08 |
| $^{190}$Pt | 6.89391e-09 | 5.42408e-09 | 4.38210e-09 | 3.02155e-09 | 1.54209e-09 | 6.80819e-10 | 1.58590e-09 | 2.86060e-10 | 2.24963e-10 | 3.49481e-09 | 2.08891e-11 | 1.33594e-11 | 2.51136e-09 |
| $^{196}$Hg | 4.40220e-07 | 1.99767e-07 | 2.22030e-07 | 6.62471e-07 | 4.75632e-07 | 1.69965e-07 | 1.10202e-07 | 1.84403e-07 | 7.76485e-08 | 2.26518e-07 | 1.78394e-07 | 1.22453e-07 | 1.36948e-07 |
| **20** $M_{\odot}$ | | | | | | | | | | | | | |
| $^{16}$O | 3.48852e+01 | 3.39332e+01 | 3.15883e+01 | 3.48471e+01 | 3.83496e+01 | 3.69048e+01 | 3.32367e+01 | 3.60103e+01 | 3.67319e+01 | 3.53811e+01 | 3.27793e+01 | 3.31798e+01 | 2.98093e+01 |
| $^{56}$Fe | 6.62378e+00 | 8.46595e+00 | 9.65713e+00 | 5.91507e+00 | 6.01645e+00 | 8.55428e+00 | 1.27397e+01 | 1.20039e+01 | 1.19846e+01 | 1.04903e+01 | 1.16206e+01 | 1.05652e+01 | 1.26535e+01 |
| $^{74}$Se | 7.05812e-06 | 1.64490e-05 | 2.01160e-05 | 8.53491e-06 | 1.95614e-06 | 4.78159e-06 | 3.92785e-06 | 2.21459e-06 | 2.51135e-07 | 4.72218e-08 | 2.93044e-08 | 2.25069e-08 | 2.32865e-08 |
| $^{78}$Kr | 1.33615e-06 | 1.57622e-06 | 5.86286e-06 | 1.59793e-06 | 5.40582e-07 | 8.45240e-07 | 5.69811e-07 | 1.18710e-07 | 2.18302e-08 | 8.92446e-09 | 2.97460e-09 | 1.86767e-09 | 1.25991e-09 |
| $^{84}$Sr | 2.56604e-06 | 9.70417e-07 | 5.40746e-06 | 2.23973e-06 | 2.45612e-06 | 1.58439e-06 | 6.51106e-07 | 8.06207e-08 | 2.47108e-08 | 1.96514e-08 | 4.03547e-09 | 4.34311e-09 | 2.25315e-09 |
| $^{92}$Mo | 2.06523e-06 | 1.76777e-06 | 1.26688e-06 | 8.51339e-07 | 5.24635e-07 | 3.40054e-07 | 2.30722e-07 | 1.60201e-07 | 5.38809e-08 | 2.18640e-08 | 8.29872e-09 | 5.12888e-09 | 4.52776e-09 |
| $^{94}$Mo | 1.46016e-06 | 1.17941e-06 | 8.92551e-07 | 5.79059e-07 | 3.63856e-07 | 2.36327e-07 | 1.48684e-07 | 1.01034e-07 | 3.24964e-08 | 1.10337e-08 | 3.94704e-09 | 1.41414e-09 | 2.04257e-10 |
| $^{96}$Ru | 5.85893e-07 | 5.12982e-07 | 3.58935e-07 | 2.41113e-07 | 1.46893e-07 | 9.46705e-08 | 6.51110e-08 | 4.68790e-08 | 1.61875e-08 | 6.48940e-09 | 2.73314e-09 | 1.67476e-09 | 1.25201e-09 |
| $^{98}$Ru | 2.47784e-07 | 2.17984e-07 | 1.40695e-07 | 9.21644e-08 | 5.32548e-08 | 3.68584e-08 | 2.43311e-08 | 1.96381e-08 | 6.36340e-09 | 2.61321e-09 | 1.04932e-09 | 6.91464e-10 | 5.20396e-10 |
| $^{102}$Pd | 1.38581e-07 | 1.76862e-07 | 6.46621e-08 | 5.40146e-08 | 1.91657e-08 | 1.26996e-08 | 1.40362e-08 | 1.77158e-08 | 6.24322e-09 | 3.82309e-09 | 1.80019e-09 | 1.81200e-09 | 1.87489e-09 |
| $^{106}$Cd | 2.23977e-07 | 2.65171e-07 | 9.24034e-08 | 8.30073e-08 | 2.75251e-08 | 1.62824e-08 | 1.90770e-08 | 2.66380e-08 | 1.00943e-08 | 5.94491e-09 | 3.34618e-09 | 3.03289e-09 | 2.53720e-09 |
| $^{108}$Cd | 2.39797e-07 | 1.91320e-07 | 1.15465e-07 | 7.42245e-08 | 3.61742e-08 | 2.31342e-08 | 1.29072e-08 | 1.40182e-08 | 4.17564e-09 | 1.99922e-09 | 9.33829e-10 | 7.54180e-10 | 3.55203e-10 |
| $^{113}$In | 8.70397e-08 | 7.10951e-08 | 3.95993e-08 | 2.49506e-08 | 1.27588e-08 | 7.14738e-09 | 4.22068e-09 | 4.62135e-09 | 1.47211e-09 | 5.14426e-10 | 2.27916e-10 | 1.72125e-10 | 6.23803e-11 |
| $^{112}$Sn | 5.84484e-07 | 5.55653e-07 | 1.81905e-07 | 1.64684e-07 | 5.51351e-08 | 3.22952e-08 | 2.59700e-08 | 5.15010e-08 | 1.86339e-08 | 8.30610e-09 | 4.67883e-09 | 3.86448e-09 | 1.98005e-09 |
| $^{114}$Sn | 4.17007e-07 | 3.55760e-07 | 1.88139e-07 | 1.10724e-07 | 4.15636e-08 | 3.14189e-08 | 1.59414e-08 | 2.78765e-08 | 9.15182e-09 | 3.39408e-09 | 1.66721e-09 | 1.13262e-09 | 2.76638e-10 |
| $^{115}$Sn | 1.24186e-07 | 9.51821e-08 | 6.74726e-08 | 4.15896e-08 | 2.11870e-08 | 1.21444e-08 | 6.74704e-09 | 4.50823e-09 | 1.17720e-09 | 3.28255e-10 | 1.08036e-10 | 4.32705e-11 | 5.92380e-12 |
| $^{120}$Te | 9.45012e-08 | 8.12114e-08 | 5.27363e-08 | 2.31139e-08 | 8.73224e-09 | 8.63024e-09 | 4.27888e-09 | 7.31202e-09 | 2.60282e-09 | 6.93083e-10 | 4.33309e-10 | 2.15474e-10 | 2.94466e-11 |
| $^{124}$Xe | 2.82076e-07 | 2.34216e-07 | 8.87572e-08 | 4.91688e-08 | 1.35501e-08 | 1.43111e-08 | 6.22022e-09 | 1.98276e-08 | 6.35565e-09 | 2.12623e-09 | 1.40499e-09 | 9.30330e-10 | 1.73758e-10 |
| $^{126}$Xe | 2.02632e-07 | 1.77303e-07 | 1.21574e-07 | 5.39070e-08 | 1.59690e-08 | 2.13328e-08 | 1.08039e-08 | 1.72084e-08 | 5.58569e-09 | 1.86746e-09 | 1.35147e-09 | 8.07450e-10 | 1.01659e-10 |
| $^{130}$Ba | 3.54494e-07 | 3.22594e-07 | 2.71830e-07 | 6.72270e-08 | 1.58155e-08 | 4.94308e-08 | 1.89183e-08 | 3.70114e-08 | 1.04694e-08 | 2.52795e-09 | 9.87167e-10 | 4.98619e-10 | 5.73240e-11 |
| $^{132}$Ba | 2.39318e-07 | 2.83863e-07 | 2.45680e-07 | 6.40712e-08 | 3.66505e-08 | 4.97076e-08 | 3.41365e-08 | 3.08966e-08 | 8.78109e-09 | 2.16869e-09 | 9.77980e-10 | 4.52099e-10 | 5.72602e-11 |
| $^{138}$La | 5.43723e-08 | 5.05609e-08 | 4.22446e-08 | 2.26002e-08 | 1.49322e-08 | 1.02702e-08 | 6.44849e-09 | 4.57584e-09 | 1.16167e-09 | 1.75214e-10 | 2.84634e-11 | 4.68635e-12 | 4.22311e-13 |
| $^{136}$Ce | 7.39344e-08 | 6.12355e-08 | 4.78941e-08 | 1.63695e-08 | 6.22318e-09 | 9.81925e-09 | 4.22475e-09 | 5.97943e-09 | 1.26524e-09 | 4.32668e-10 | 2.69207e-10 | 2.30559e-10 | 3.21611e-11 |
| $^{138}$Ce | 8.64196e-08 | 8.50156e-08 | 6.66023e-08 | 2.28334e-08 | 1.46028e-08 | 1.52011e-08 | 9.78552e-09 | 8.16010e-09 | 1.54120e-09 | 2.56160e-10 | 1.38423e-10 | 6.14359e-11 | 7.63205e-12 |
| $^{144}$Sm | 3.32135e-07 | 2.99364e-07 | 1.43032e-07 | 7.26167e-08 | 2.78406e-08 | 2.66408e-08 | 1.53019e-08 | 3.53340e-08 | 2.09106e-08 | 1.42695e-08 | 1.12407e-08 | 9.48100e-09 | 6.66845e-10 |

Table 1—Continued

| $Z$ | 2.425e-02 | 1.930e-02 | 1.530e-02 | 9.655e-03 | 6.092e-03 | 3.844e-03 | 2.425e-03 | 1.530e-03 | 4.839e-04 | 1.530e-04 | 4.839e-05 | 1.530e-05 | 1.530e-06 |
|---|---|---|---|---|---|---|---|---|---|---|---|---|---|
| $^{152}$Gd | 2.66890e-08 | 1.67558e-08 | 9.54015e-09 | 5.55082e-09 | 2.75072e-09 | 1.34885e-09 | 6.75608e-10 | 4.22943e-10 | 8.36504e-11 | 2.39430e-11 | 9.76215e-12 | 4.28073e-12 | 6.65271e-13 |
| $^{156}$Dy | 6.22067e-09 | 5.82046e-09 | 4.03183e-09 | 2.09413e-09 | 1.42297e-09 | 9.28465e-10 | 6.51784e-10 | 4.46736e-10 | 1.02945e-10 | 5.44368e-11 | 2.44683e-11 | 2.22396e-11 | 2.40043e-12 |
| $^{158}$Dy | 1.14420e-08 | 1.05868e-08 | 7.02061e-09 | 3.58106e-09 | 2.47289e-09 | 1.54703e-09 | 1.04593e-09 | 7.52388e-10 | 1.73831e-10 | 6.14728e-11 | 4.32812e-11 | 2.67220e-11 | 3.87128e-12 |
| $^{162}$Er | 1.49946e-08 | 1.36200e-08 | 1.24243e-08 | 4.11640e-09 | 2.74557e-09 | 2.35472e-09 | 1.62747e-09 | 1.16758e-09 | 5.65206e-10 | 3.13449e-10 | 2.98644e-10 | 1.86834e-10 | 1.69277e-11 |
| $^{164}$Er | 9.91233e-08 | 7.67092e-08 | 5.77454e-08 | 2.60533e-08 | 1.41787e-08 | 8.62271e-09 | 5.19184e-09 | 3.27107e-09 | 1.18662e-09 | 5.24977e-10 | 4.65905e-10 | 2.66509e-10 | 4.84372e-11 |
| $^{168}$Yb | 1.72154e-08 | 1.66330e-08 | 1.26739e-08 | 5.46601e-09 | 3.60869e-09 | 2.72214e-09 | 1.53610e-09 | 1.45794e-09 | 5.22134e-10 | 2.59197e-10 | 2.58743e-10 | 1.32283e-10 | 1.77013e-11 |
| $^{174}$Hf | 1.93153e-08 | 1.94601e-08 | 1.64012e-08 | 5.69078e-09 | 3.95922e-09 | 2.89267e-09 | 1.92900e-09 | 2.26746e-09 | 8.74601e-10 | 4.51849e-10 | 4.98043e-10 | 1.45393e-10 | 1.18506e-11 |
| $^{180}$Ta | 2.60310e-09 | 1.65934e-09 | 1.33642e-09 | 4.01280e-10 | 4.50683e-10 | 2.62179e-10 | 1.05150e-10 | 2.14155e-10 | 5.78999e-11 | 4.24627e-11 | 3.01002e-11 | 5.45721e-12 | 1.00187e-13 |
| $^{180}$W | 1.61534e-08 | 1.69477e-08 | 1.43741e-08 | 5.78189e-09 | 3.43716e-09 | 2.50671e-09 | 1.99033e-09 | 3.18146e-09 | 1.53196e-09 | 1.12669e-09 | 6.49319e-10 | 2.06845e-10 | 5.50834e-11 |
| $^{184}$Os | 1.23630e-08 | 1.31833e-08 | 1.01654e-08 | 3.98349e-09 | 2.62202e-09 | 1.83480e-09 | 1.21160e-09 | 1.44657e-09 | 7.07235e-10 | 5.18616e-10 | 1.49088e-10 | 3.75678e-11 | 1.00623e-11 |
| $^{190}$Pt | 7.12779e-09 | 7.37501e-09 | 5.05656e-09 | 2.72682e-09 | 1.44748e-09 | 8.88175e-10 | 6.62413e-10 | 6.12294e-10 | 3.33278e-10 | 1.80328e-10 | 3.75674e-11 | 2.39381e-11 | 7.91457e-12 |
| $^{196}$Hg | 1.18602e-06 | 1.33280e-06 | 9.08433e-07 | 3.44516e-07 | 1.66619e-07 | 3.00378e-07 | 2.54100e-07 | 5.03926e-07 | 4.00067e-07 | 2.71599e-07 | 3.52382e-07 | 2.57463e-07 | 1.13810e-07 |
| **22** $M_\odot$ | | | | | | | | | | | | | |
| $^{16}$O | 4.16717e+01 | 4.13748e+01 | 3.94105e+01 | 3.95705e+01 | 4.43583e+01 | 4.49804e+01 | 4.78736e+01 | 5.01608e+01 | 5.03453e+01 | 4.76760e+01 | 4.61009e+01 | 4.93970e+01 | 3.90164e+01 |
| $^{56}$Fe | 7.74002e+00 | 7.27638e+00 | 1.12273e+01 | 1.26016e+01 | 1.13805e+01 | 1.08754e+01 | 7.09363e+00 | 6.56342e+00 | 6.76697e+00 | 7.39099e+00 | 7.95636e+00 | 6.96711e+00 | 1.06170e+01 |
| $^{74}$Se | 3.70831e-05 | 2.85373e-05 | 3.28388e-05 | 2.18335e-05 | 1.70501e-05 | 1.35142e-05 | 1.53662e-05 | 7.95526e-06 | 6.96253e-07 | 7.16138e-08 | 3.06947e-08 | 2.19012e-08 | 2.07011e-08 |
| $^{78}$Kr | 1.25512e-05 | 8.92477e-06 | 9.14444e-06 | 4.71382e-06 | 1.04671e-06 | 1.45667e-06 | 9.64599e-07 | 4.86146e-07 | 4.59911e-08 | 8.91599e-09 | 3.57778e-09 | 1.83096e-09 | 1.22247e-09 |
| $^{84}$Sr | 2.87215e-05 | 1.95183e-05 | 5.52070e-06 | 4.18013e-06 | 1.61971e-06 | 2.06439e-06 | 1.25005e-06 | 5.60022e-07 | 5.14668e-08 | 1.36768e-08 | 5.24851e-09 | 2.44036e-09 | 3.19753e-09 |
| $^{92}$Mo | 2.14552e-06 | 1.72486e-06 | 1.41669e-06 | 9.00278e-07 | 6.82355e-07 | 4.12459e-07 | 3.59022e-07 | 2.22855e-07 | 6.91370e-08 | 2.77440e-08 | 1.78970e-08 | 1.09464e-08 | 5.17997e-09 |
| $^{94}$Mo | 1.48909e-06 | 1.19312e-06 | 9.48359e-07 | 6.35567e-07 | 4.40313e-07 | 2.75908e-07 | 3.89767e-07 | 2.09238e-07 | 4.84840e-08 | 1.46106e-08 | 5.88053e-09 | 2.67338e-09 | 2.74054e-10 |
| $^{96}$Ru | 6.07799e-07 | 4.88224e-07 | 4.01702e-07 | 2.53537e-07 | 1.96792e-07 | 1.15569e-07 | 8.23856e-08 | 5.30513e-08 | 1.77039e-08 | 7.43548e-09 | 4.32106e-09 | 2.56720e-09 | 1.74469e-09 |
| $^{98}$Ru | 2.17756e-07 | 1.74822e-07 | 1.46269e-07 | 1.08343e-07 | 9.41766e-08 | 5.24413e-08 | 1.01014e-07 | 5.17313e-08 | 1.16493e-08 | 4.37772e-09 | 3.30330e-09 | 2.15910e-09 | 7.88332e-10 |
| $^{102}$Pd | 1.06147e-07 | 8.09014e-08 | 8.64494e-08 | 5.30630e-08 | 9.57819e-08 | 3.77381e-08 | 7.41771e-08 | 5.12843e-08 | 1.33199e-08 | 6.98399e-09 | 8.50107e-09 | 6.46060e-09 | 2.77589e-09 |
| $^{106}$Cd | 1.69157e-07 | 1.26661e-07 | 1.29308e-07 | 7.01291e-08 | 1.41699e-07 | 5.44814e-08 | 5.31316e-08 | 3.87916e-08 | 1.27176e-08 | 7.52423e-09 | 8.26865e-09 | 6.79022e-09 | 4.45141e-09 |
| $^{108}$Cd | 1.92744e-07 | 1.47708e-07 | 1.14625e-07 | 8.71459e-08 | 8.46781e-08 | 4.01161e-08 | 1.28663e-07 | 8.23797e-08 | 1.64327e-08 | 5.82395e-09 | 6.31201e-09 | 4.73650e-09 | 1.03116e-09 |
| $^{113}$In | 7.58316e-08 | 5.70218e-08 | 4.30502e-08 | 2.49275e-08 | 2.54181e-08 | 9.96447e-09 | 2.29925e-08 | 1.87155e-08 | 4.67934e-09 | 1.34769e-09 | 1.18170e-09 | 1.13494e-09 | 2.69377e-10 |
| $^{112}$Sn | 3.38104e-07 | 2.52781e-07 | 2.16039e-07 | 1.33227e-07 | 2.79575e-07 | 1.12635e-07 | 1.82901e-07 | 1.49969e-07 | 4.83450e-08 | 1.73531e-08 | 1.91480e-08 | 1.61221e-08 | 5.38285e-09 |
| $^{114}$Sn | 2.45725e-07 | 1.84119e-07 | 1.43475e-07 | 1.53655e-07 | 1.60687e-07 | 7.88763e-08 | 3.94376e-07 | 2.38388e-07 | 6.41584e-08 | 2.25397e-08 | 1.42191e-08 | 1.07820e-08 | 1.43393e-09 |
| $^{115}$Sn | 1.26414e-07 | 9.52579e-08 | 7.08336e-08 | 4.21516e-08 | 2.62325e-08 | 1.41074e-08 | 2.08923e-08 | 1.21531e-08 | 2.88947e-09 | 8.94681e-10 | 4.09309e-10 | 3.03228e-10 | 3.88837e-11 |
| $^{120}$Te | 4.92611e-08 | 3.67089e-08 | 2.93284e-08 | 4.34847e-08 | 3.82119e-08 | 1.95996e-08 | 3.98722e-08 | 2.26060e-08 | 9.39233e-09 | 3.66163e-09 | 1.78273e-09 | 1.09178e-09 | 1.95353e-10 |
| $^{124}$Xe | 8.20830e-08 | 6.06997e-08 | 4.87967e-08 | 8.29364e-08 | 1.03256e-07 | 4.76640e-08 | 1.28837e-07 | 8.21107e-08 | 3.43218e-08 | 1.41892e-08 | 1.05933e-08 | 7.45069e-09 | 1.08724e-09 |
| $^{126}$Xe | 7.47596e-08 | 5.49550e-08 | 5.31419e-08 | 9.71045e-08 | 8.60832e-08 | 5.15386e-08 | 2.97029e-07 | 1.62545e-07 | 5.21859e-08 | 1.74702e-08 | 9.96975e-09 | 6.58454e-09 | 7.74415e-10 |
| $^{130}$Ba | 6.20068e-08 | 4.47741e-08 | 4.20213e-08 | 2.59704e-07 | 2.02287e-07 | 1.01330e-07 | 5.67289e-07 | 3.04287e-07 | 1.00233e-07 | 3.33846e-08 | 1.17794e-08 | 6.13070e-09 | 5.69010e-10 |
| $^{132}$Ba | 6.17673e-08 | 4.47353e-08 | 1.00557e-07 | 2.26721e-07 | 1.57361e-07 | 8.54871e-08 | 7.21567e-07 | 4.25795e-07 | 1.16263e-07 | 2.82045e-08 | 7.84401e-09 | 3.54194e-09 | 4.34042e-10 |
| $^{138}$La | 5.83530e-08 | 4.53713e-08 | 4.37552e-08 | 3.35732e-08 | 2.48281e-08 | 1.49211e-08 | 3.52990e-09 | 2.40973e-09 | 5.48911e-10 | 8.51734e-11 | 1.06675e-11 | 4.57627e-12 | 3.95850e-13 |

# Table 1—Continued

| $Z$ | 2.425e-02 | 1.930e-02 | 1.530e-02 | 9.655e-03 | 6.092e-03 | 3.844e-03 | 2.425e-03 | 1.530e-03 | 4.839e-04 | 1.530e-04 | 4.839e-05 | 1.530e-05 | 1.530e-06 |
|---|---|---|---|---|---|---|---|---|---|---|---|---|---|
| $^{136}$Ce | 3.11269e-08 | 2.30769e-08 | 1.94838e-08 | 4.06366e-08 | 3.76653e-08 | 1.91521e-08 | 1.21865e-07 | 6.07483e-08 | 1.50730e-08 | 4.81933e-09 | 4.50027e-09 | 3.71922e-09 | 2.76226e-10 |
| $^{138}$Ce | 4.37656e-08 | 3.24852e-08 | 4.81635e-08 | 5.66293e-08 | 4.83745e-08 | 2.62112e-08 | 1.47289e-07 | 8.21057e-08 | 1.53760e-08 | 2.61629e-09 | 1.65864e-09 | 1.16069e-09 | 5.67035e-11 |
| $^{144}$Sm | 1.33323e-07 | 9.93358e-08 | 9.23141e-08 | 1.21367e-07 | 1.49521e-07 | 6.76273e-08 | 8.89482e-07 | 6.46929e-07 | 4.85056e-07 | 4.48792e-07 | 4.98797e-07 | 4.13462e-07 | 1.23728e-08 |
| $^{152}$Gd | 2.41504e-08 | 1.63979e-08 | 1.01590e-08 | 5.33186e-09 | 3.05190e-09 | 1.67391e-09 | 1.14566e-09 | 7.83590e-10 | 3.72225e-10 | 1.98301e-10 | 2.06307e-10 | 2.63399e-10 | 1.13997e-12 |
| $^{156}$Dy | 4.19445e-09 | 3.15238e-09 | 4.69187e-09 | 3.43408e-09 | 2.83644e-09 | 1.69114e-09 | 5.43165e-08 | 5.28275e-08 | 3.92949e-08 | 2.37519e-08 | 2.65583e-08 | 3.42048e-08 | 2.47384e-11 |
| $^{158}$Dy | 8.32503e-09 | 6.61580e-09 | 8.91687e-09 | 5.98584e-09 | 5.05019e-09 | 2.84700e-09 | 2.05044e-08 | 1.93783e-08 | 1.38063e-08 | 8.37666e-09 | 8.06353e-09 | 1.19206e-08 | 1.77940e-11 |
| $^{162}$Er | 7.59982e-09 | 5.81634e-09 | 1.18010e-08 | 9.05863e-09 | 5.88245e-09 | 3.43799e-09 | 6.46405e-08 | 6.08002e-08 | 4.48850e-08 | 3.03685e-08 | 3.25709e-08 | 3.88116e-08 | 1.93301e-10 |
| $^{164}$Er | 9.62514e-08 | 6.93356e-08 | 5.89738e-08 | 3.53097e-08 | 2.01697e-08 | 1.09960e-08 | 2.01085e-08 | 1.65999e-08 | 1.16998e-08 | 8.82226e-09 | 8.12963e-09 | 9.89750e-09 | 2.44494e-10 |
| $^{168}$Yb | 9.84959e-09 | 8.08695e-09 | 1.32715e-08 | 8.57475e-09 | 8.70463e-09 | 4.83632e-09 | 4.07824e-08 | 3.55563e-08 | 2.70727e-08 | 2.12612e-08 | 2.08817e-08 | 2.39053e-08 | 1.02979e-10 |
| $^{174}$Hf | 1.78802e-08 | 1.34551e-08 | 1.68038e-08 | 1.11678e-08 | 1.10111e-08 | 5.40180e-09 | 1.44650e-08 | 1.31213e-08 | 1.01538e-08 | 7.98061e-09 | 6.72242e-09 | 8.95302e-09 | 7.32002e-11 |
| $^{180}$Ta | 2.31285e-09 | 1.56865e-09 | 1.18304e-09 | 9.33261e-10 | 2.03825e-09 | 9.62965e-10 | 7.20272e-11 | 6.00545e-11 | 4.41017e-11 | 3.13141e-11 | 3.02602e-11 | 3.70030e-11 | 4.92603e-13 |
| $^{180}$W | 1.27356e-08 | 9.20788e-09 | 1.53273e-08 | 9.97522e-09 | 1.24408e-08 | 6.12082e-09 | 7.27991e-09 | 6.24397e-09 | 4.82668e-09 | 4.03642e-09 | 2.95007e-09 | 4.24627e-09 | 1.35183e-10 |
| $^{184}$Os | 9.68702e-09 | 7.71212e-09 | 1.09248e-08 | 8.40133e-09 | 8.03985e-09 | 4.19429e-09 | 3.06289e-08 | 2.87319e-08 | 2.27553e-08 | 1.78501e-08 | 1.26984e-08 | 1.97371e-08 | 2.19171e-11 |
| $^{190}$Pt | 6.85301e-09 | 5.28604e-09 | 5.58936e-09 | 3.64944e-09 | 3.58873e-09 | 1.91188e-09 | 3.07743e-09 | 3.26405e-09 | 2.22437e-09 | 1.43006e-09 | 9.52525e-10 | 1.82412e-09 | 7.40116e-12 |
| $^{196}$Hg | 4.88617e-08 | 2.86240e-08 | 3.80445e-07 | 1.05971e-06 | 1.14861e-06 | 7.80764e-07 | 6.87687e-07 | 6.62346e-07 | 5.45300e-07 | 4.89205e-07 | 3.44417e-07 | 4.64964e-07 | 2.80201e-07 |
| **25** $M_\odot$ | | | | | | | | | | | | | |
| $^{16}$O | 5.21676e+01 | 5.44693e+01 | 5.46062e+01 | 5.87423e+01 | 6.04172e+01 | 6.01181e+01 | 6.06634e+01 | 6.08070e+01 | 6.06864e+01 | 6.30771e+01 | 6.15922e+01 | 5.95609e+01 | 6.08842e+01 |
| $^{56}$Fe | 1.18646e+01 | 1.06578e+01 | 8.24455e+00 | 6.53270e+00 | 6.55681e+00 | 7.24817e+00 | 8.97689e+00 | 9.59573e+00 | 9.21586e+00 | 1.11455e+01 | 9.39098e+00 | 8.51757e+00 | 4.82630e+00 |
| $^{74}$Se | 4.44637e-05 | 3.14929e-05 | 5.14964e-05 | 1.20121e-04 | 6.68142e-05 | 5.60131e-05 | 1.67243e-05 | 6.70113e-06 | 5.87141e-07 | 7.55273e-08 | 2.95150e-08 | 2.29355e-08 | 1.85445e-08 |
| $^{78}$Kr | 1.14447e-05 | 7.37343e-06 | 7.68700e-06 | 1.51155e-05 | 1.09573e-05 | 7.38645e-06 | 2.96044e-06 | 1.13149e-06 | 9.56191e-08 | 1.03841e-08 | 4.36615e-09 | 2.67227e-09 | 1.16243e-09 |
| $^{84}$Sr | 1.47862e-05 | 1.27005e-05 | 1.46944e-05 | 2.89291e-05 | 2.32692e-05 | 1.26763e-05 | 6.68894e-06 | 2.14785e-06 | 1.55968e-07 | 1.77560e-08 | 1.15761e-08 | 5.06681e-09 | 1.28318e-09 |
| $^{92}$Mo | 2.58630e-06 | 2.04914e-06 | 1.72868e-06 | 1.67931e-06 | 9.90827e-07 | 7.98375e-07 | 3.40208e-07 | 1.98561e-07 | 6.63051e-08 | 2.14903e-08 | 9.49085e-09 | 1.39358e-08 | 8.40483e-09 |
| $^{94}$Mo | 1.83599e-06 | 1.47141e-06 | 1.16344e-06 | 8.21500e-07 | 5.43649e-07 | 3.77290e-07 | 2.04924e-07 | 1.24151e-07 | 3.94407e-08 | 1.28394e-08 | 4.54395e-09 | 1.86500e-09 | 1.12006e-09 |
| $^{96}$Ru | 7.20957e-07 | 5.66096e-07 | 4.84895e-07 | 4.31068e-07 | 2.53044e-07 | 1.94942e-07 | 9.09358e-08 | 5.46896e-08 | 1.88787e-08 | 6.52062e-09 | 2.85583e-09 | 4.20591e-09 | 2.05606e-09 |
| $^{98}$Ru | 3.47260e-07 | 2.73436e-07 | 2.17669e-07 | 2.64197e-07 | 1.52188e-07 | 1.40812e-07 | 4.57202e-08 | 2.40419e-08 | 7.46429e-09 | 2.47847e-09 | 1.09907e-09 | 2.37376e-09 | 1.94907e-09 |
| $^{102}$Pd | 2.59170e-07 | 1.72039e-07 | 1.66988e-07 | 4.85193e-07 | 2.09469e-07 | 2.53732e-07 | 4.30925e-08 | 1.76275e-08 | 6.45281e-09 | 2.35119e-09 | 1.73351e-09 | 1.00872e-08 | 6.52081e-09 |
| $^{106}$Cd | 4.12668e-07 | 2.27268e-07 | 2.40281e-07 | 5.95172e-07 | 2.44533e-07 | 3.11085e-07 | 5.37547e-08 | 2.52198e-08 | 9.84488e-09 | 4.36709e-09 | 3.03016e-09 | 1.69700e-08 | 7.80079e-09 |
| $^{108}$Cd | 3.63843e-07 | 2.49536e-07 | 2.06159e-07 | 4.37266e-07 | 2.02995e-07 | 2.50098e-07 | 3.99807e-08 | 1.63956e-08 | 4.51906e-09 | 1.43894e-09 | 8.02961e-10 | 5.68529e-09 | 6.73948e-09 |
| $^{113}$In | 1.34888e-07 | 9.19596e-08 | 6.12718e-08 | 1.32510e-07 | 5.23878e-08 | 9.25192e-08 | 9.60128e-09 | 4.09842e-09 | 1.27820e-09 | 4.70249e-10 | 1.87570e-10 | 3.11749e-09 | 2.74616e-09 |
| $^{112}$Sn | 1.27870e-06 | 8.03574e-07 | 4.96493e-07 | 1.13554e-06 | 4.20189e-07 | 8.12127e-07 | 9.43480e-08 | 4.63453e-08 | 1.70106e-08 | 8.32460e-09 | 4.30187e-09 | 4.84544e-08 | 2.24090e-08 |
| $^{114}$Sn | 7.88521e-07 | 5.98217e-07 | 3.99581e-07 | 6.81283e-07 | 3.06027e-07 | 5.13914e-07 | 6.85110e-08 | 3.06906e-08 | 8.98298e-09 | 4.26836e-09 | 1.62473e-09 | 2.04585e-08 | 1.64204e-08 |
| $^{115}$Sn | 1.56496e-07 | 1.17522e-07 | 9.34414e-08 | 8.62118e-08 | 4.44535e-08 | 4.41814e-08 | 1.10251e-08 | 5.62310e-09 | 1.44156e-09 | 4.21660e-10 | 1.56079e-10 | 7.01713e-10 | 8.19373e-10 |
| $^{120}$Te | 1.81271e-07 | 1.44349e-07 | 7.76429e-08 | 5.39052e-08 | 3.25938e-08 | 3.41186e-08 | 1.18456e-08 | 6.46567e-09 | 2.12545e-09 | 1.12754e-09 | 3.48403e-10 | 1.73123e-09 | 7.55805e-10 |
| $^{124}$Xe | 5.48765e-07 | 4.29101e-07 | 1.81689e-07 | 1.77501e-07 | 8.95999e-08 | 1.43933e-07 | 3.05195e-08 | 1.74891e-08 | 5.82828e-09 | 2.99602e-09 | 1.24413e-09 | 1.58006e-08 | 5.93394e-09 |
| $^{126}$Xe | 3.90299e-07 | 3.17626e-07 | 2.10772e-07 | 1.78376e-07 | 1.03325e-07 | 1.25379e-07 | 3.47540e-08 | 1.91743e-08 | 6.00803e-09 | 2.88384e-09 | 1.24077e-09 | 5.65498e-09 | 5.40022e-09 |

Table 1—Continued

| $Z$ | 2.425e-02 | 1.930e-02 | 1.530e-02 | 9.655e-03 | 6.092e-03 | 3.844e-03 | 2.425e-03 | 1.530e-03 | 4.839e-04 | 1.530e-04 | 4.839e-05 | 1.530e-05 | 1.530e-06 |
|---|---|---|---|---|---|---|---|---|---|---|---|---|---|
| $^{130}$Ba | 9.19134e-07 | 7.32048e-07 | 3.27254e-07 | 2.51504e-07 | 1.62420e-07 | 1.99340e-07 | 6.41914e-08 | 3.39085e-08 | 9.83025e-09 | 4.06188e-09 | 9.72423e-10 | 8.11895e-09 | 5.28830e-09 |
| $^{132}$Ba | 6.73468e-07 | 5.41186e-07 | 3.30578e-07 | 2.54639e-07 | 1.71875e-07 | 1.59865e-07 | 5.94501e-08 | 3.04678e-08 | 8.48170e-09 | 3.82281e-09 | 9.59879e-10 | 3.24025e-09 | 2.63159e-09 |
| $^{138}$La | 1.13040e-07 | 9.11904e-08 | 6.08374e-08 | 2.79854e-08 | 2.10025e-08 | 1.13113e-08 | 9.11174e-09 | 5.90381e-09 | 1.47736e-09 | 4.12614e-10 | 4.93177e-11 | 4.59667e-12 | 6.35150e-12 |
| $^{136}$Ce | 1.73461e-07 | 1.33888e-07 | 6.89589e-08 | 5.60649e-08 | 3.72110e-08 | 4.43312e-08 | 1.35739e-08 | 6.70891e-09 | 1.58523e-09 | 5.48043e-10 | 2.92456e-10 | 3.69607e-09 | 3.19192e-09 |
| $^{138}$Ce | 2.05424e-07 | 1.61213e-07 | 1.02553e-07 | 6.64099e-08 | 4.81737e-08 | 3.95411e-08 | 1.73480e-08 | 8.98075e-09 | 1.74544e-09 | 4.55069e-10 | 1.01193e-10 | 6.23409e-10 | 7.06450e-10 |
| $^{144}$Sm | 7.21597e-07 | 5.34503e-07 | 3.04125e-07 | 3.85557e-07 | 2.11949e-07 | 4.22198e-07 | 6.52492e-08 | 3.33490e-08 | 1.91517e-08 | 1.74214e-08 | 1.02351e-08 | 2.51150e-07 | 1.94938e-07 |
| $^{152}$Gd | 2.41406e-08 | 1.69738e-08 | 1.25816e-08 | 6.61967e-09 | 3.33642e-09 | 1.84149e-09 | 9.95780e-10 | 5.20573e-10 | 9.62012e-11 | 2.50846e-11 | 9.94148e-12 | 8.38309e-12 | 6.44267e-11 |
| $^{156}$Dy | 1.11628e-08 | 9.66640e-09 | 7.41973e-09 | 5.38433e-09 | 3.80208e-09 | 3.23559e-09 | 1.29299e-09 | 6.56145e-10 | 1.36991e-10 | 5.31110e-11 | 2.80780e-11 | 4.75138e-10 | 8.22193e-09 |
| $^{158}$Dy | 2.09403e-08 | 1.83810e-08 | 1.34065e-08 | 7.86495e-09 | 5.64098e-09 | 3.78202e-09 | 2.01753e-09 | 1.04935e-09 | 1.93304e-10 | 8.90012e-11 | 4.03654e-11 | 1.12554e-10 | 3.38494e-09 |
| $^{162}$Er | 2.75723e-08 | 2.12650e-08 | 1.58358e-08 | 1.07199e-08 | 7.70698e-09 | 6.33174e-09 | 2.75049e-09 | 1.44348e-09 | 5.20352e-10 | 4.91678e-10 | 2.58721e-10 | 9.82853e-10 | 1.61747e-08 |
| $^{164}$Er | 1.38612e-07 | 1.00625e-07 | 7.11438e-08 | 3.60718e-08 | 2.12466e-08 | 1.24237e-08 | 6.88835e-09 | 3.90390e-09 | 1.06100e-09 | 8.47975e-10 | 4.15544e-10 | 8.44517e-10 | 5.48559e-09 |
| $^{168}$Yb | 3.75789e-08 | 3.06520e-08 | 2.34081e-08 | 1.41905e-08 | 1.04604e-08 | 7.52739e-09 | 3.73881e-09 | 1.90324e-09 | 4.56674e-10 | 3.75588e-10 | 2.08526e-10 | 5.63832e-10 | 1.32121e-08 |
| $^{174}$Hf | 4.32180e-08 | 3.63149e-08 | 2.37699e-08 | 1.23624e-08 | 9.34581e-09 | 6.30439e-09 | 4.06118e-09 | 2.70017e-09 | 5.30065e-10 | 8.59740e-10 | 3.73936e-10 | 2.67565e-10 | 5.55615e-09 |
| $^{180}$Ta | 4.86977e-09 | 4.43454e-09 | 1.76313e-09 | 6.11150e-10 | 5.19921e-10 | 2.80691e-10 | 4.92240e-10 | 4.16975e-10 | 6.35775e-11 | 5.74528e-11 | 5.57153e-11 | 3.24021e-12 | 1.85420e-11 |
| $^{180}$W | 4.14769e-08 | 3.46711e-08 | 2.18157e-08 | 1.00279e-08 | 7.64716e-09 | 4.84972e-09 | 3.72290e-09 | 2.89988e-09 | 8.05973e-10 | 1.66584e-09 | 6.33000e-10 | 3.07015e-10 | 3.77022e-09 |
| $^{184}$Os | 3.79094e-08 | 2.92489e-08 | 1.66348e-08 | 8.08966e-09 | 5.87409e-09 | 3.88707e-09 | 2.17379e-09 | 1.25537e-09 | 3.30434e-10 | 6.72367e-10 | 1.75847e-10 | 2.24945e-10 | 1.32195e-08 |
| $^{190}$Pt | 1.77603e-08 | 1.42280e-08 | 8.52859e-09 | 4.44591e-09 | 2.97270e-09 | 1.84433e-09 | 1.09553e-09 | 6.16820e-10 | 1.72943e-10 | 2.58845e-10 | 5.19566e-11 | 2.07465e-11 | 1.93626e-09 |
| $^{196}$Hg | 3.27397e-06 | 2.75119e-06 | 1.52411e-06 | 1.11737e-06 | 8.63846e-07 | 1.03593e-06 | 5.66228e-07 | 4.69254e-07 | 3.58983e-07 | 4.85523e-07 | 3.17931e-07 | 7.57384e-07 | 7.12994e-07 |
| **30** $M_{\odot}$ | | | | | | | | | | | | | |
| $^{16}$O | 8.05595e+01 | 8.37527e+01 | 8.33730e+01 | 8.80350e+01 | 9.11110e+01 | 8.67185e+01 | 1.22414e-01 | 4.11828e-02 | 5.80398e-03 | 1.13581e-03 | 2.31076e-04 | 4.64366e-05 | 8.79101e+01 |
| $^{56}$Fe | 9.08993e+00 | 9.40085e+00 | 9.23433e+00 | 1.16493e+01 | 1.35324e+01 | 1.34309e+01 | 1.79056e-02 | 5.20518e-03 | 5.74514e-04 | 8.97145e-05 | 1.46397e-05 | 2.36708e-06 | 1.37728e+01 |
| $^{74}$Se | 2.08509e-05 | 2.28760e-05 | 2.65306e-05 | 3.64622e-05 | 2.52657e-05 | 1.47648e-05 | 6.73885e-08 | 1.96252e-08 | 2.13128e-09 | 3.31291e-10 | 5.47150e-11 | 9.08456e-12 | 2.56553e-08 |
| $^{78}$Kr | 3.23751e-06 | 1.38129e-06 | 1.29790e-06 | 2.20485e-06 | 1.89911e-06 | 1.05647e-06 | 2.52429e-08 | 7.35137e-09 | 7.98352e-10 | 1.24097e-10 | 2.04955e-11 | 3.40296e-12 | 1.19380e-09 |
| $^{84}$Sr | 6.48882e-06 | 1.82268e-06 | 7.91026e-07 | 8.33618e-07 | 1.12469e-06 | 7.18245e-07 | 1.87917e-08 | 5.47263e-09 | 5.94323e-10 | 9.23828e-11 | 1.52577e-11 | 2.53329e-12 | 7.76884e-10 |
| $^{92}$Mo | 2.89335e-06 | 2.33725e-06 | 1.93081e-06 | 1.28689e-06 | 8.87738e-07 | 5.79805e-07 | 8.11830e-08 | 2.47926e-08 | 2.98000e-09 | 5.01483e-10 | 8.80186e-11 | 1.52998e-11 | 3.17222e-09 |
| $^{94}$Mo | 1.99898e-06 | 1.58607e-06 | 1.28076e-06 | 8.45090e-07 | 5.94238e-07 | 3.86774e-07 | 5.32225e-08 | 1.62441e-08 | 1.95021e-09 | 3.27912e-10 | 5.75194e-11 | 9.99396e-12 | 3.05622e-10 |
| $^{96}$Ru | 8.24208e-07 | 6.71586e-07 | 5.60482e-07 | 3.79052e-07 | 2.61219e-07 | 1.70278e-07 | 2.35709e-08 | 7.19837e-09 | 8.65224e-10 | 1.45603e-10 | 2.55557e-11 | 4.44220e-12 | 1.50782e-09 |
| $^{98}$Ru | 3.51739e-07 | 2.95311e-07 | 2.42331e-07 | 1.76780e-07 | 1.45646e-07 | 9.25420e-08 | 8.28288e-09 | 2.52952e-09 | 3.04041e-10 | 5.11650e-11 | 8.98031e-12 | 1.56100e-12 | 6.38212e-10 |
| $^{102}$Pd | 1.98836e-07 | 2.14164e-07 | 1.90749e-07 | 1.54293e-07 | 1.26064e-07 | 7.68211e-08 | 2.58788e-09 | 7.28063e-10 | 7.25924e-11 | 1.04088e-11 | 1.59858e-12 | 2.49299e-13 | 2.27511e-09 |
| $^{106}$Cd | 2.89289e-07 | 3.71645e-07 | 3.19922e-07 | 2.39521e-07 | 1.73186e-07 | 1.05369e-07 | 3.38416e-09 | 9.06167e-10 | 7.84065e-11 | 9.49822e-12 | 1.20008e-12 | 1.50166e-13 | 4.01760e-09 |
| $^{108}$Cd | 2.97197e-07 | 2.79039e-07 | 2.20820e-07 | 1.56962e-07 | 1.23098e-07 | 7.33990e-08 | 2.50162e-09 | 6.69782e-10 | 5.79505e-11 | 7.01995e-12 | 8.86931e-13 | 1.10980e-13 | 8.30052e-10 |
| $^{113}$In | 1.22005e-07 | 1.05201e-07 | 7.93502e-08 | 4.84375e-08 | 3.09850e-08 | 1.83265e-08 | 1.49100e-09 | 3.99239e-10 | 3.45443e-11 | 4.18473e-12 | 5.28732e-13 | 6.61602e-14 | 2.79105e-10 |
| $^{112}$Sn | 8.47878e-07 | 9.00557e-07 | 6.87186e-07 | 4.63673e-07 | 3.25689e-07 | 1.93840e-07 | 6.72123e-09 | 1.79972e-09 | 1.55722e-10 | 1.88643e-11 | 2.38346e-12 | 2.98243e-13 | 5.00589e-09 |
| $^{114}$Sn | 5.96767e-07 | 5.06039e-07 | 3.83376e-07 | 2.71321e-07 | 2.19070e-07 | 1.30770e-07 | 4.72594e-09 | 1.26545e-09 | 1.09494e-10 | 1.32641e-11 | 1.67590e-12 | 2.09705e-13 | 1.18429e-09 |
| $^{115}$Sn | 1.71198e-07 | 1.29102e-07 | 9.76543e-08 | 5.63117e-08 | 3.39301e-08 | 1.99751e-08 | 2.47357e-09 | 6.62313e-10 | 5.73012e-11 | 6.94082e-12 | 8.76872e-13 | 1.09712e-13 | 4.00745e-11 |

Table 1—Continued

| $Z$ | 2.425e-02 | 1.930e-02 | 1.530e-02 | 9.655e-03 | 6.092e-03 | 3.844e-03 | 2.425e-03 | 1.530e-03 | 4.839e-04 | 1.530e-04 | 4.839e-05 | 1.530e-05 | 1.530e-06 |
|---|---|---|---|---|---|---|---|---|---|---|---|---|---|
| $^{120}$Te | 1.31094e-07 | 1.04230e-07 | 8.17861e-08 | 6.00853e-08 | 4.66499e-08 | 2.75784e-08 | 9.92846e-10 | 2.65852e-10 | 2.30030e-11 | 2.78660e-12 | 3.52081e-13 | 4.40559e-14 | 2.17020e-10 |
| $^{124}$Xe | 3.84452e-07 | 3.29007e-07 | 2.46000e-07 | 1.63281e-07 | 1.14831e-07 | 6.75418e-08 | 1.65354e-09 | 4.42766e-10 | 3.83105e-11 | 4.64097e-12 | 5.86377e-13 | 7.33733e-14 | 1.02072e-09 |
| $^{126}$Xe | 2.79561e-07 | 2.33198e-07 | 1.88064e-07 | 1.33884e-07 | 1.00704e-07 | 6.06438e-08 | 1.48047e-09 | 3.96421e-10 | 3.43006e-11 | 4.15520e-12 | 5.25000e-13 | 6.56933e-14 | 7.06933e-10 |
| $^{130}$Ba | 4.94747e-07 | 3.96284e-07 | 3.12889e-07 | 2.64383e-07 | 2.04968e-07 | 1.24562e-07 | 1.22400e-09 | 3.27749e-10 | 2.83587e-11 | 3.43539e-12 | 4.34055e-13 | 5.43133e-14 | 4.69658e-10 |
| $^{132}$Ba | 4.45806e-07 | 3.62004e-07 | 2.88509e-07 | 2.23113e-07 | 1.75927e-07 | 1.06644e-07 | 1.20709e-09 | 3.23220e-10 | 2.79668e-11 | 3.38792e-12 | 4.28057e-13 | 5.35627e-14 | 4.24876e-10 |
| $^{138}$La | 1.09534e-07 | 9.08865e-08 | 7.10027e-08 | 5.11528e-08 | 3.62299e-08 | 2.16266e-08 | 1.21875e-10 | 3.26344e-11 | 2.82372e-12 | 3.42068e-13 | 4.32196e-14 | 5.40808e-15 | 4.77266e-12 |
| $^{136}$Ce | 1.07588e-07 | 8.16123e-08 | 6.46045e-08 | 5.69486e-08 | 4.63256e-08 | 2.79513e-08 | 6.21401e-10 | 1.66391e-10 | 1.43970e-11 | 1.74407e-12 | 2.20360e-13 | 2.75736e-14 | 2.04512e-10 |
| $^{138}$Ce | 1.51556e-07 | 1.18540e-07 | 9.43493e-08 | 7.97895e-08 | 6.77325e-08 | 4.10045e-08 | 8.59947e-10 | 2.30266e-10 | 1.99238e-11 | 2.41359e-12 | 3.04952e-13 | 3.81586e-14 | 8.45706e-11 |
| $^{144}$Sm | 4.88188e-07 | 4.24814e-07 | 3.20603e-07 | 2.31540e-07 | 1.79298e-07 | 1.10023e-07 | 2.59518e-09 | 6.94905e-10 | 6.01269e-11 | 7.28382e-12 | 9.20296e-13 | 1.15157e-13 | 1.27789e-08 |
| $^{152}$Gd | 3.12081e-08 | 2.35140e-08 | 1.43983e-08 | 7.01276e-09 | 3.77126e-09 | 2.19124e-09 | 1.83248e-10 | 4.54172e-11 | 3.30773e-12 | 3.46506e-13 | 3.88495e-14 | 4.41796e-15 | 1.45641e-12 |
| $^{156}$Dy | 1.10619e-08 | 8.48549e-09 | 6.70886e-09 | 5.29100e-09 | 4.46171e-09 | 2.79296e-09 | 8.42461e-11 | 2.25585e-11 | 1.95189e-12 | 2.36454e-13 | 2.98755e-14 | 3.73832e-15 | 1.05579e-11 |
| $^{158}$Dy | 1.98686e-08 | 1.65740e-08 | 1.31177e-08 | 9.86064e-09 | 9.99123e-09 | 8.05531e-09 | 1.46607e-10 | 3.92567e-11 | 3.39671e-12 | 4.11481e-13 | 5.19897e-14 | 6.50547e-15 | 2.66380e-11 |
| $^{162}$Er | 2.46408e-08 | 1.92952e-08 | 1.48535e-08 | 1.15857e-08 | 9.79305e-09 | 6.12305e-09 | 1.46129e-10 | 3.91289e-11 | 3.38566e-12 | 4.10141e-13 | 5.18205e-14 | 6.48430e-15 | 2.56994e-10 |
| $^{164}$Er | 1.46448e-07 | 1.07112e-07 | 7.78870e-08 | 4.69729e-08 | 4.11639e-08 | 2.98479e-08 | 1.20290e-09 | 2.97211e-10 | 2.14231e-11 | 2.22083e-12 | 2.46545e-13 | 2.77829e-14 | 3.63084e-10 |
| $^{168}$Yb | 3.05628e-08 | 2.51607e-08 | 1.96139e-08 | 1.60566e-08 | 2.11740e-08 | 2.05481e-08 | 1.36528e-10 | 3.65580e-11 | 3.16320e-12 | 3.83193e-13 | 4.84156e-14 | 6.05825e-15 | 1.72141e-10 |
| $^{174}$Hf | 3.21194e-08 | 3.21829e-08 | 2.49666e-08 | 3.05213e-08 | 1.12361e-07 | 9.74335e-08 | 1.17122e-10 | 3.13615e-11 | 2.71357e-12 | 3.28724e-13 | 4.15336e-14 | 5.19710e-15 | 2.35955e-10 |
| $^{180}$Ta | 4.79097e-09 | 3.66156e-09 | 2.68614e-09 | 2.84093e-09 | 2.35533e-09 | 1.24779e-09 | 1.28409e-12 | 3.43841e-13 | 2.97510e-14 | 3.60407e-15 | 4.55367e-16 | 5.69800e-17 | 7.23815e-13 |
| $^{180}$W | 2.67969e-08 | 2.49438e-08 | 2.00317e-08 | 3.16139e-08 | 1.13978e-07 | 8.50347e-08 | 8.15010e-11 | 2.18233e-11 | 1.88827e-12 | 2.28747e-13 | 2.89017e-14 | 3.61647e-15 | 2.33967e-10 |
| $^{184}$Os | 2.02273e-08 | 1.97457e-08 | 1.55229e-08 | 1.43989e-08 | 4.03562e-08 | 3.51199e-08 | 6.98325e-11 | 1.86989e-11 | 1.61793e-12 | 1.95998e-13 | 2.47639e-14 | 3.09871e-15 | 4.27389e-11 |
| $^{190}$Pt | 1.04274e-08 | 9.96986e-09 | 7.80980e-09 | 5.74063e-09 | 8.67564e-09 | 8.34253e-09 | 9.67606e-11 | 2.59094e-11 | 2.24182e-12 | 2.71576e-13 | 3.43131e-14 | 4.29360e-15 | 3.99193e-11 |
| $^{196}$Hg | 1.90613e-06 | 1.76201e-06 | 1.46249e-06 | 1.35449e-06 | 1.20361e-06 | 9.08922e-07 | 4.13599e-10 | 1.10748e-10 | 9.58256e-12 | 1.16084e-12 | 1.46670e-13 | 1.83528e-14 | 4.21016e-07 |

Table 2. NUGRID ccSN model: ejected mass

| $Z$ | 2.00E-02 | 1.00E-02 | r2 2.00E-02 |
|---|---|---|---|
| **15** $M_{\odot}$ | | | |
| $^{16}$O | 2.986E-01 | 2.187E-01 | 3.171E-01 |
| $^{56}$Fe | 1.915E-01 | 1.856E-01 | 1.681E-01 |
| $^{74}$Se | 6.317E-05 | 1.249E-04 | 1.122E-06 |
| $^{78}$Kr | 2.331E-06 | 4.761E-06 | 2.355E-07 |
| $^{84}$Sr | 1.291E-06 | 3.055E-06 | 1.004E-07 |
| $^{92}$Mo | 1.849E-06 | 4.088E-06 | 1.162E-08 |
| $^{94}$Mo | 2.150E-08 | 3.238E-08 | 7.419E-09 |
| $^{96}$Ru | 3.263E-09 | 1.783E-09 | 3.430E-09 |
| $^{98}$Ru | 1.191E-09 | 6.130E-10 | 1.254E-09 |
| $^{102}$Pd | 5.976E-10 | 3.999E-10 | 7.600E-10 |
| $^{106}$Cd | 1.034E-09 | 6.582E-10 | 1.152E-09 |
| $^{108}$Cd | 9.370E-10 | 4.009E-10 | 8.760E-10 |
| $^{113}$In | 2.892E-10 | 1.428E-10 | 2.861E-10 |
| $^{112}$Sn | 2.559E-09 | 1.072E-09 | 2.296E-09 |
| $^{114}$Sn | 1.425E-09 | 6.550E-10 | 1.347E-09 |
| $^{115}$Sn | 4.812E-10 | 2.367E-10 | 4.766E-10 |
| $^{120}$Te | 2.730E-10 | 1.392E-10 | 2.729E-10 |
| $^{124}$Xe | 5.187E-10 | 2.950E-10 | 5.117E-10 |
| $^{126}$Xe | 4.924E-10 | 2.916E-10 | 5.180E-10 |
| $^{130}$Ba | 9.539E-10 | 4.483E-10 | 9.977E-10 |
| $^{132}$Ba | 8.556E-10 | 4.111E-10 | 8.548E-10 |
| $^{138}$La | 2.058E-11 | 1.356E-11 | 2.389E-11 |
| $^{136}$Ce | 1.776E-10 | 1.055E-10 | 1.962E-10 |
| $^{138}$Ce | 2.441E-10 | 1.552E-10 | 2.489E-10 |
| $^{144}$Sm | 1.038E-09 | 6.124E-10 | 1.151E-09 |
| $^{152}$Gd | 7.070E-11 | 2.416E-11 | 5.871E-11 |
| $^{156}$Dy | 1.282E-11 | 8.341E-12 | 1.450E-11 |
| $^{158}$Dy | 2.271E-11 | 1.675E-11 | 2.371E-11 |
| $^{162}$Er | 3.844E-11 | 2.110E-11 | 3.579E-11 |
| $^{164}$Er | 3.150E-10 | 1.333E-10 | 2.980E-10 |
| $^{168}$Yb | 4.972E-11 | 3.191E-11 | 4.785E-11 |

Table 2—Continued

| $Z$ | 2.00E-02 | 1.00E-02 | r2 2.00E-02 |
|---|---|---|---|
| $^{174}$Hf | 3.994E-11 | 3.496E-11 | 5.319E-11 |
| $^{180}$Ta | 1.504E-12 | 7.386E-13 | 1.475E-12 |
| $^{180}$W | 5.173E-11 | 4.576E-11 | 7.528E-11 |
| $^{184}$Os | 3.243E-11 | 4.044E-11 | 7.935E-11 |
| $^{190}$Pt | 1.316E-11 | 1.065E-11 | 2.924E-11 |
| $^{196}$Hg | 1.332E-10 | 8.162E-11 | 2.625E-10 |
| **20** $M_{\odot}$ | | | |
| $^{16}$O | 1.266E+00 | 1.443E+00 | |
| $^{56}$Fe | 2.679E-02 | 1.353E-02 | |
| $^{74}$Se | 1.315E-06 | 5.735E-07 | |
| $^{78}$Kr | 1.467E-07 | 7.888E-08 | |
| $^{84}$Sr | 5.166E-08 | 2.492E-08 | |
| $^{92}$Mo | 1.915E-08 | 8.132E-09 | |
| $^{94}$Mo | 1.019E-08 | 5.242E-09 | |
| $^{96}$Ru | 5.655E-09 | 2.464E-09 | |
| $^{98}$Ru | 2.572E-09 | 1.124E-09 | |
| $^{102}$Pd | 2.678E-09 | 1.614E-09 | |
| $^{106}$Cd | 3.814E-09 | 2.829E-09 | |
| $^{108}$Cd | 2.537E-09 | 1.125E-09 | |
| $^{113}$In | 4.002E-10 | 1.934E-10 | |
| $^{112}$Sn | 1.037E-08 | 4.089E-09 | |
| $^{114}$Sn | 5.694E-09 | 2.411E-09 | |
| $^{115}$Sn | 6.028E-10 | 3.056E-10 | |
| $^{120}$Te | 1.425E-09 | 5.980E-10 | |
| $^{124}$Xe | 4.333E-09 | 1.578E-09 | |
| $^{126}$Xe | 3.793E-09 | 1.575E-09 | |
| $^{130}$Ba | 7.406E-09 | 2.707E-09 | |
| $^{132}$Ba | 5.220E-09 | 2.344E-09 | |
| $^{138}$La | 5.850E-11 | 2.504E-11 | |
| $^{136}$Ce | 1.087E-09 | 3.898E-10 | |
| $^{138}$Ce | 1.561E-09 | 6.830E-10 | |
| $^{144}$Sm | 9.942E-09 | 3.930E-09 | |

Table 2—Continued

| $Z$ | 2.00E-02 | 1.00E-02 | r2 2.00E-02 |
|---|---|---|---|
| $^{152}$Gd | 6.764E-11 | 3.512E-11 | |
| $^{156}$Dy | 8.077E-11 | 4.382E-11 | |
| $^{158}$Dy | 1.196E-10 | 5.713E-11 | |
| $^{162}$Er | 1.217E-10 | 6.252E-11 | |
| $^{164}$Er | 9.115E-10 | 4.382E-10 | |
| $^{168}$Yb | 4.135E-10 | 1.439E-10 | |
| $^{174}$Hf | 3.188E-10 | 1.231E-10 | |
| $^{180}$Ta | 4.629E-12 | 2.507E-12 | |
| $^{180}$W | 5.375E-10 | 1.662E-10 | |
| $^{184}$Os | 3.473E-10 | 1.570E-10 | |
| $^{190}$Pt | 7.964E-11 | 3.862E-11 | |
| $^{196}$Hg | 2.440E-09 | 9.964E-10 | |
| **25** $M_\odot$ | | | |
| $^{16}$O | 8.163E-01 | 6.345E-01 | |
| $^{56}$Fe | 2.351E-02 | 1.155E-02 | |
| $^{74}$Se | 2.838E-08 | 4.269E-07 | |
| $^{78}$Kr | 6.479E-09 | 6.543E-09 | |
| $^{84}$Sr | 5.285E-09 | 3.073E-09 | |
| $^{92}$Mo | 1.740E-08 | 8.802E-09 | |
| $^{94}$Mo | 1.332E-08 | 6.866E-09 | |
| $^{96}$Ru | 4.893E-09 | 2.436E-09 | |
| $^{98}$Ru | 1.835E-09 | 1.077E-09 | |
| $^{102}$Pd | 6.960E-10 | 4.703E-10 | |
| $^{106}$Cd | 1.046E-09 | 5.585E-10 | |
| $^{108}$Cd | 1.703E-09 | 1.052E-09 | |
| $^{113}$In | 4.495E-10 | 2.282E-10 | |
| $^{112}$Sn | 2.039E-09 | 1.209E-09 | |
| $^{114}$Sn | 1.593E-09 | 1.490E-09 | |
| $^{115}$Sn | 7.779E-10 | 3.874E-10 | |
| $^{120}$Te | 3.535E-10 | 3.474E-10 | |
| $^{124}$Xe | 3.648E-10 | 7.498E-10 | |
| $^{126}$Xe | 6.589E-10 | 6.437E-10 | |

Table 2—Continued

| $Z$ | 2.00E-02 | 1.00E-02 | r2 2.00E-02 |
|---|---|---|---|
| $^{130}$Ba | 7.492E-10 | 2.029E-09 | |
| $^{132}$Ba | 2.452E-09 | 1.264E-09 | |
| $^{138}$La | 3.810E-11 | 2.209E-11 | |
| $^{136}$Ce | 1.903E-10 | 2.589E-10 | |
| $^{138}$Ce | 4.879E-10 | 3.016E-10 | |
| $^{144}$Sm | 9.939E-10 | 2.555E-09 | |
| $^{152}$Gd | 2.236E-10 | 1.018E-10 | |
| $^{156}$Dy | 6.630E-11 | 2.957E-11 | |
| $^{158}$Dy | 7.766E-11 | 4.142E-11 | |
| $^{162}$Er | 1.402E-10 | 3.984E-11 | |
| $^{164}$Er | 4.471E-10 | 1.822E-10 | |
| $^{168}$Yb | 1.339E-10 | 6.812E-11 | |
| $^{174}$Hf | 1.024E-10 | 4.538E-11 | |
| $^{180}$Ta | 2.128E-12 | 9.543E-13 | |
| $^{180}$W | 7.947E-11 | 3.225E-11 | |
| $^{184}$Os | 1.495E-10 | 5.190E-11 | |
| $^{190}$Pt | 2.874E-11 | 1.246E-11 | |
| $^{196}$Hg | 6.744E-10 | 2.724E-10 | |

Table 3. Chemical evolution of $p$-nuclei with KEPLER and NUGRID models

| Isotope | Solar System[a] | xi45[b] | xi25[c] | nocutoff[d] | ertl[e] | Nugrid[f] |
|---|---|---|---|---|---|---|
| $^{74}$Se | 1.030D-09 | 4.694D-10 | 2.900D-10 | 4.696D-10 | 3.217D-10 | 1.883D-09 |
| $^{78}$Kr | 4.273D-10 | 8.208D-11 | 5.191D-11 | 8.209D-11 | 5.790D-11 | 3.276D-10 |
| $^{84}$Sr | 2.991D-10 | 1.274D-10 | 8.025D-11 | 1.274D-10 | 9.376D-11 | 1.338D-10 |
| $^{92}$Mo | 9.722D-10 | 4.817D-11 | 4.529D-11 | 4.818D-11 | 4.449D-11 | 2.306D-11 |
| $^{94}$Mo | 6.258D-10 | 3.178D-11 | 2.981D-11 | 3.178D-11 | 2.924D-11 | 1.534D-11 |
| $^{96}$Ru | 2.603D-10 | 1.331D-11 | 1.259D-11 | 1.331D-11 | 1.235D-11 | 6.775D-12 |
| $^{98}$Ru | 8.858D-11 | 5.248D-12 | 4.804D-12 | 5.249D-12 | 4.701D-12 | 2.718D-12 |
| $^{102}$Pd | 3.935D-11 | 3.486D-12 | 3.034D-12 | 3.488D-12 | 3.003D-12 | 2.202D-12 |
| $^{106}$Cd | 5.807D-11 | 5.140D-12 | 4.588D-12 | 5.142D-12 | 4.501D-12 | 3.529D-12 |
| $^{108}$Cd | 4.142D-11 | 4.217D-12 | 3.689D-12 | 4.218D-12 | 3.628D-12 | 2.273D-12 |
| $^{113}$In | 2.476D-11 | 1.495D-12 | 1.356D-12 | 1.495D-12 | 1.331D-12 | 5.685D-13 |
| $^{112}$Sn | 1.074D-10 | 1.060D-11 | 9.106D-12 | 1.061D-11 | 8.827D-12 | 6.013D-12 |
| $^{114}$Sn | 7.494D-11 | 7.014D-12 | 5.942D-12 | 7.016D-12 | 5.829D-12 | 3.903D-12 |
| $^{115}$Sn | 3.780D-11 | 2.055D-12 | 1.942D-12 | 2.055D-12 | 1.903D-12 | 9.390D-13 |
| $^{120}$Te | 1.643D-11 | 1.434D-12 | 1.241D-12 | 1.435D-12 | 1.187D-12 | 8.873D-13 |
| $^{124}$Xe | 2.377D-11 | 3.507D-12 | 2.913D-12 | 3.508D-12 | 2.740D-12 | 2.024D-12 |
| $^{126}$Xe | 2.071D-11 | 2.911D-12 | 2.406D-12 | 2.912D-12 | 2.332D-12 | 1.979D-12 |
| $^{130}$Ba | 1.740D-11 | 4.755D-12 | 3.624D-12 | 4.757D-12 | 3.456D-12 | 3.958D-12 |
| $^{132}$Ba | 1.683D-11 | 3.713D-12 | 2.790D-12 | 3.715D-12 | 2.707D-12 | 3.186D-12 |
| $^{138}$La | 1.428D-12 | 5.518D-13 | 4.302D-13 | 5.521D-13 | 3.806D-13 | 5.501D-14 |
| $^{136}$Ce | 8.095D-12 | 9.219D-13 | 7.293D-13 | 9.222D-13 | 7.109D-13 | 6.147D-13 |
| $^{138}$Ce | 1.104D-11 | 1.156D-12 | 9.275D-13 | 1.157D-12 | 8.976D-13 | 8.967D-13 |
| $^{144}$Sm | 3.155D-11 | 4.448D-12 | 3.522D-12 | 4.450D-12 | 3.545D-12 | 5.222D-12 |
| $^{152}$Gd | 2.898D-12 | 2.913D-13 | 2.790D-13 | 2.913D-13 | 2.686D-13 | 1.393D-13 |
| $^{156}$Dy | 9.864D-13 | 9.542D-14 | 7.213D-14 | 9.545D-14 | 8.230D-14 | 6.064D-14 |
| $^{158}$Dy | 1.695D-12 | 1.383D-13 | 1.157D-13 | 1.385D-13 | 1.148D-13 | 8.593D-14 |
| $^{162}$Er | 1.584D-12 | 1.996D-13 | 1.615D-13 | 1.997D-13 | 1.681D-13 | 1.028D-13 |
| $^{164}$Er | 1.847D-11 | 1.115D-12 | 1.048D-12 | 1.116D-12 | 1.020D-12 | 7.288D-13 |
| $^{168}$Yb | 1.381D-12 | 2.114D-13 | 1.705D-13 | 2.119D-13 | 1.692D-13 | 1.858D-13 |
| $^{174}$Hf | 1.430D-12 | 2.471D-13 | 2.049D-13 | 2.482D-13 | 1.955D-13 | 1.670D-13 |
| $^{180}$Ta | 1.282D-14 | 1.358D-14 | 9.225D-15 | 1.358D-14 | 7.716D-15 | 3.864D-15 |
| $^{180}$W | 9.861D-13 | 2.334D-13 | 1.949D-13 | 2.340D-13 | 1.848D-13 | 2.158D-13 |

Table 3—Continued

| Isotope | Solar System[a] | xi45[b] | xi25[c] | nocutoff[d] | ertl[e] | Nugrid[f] |
|---|---|---|---|---|---|---|
| $^{184}$Os | 5.040D-13 | 1.689D-13 | 1.344D-13 | 1.693D-13 | 1.324D-13 | 2.234D-13 |
| $^{190}$Pt | 1.041D-12 | 8.680D-14 | 7.500D-14 | 8.697D-14 | 7.223D-14 | 6.595D-14 |
| $^{196}$Hg | 5.369D-12 | 1.245D-11 | 9.223D-12 | 1.248D-11 | 8.476D-12 | 1.114D-12 |

(a) – Lodders et al. (2009)

(b) – Galactic chemical evolution results using yields from *xi45* KEPLER model.

(c) – Galactic chemical evolution results using yields from *xi25* KEPLER model.

(d) – Galactic chemical evolution results using yields from *nocutoff* KEPLER model.

(e) – Galactic chemical evolution results using yields from *ertl* KEPLER model.

(f) – Galactic chemical evolution results using yields from NUGRID model.

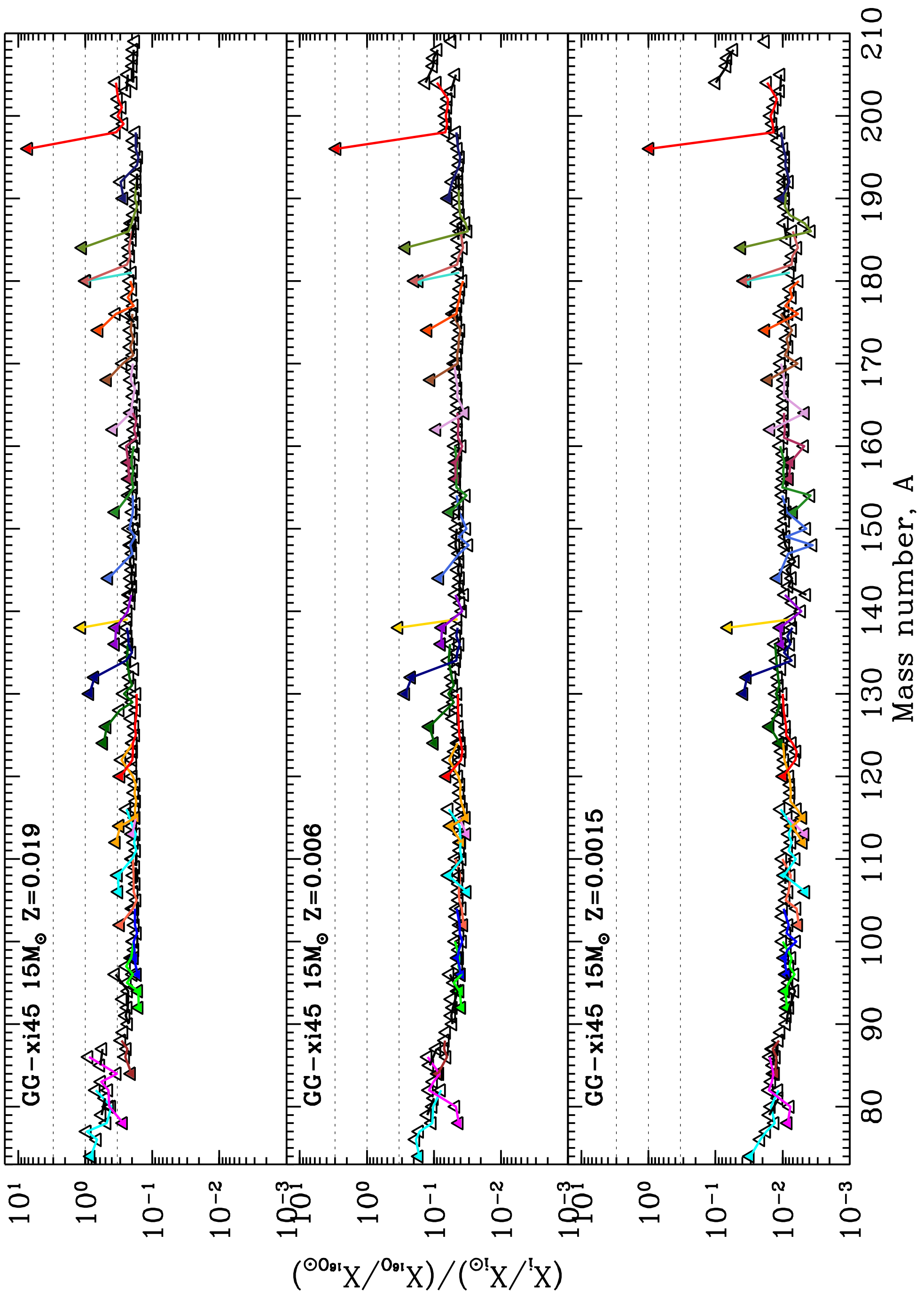

Fig. 1.—: Nucleosynthesis yields (overproduction factor normalized to $^{16}$O) for a $15M_{\odot}$ progenitor obtained with the KEPLER *xi45* ccSN model. The $p$-only isotopes are shown as *filled triangles*. *Solid lines* connect the isotopes of each element (see discussion in the text, Section 2). Three different metallicities are shown, $Z = 0.019$ (*upper panel*), $Z = 0.006$ (*middle panel*), $Z = 0.0015$ (*lower panel*).

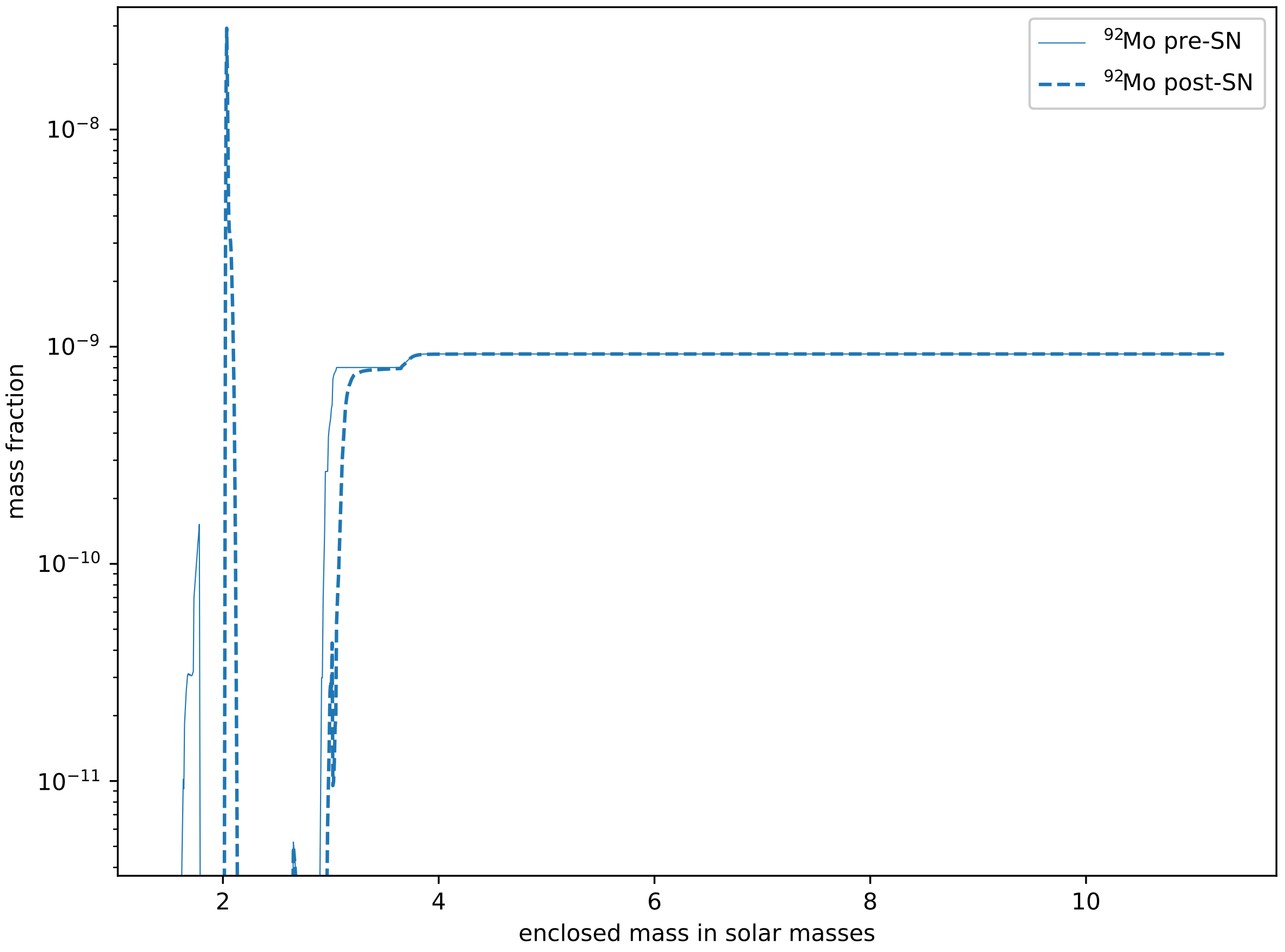

Fig. 2.—: $^{92}$Mo mass fraction as function of mass coordinate for a $15M_\odot$ *xi45* KEPLER progenitor of solar composition (see discussion in Section 2). The *thin solid line* is for the pre-explosive abundance, the *dashed line* is for the post-explosive abundance.

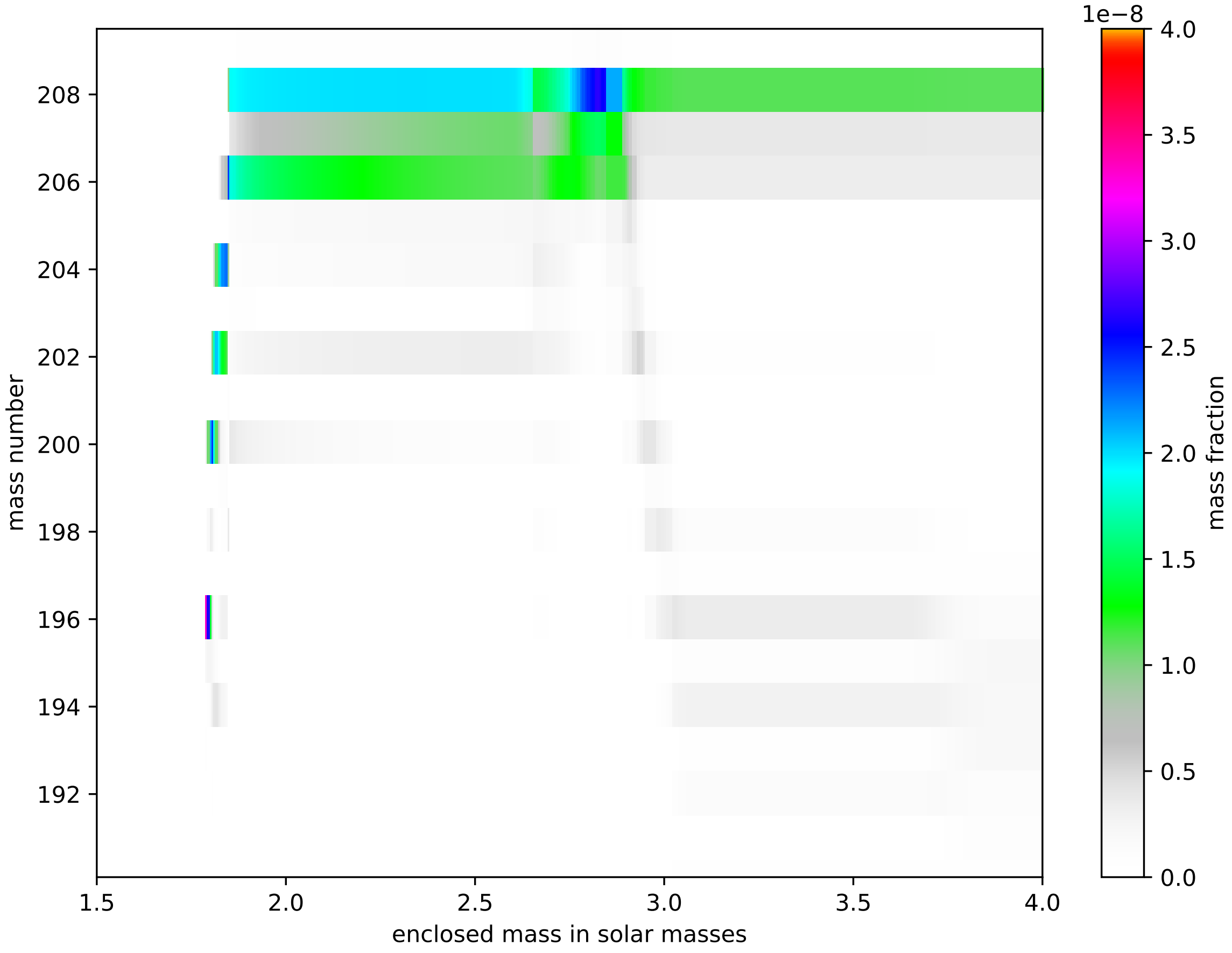

Fig. 3.—: Pre-supernova abundances of nuclides in the mass-number range $A = 190 - 209$ as function of mass coordinate in a $15 M_{\odot}$ *xi45* KEPLER model of solar initial composition (see Section 2 for details).

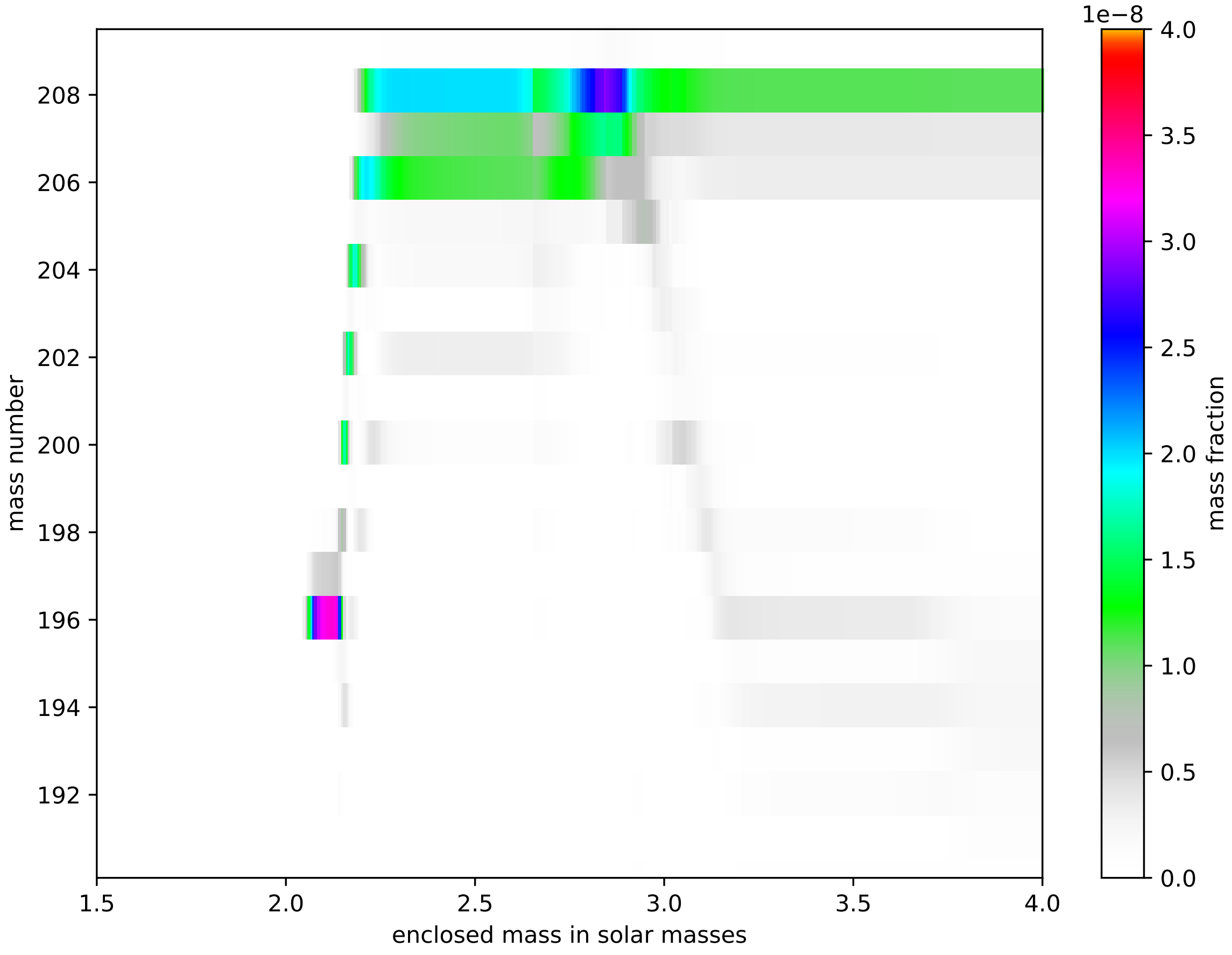

Fig. 4.—: Same as Fig. 3 but for the post-supernova phase.

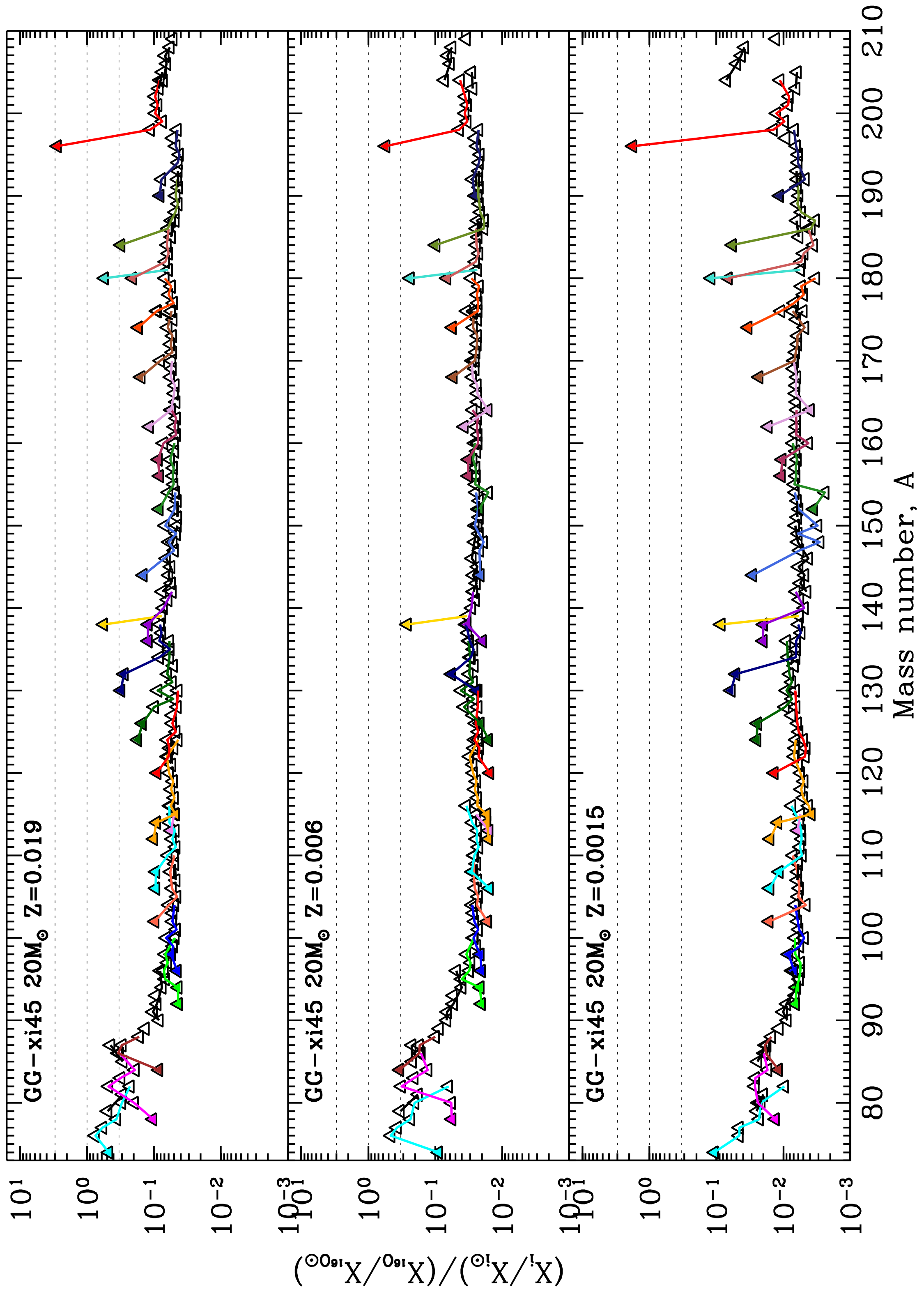

Fig. 5.—: Same as Fig. 1 for a $20M_{\odot}$ progenitor.

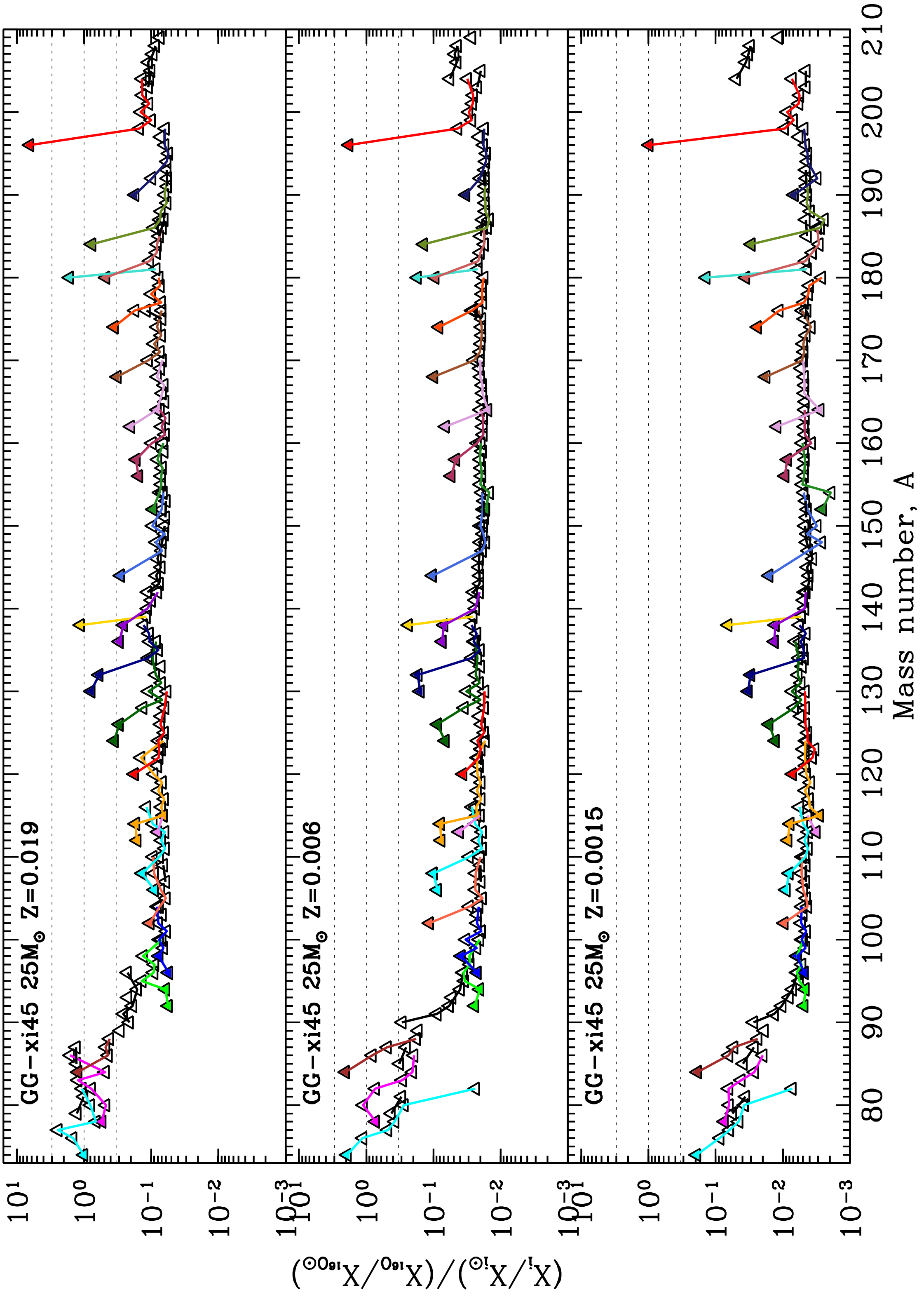

Fig. 6.—: Same as Fig. 1 for a $25 M_{\odot}$ progenitor.

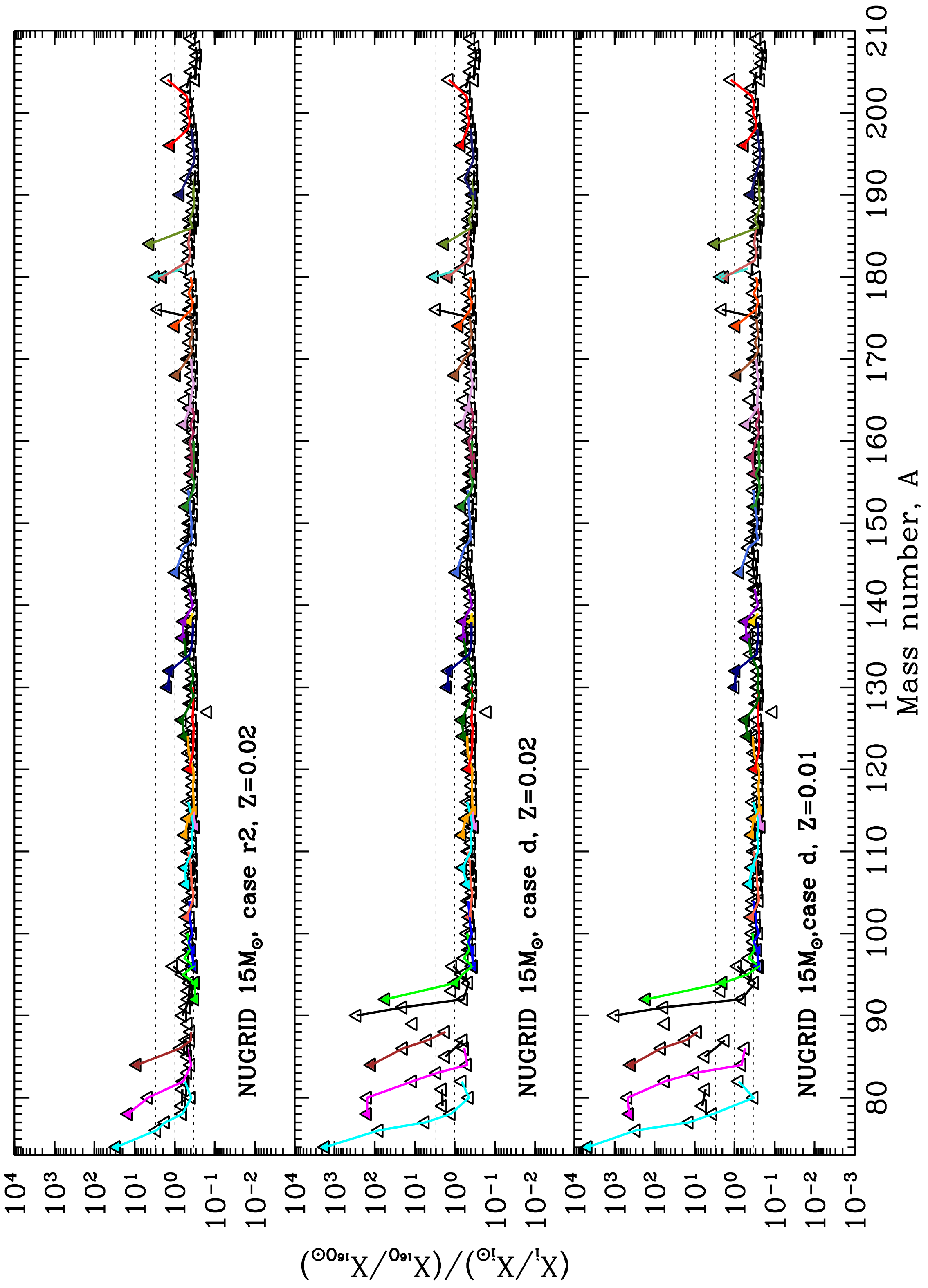

Fig. 7.—: Nucleosynthesis yields (overproduction factor normalized to $^{16}$O) for a $15 M_{\odot}$ NUGRID model (see Section 3 for details), *r2* and *d* cases. Results for two different metallicities are shown, for $Z = 0.02$ (*upper panel*) and for $Z = 0.01$ (*lower panel*). Filled triangles are for *p*-only isotopes.

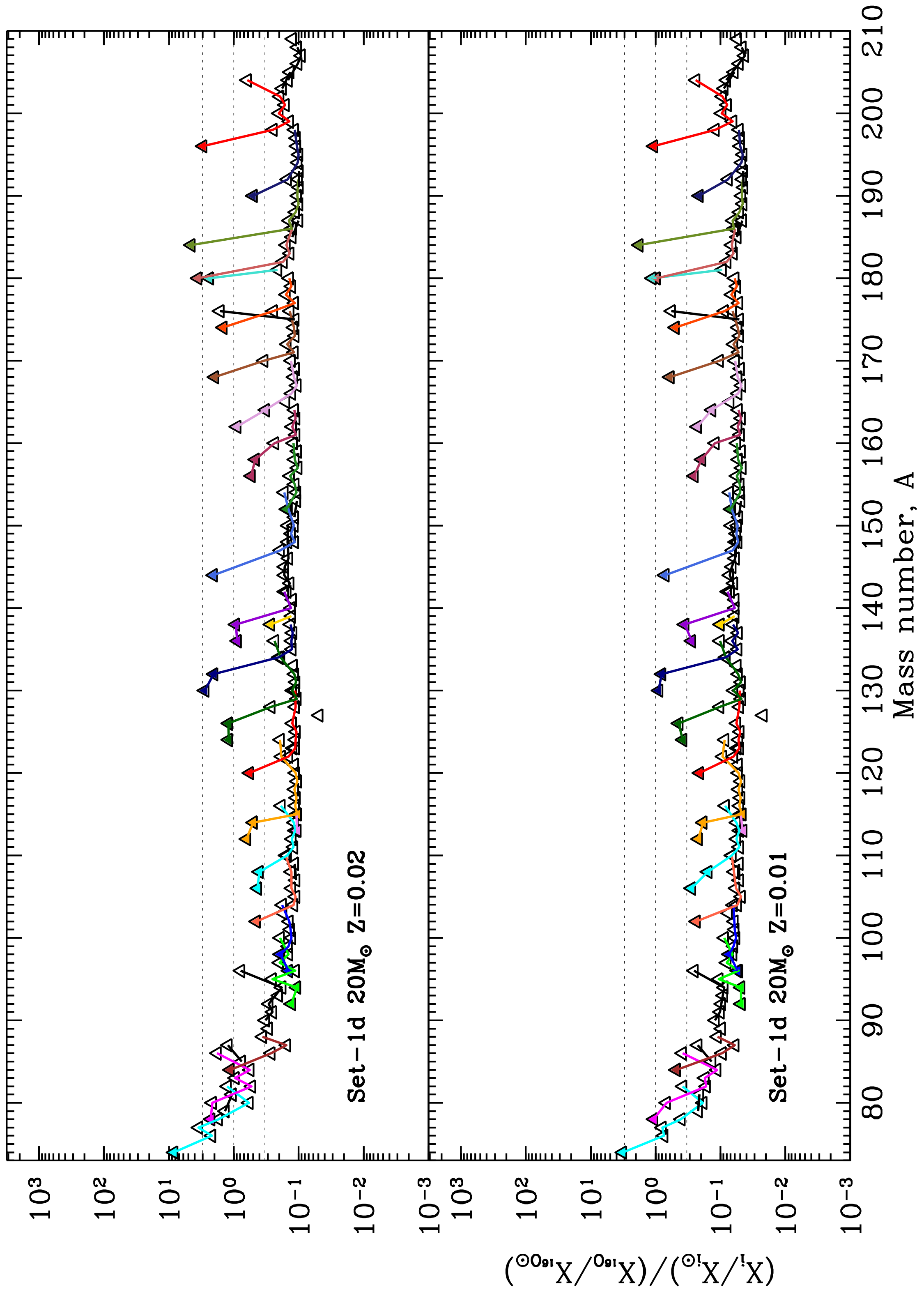

Fig. 8.—: Same as Fig. 7 for a $20M_{\odot}$ progenitor.

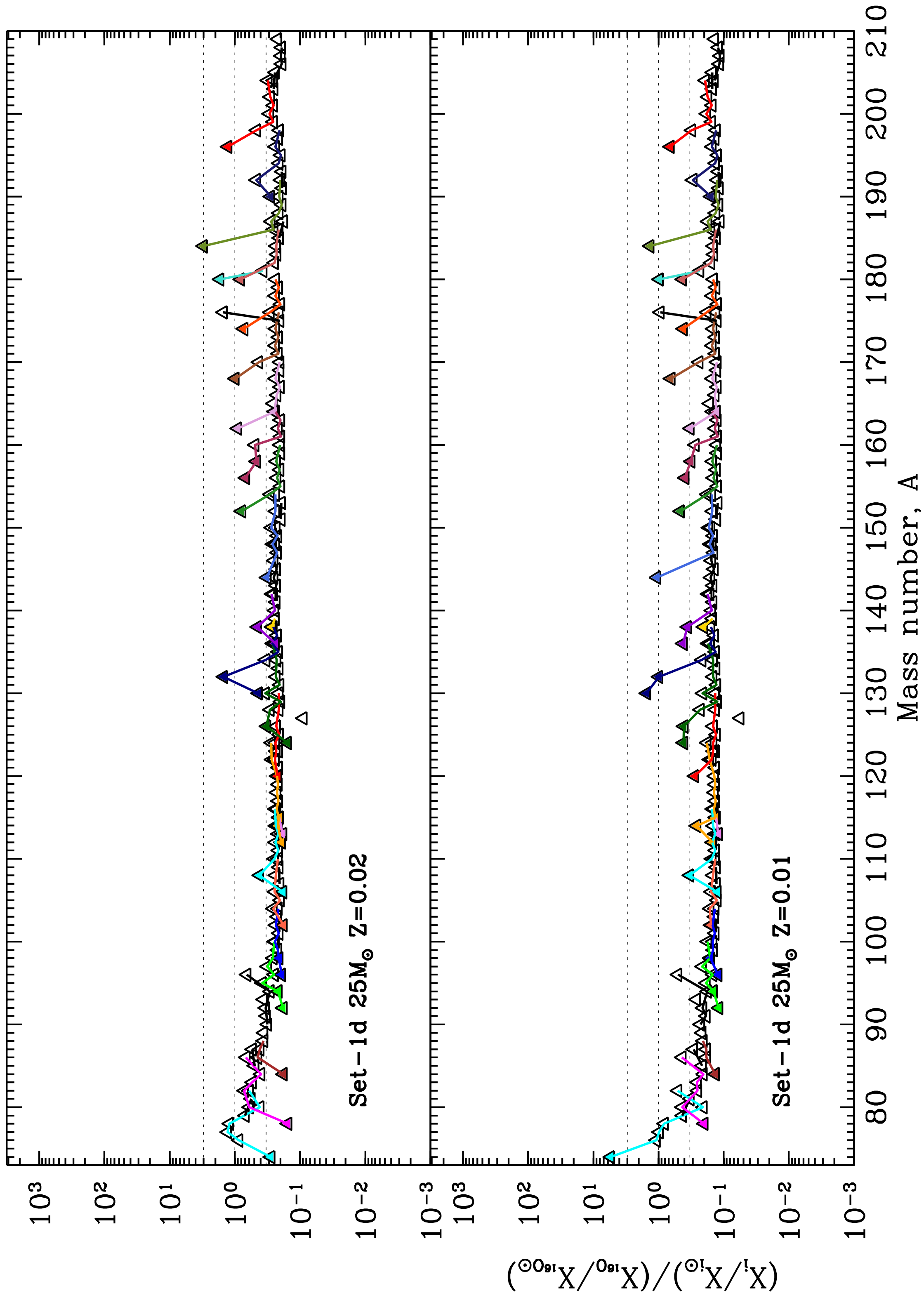

Fig. 9.—: Same as Fig. 7 for a $25M_{\odot}$ progenitor.

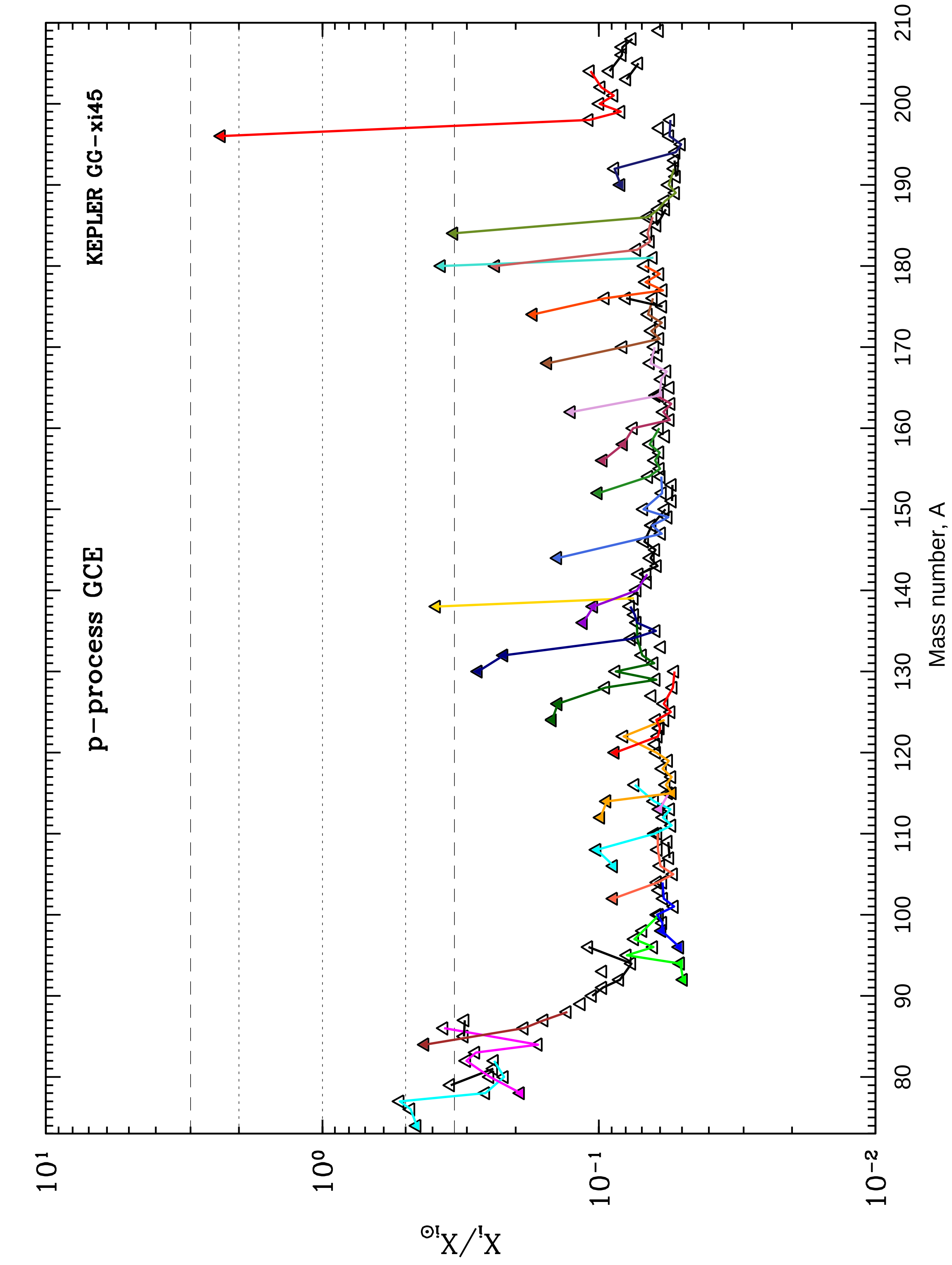

Fig. 10.—: GCE calculation using the yields from KEPLER *xi45* for the range of masses and metallicities discussed in the text. The filled triangles are *p*-only isotopes.

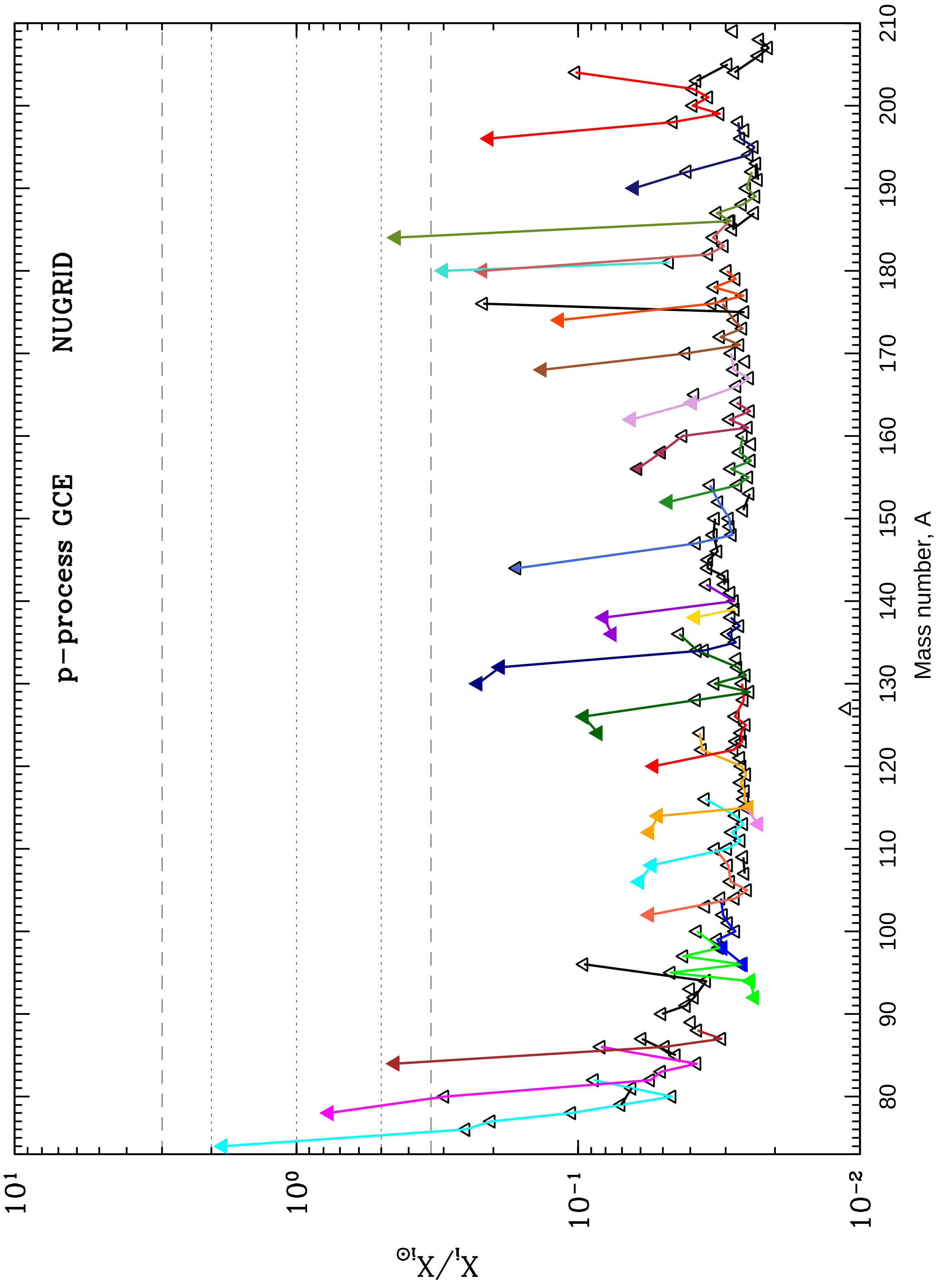

Fig. 11.—: GCE calculation using the yields from post-processed NUGRID models for the range of masses and metallicities discussed in the text. The filled triangles are *p*-only isotopes.

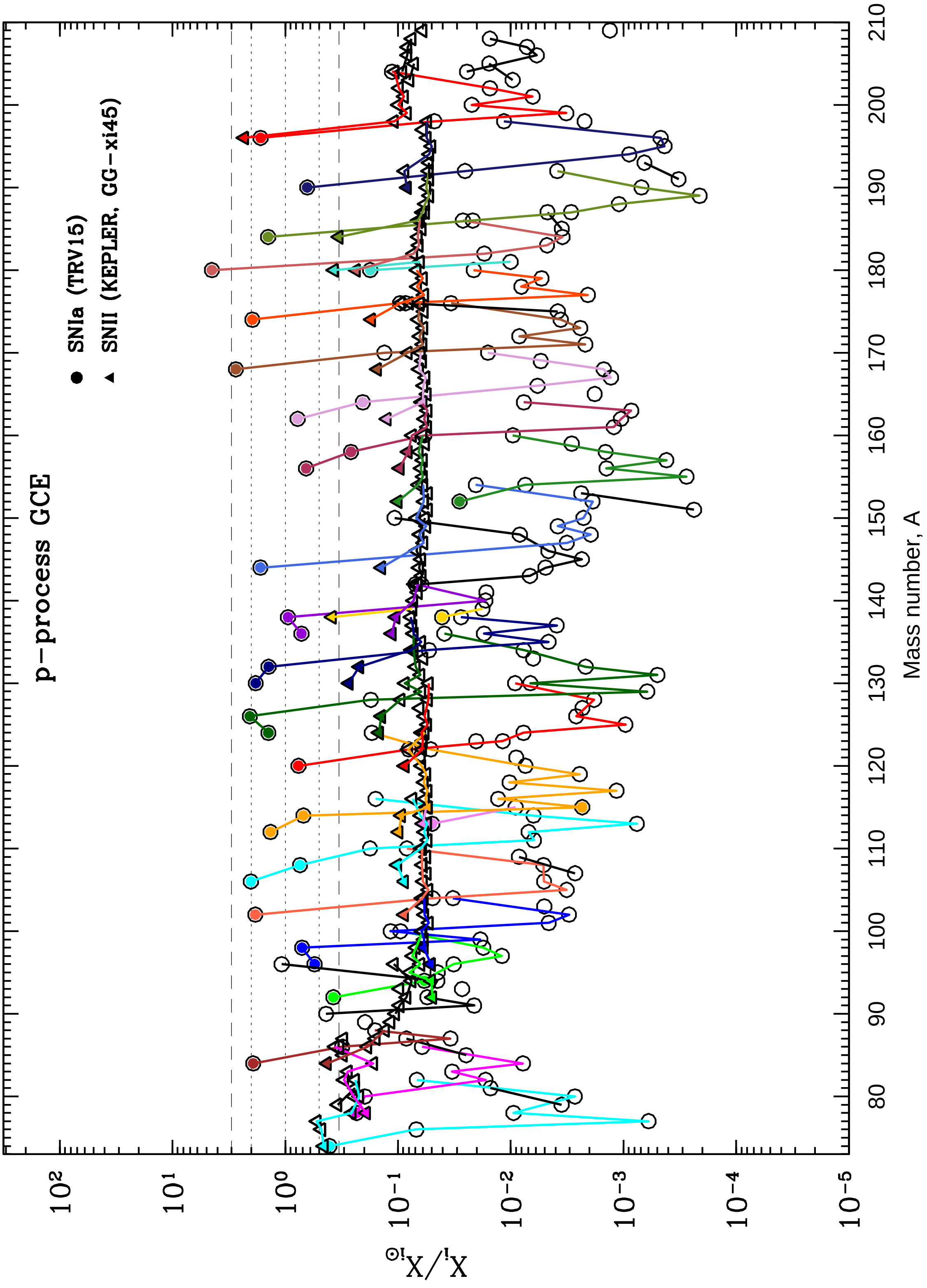

Fig. 12.—: Comparison of the GCE results when using the KEPLER *xi45* ccSN models presented in this paper (*triangles*) with the TRV15 results for SNIa (*circles*). The filled symbols are $p$-only isotopes.

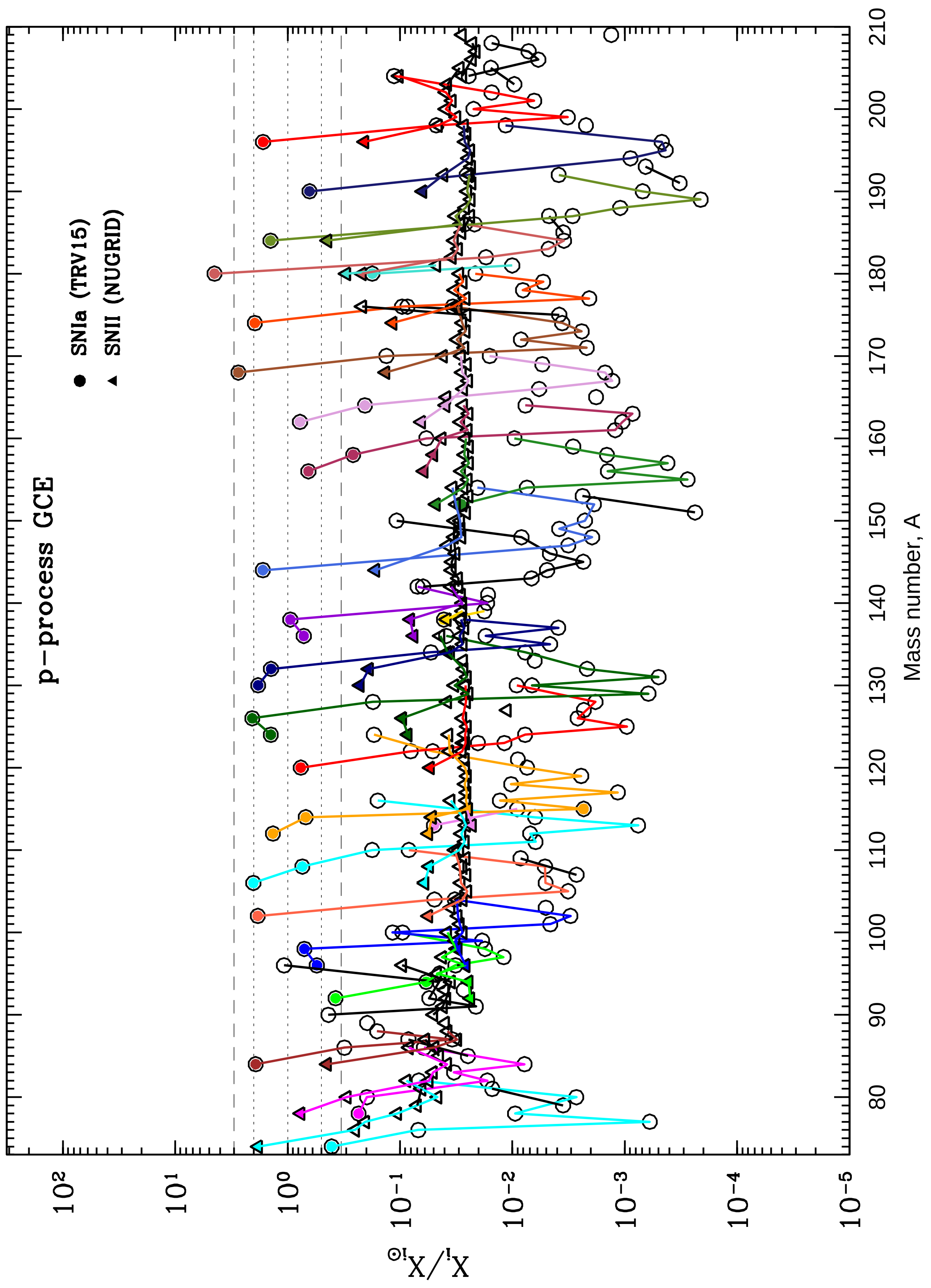

Fig. 13.—: Comparison of the GCE results when using the NUGRID ccSN models presented in this paper (*triangles*) with the TRV15 results for GCE taking into account SNIa contribution (*circles*). The filled symbols are $p$-only isotopes.